\documentclass[preprint,12pt,epsf]{aastex}
\usepackage{natbib}

\newcommand  \sii   {[\ion{S}{2}]}
\newcommand  \oiii  {[\ion{O}{3}]}
\newcommand  \ha    {H$\alpha$} 
\newcommand  \hb    {H$\beta$}
\newcommand  \hii   {\ion{H}{2}}

\newcommand  \heii   {\ion{He}{2}} 
\newcommand  \um    {$\mu$m}
\newcommand  \papillon {054004.40$-$694437.6}

\shorttitle{Young Massive Stars in N\,159}
\shortauthors{Chen et al.}

\begin{document}

\title{{\it Spitzer} View of Young Massive Stars in the LMC H\,II Complexes. II.
 N\,159}

\author{C.-H. Rosie Chen\altaffilmark{1}, Remy Indebetouw\altaffilmark{1,2},
 You-Hua Chu\altaffilmark{3}, Robert A. Gruendl\altaffilmark{3,4},
 G$\acute{e}$rard Testor\altaffilmark{5}, Fabian Heitsch\altaffilmark{6},
 Jonathan P. Seale\altaffilmark{3}, Margaret Meixner\altaffilmark{7}, and 
 Marta Sewilo\altaffilmark{7}}
\altaffiltext{1}{Department of Astronomy, University of Virginia, Charlottesville, 
 VA 22903, USA}
\altaffiltext{2}{National Radio Astronomical Observatory, Charlottesville, VA 22904, USA}
\altaffiltext{3}{Department of Astronomy, University of Illinois, Urbana, 
 IL 61801, USA}
\altaffiltext{4}{Visiting Astronomer, Cerro Tololo Inter-American Observatory.
 CTIO is operated by AURA, Inc.\ under contract to the National Science
 Foundation.}
\altaffiltext{5}{Observatoire de Paris, 92195 Meudon, France}
\altaffiltext{6}{Department of Physics and Astronomy, University of North 
Carolina at Chapel Hill, Chapel Hill, NC 27599, USA}
\altaffiltext{7}{Space Telescope Science Institute, Baltimore, MD 21218, USA}

\begin{abstract}

The \hii\ complex N\,159 in the Large Magellanic Cloud (LMC) is
used to study massive star formation in different environments,
as it contains three giant molecular clouds (GMCs) that have similar
sizes and masses but exhibit different intensities of star formation.
We identify candidate massive young stellar objects (YSOs) using infrared
photometry, and model their SEDs to constrain mass and evolutionary state.
Good fits are obtained for less evolved Type I, I/II, and II sources.
Our analysis suggests that there are massive embedded YSOs in N\,159B, 
a maser source, and several ultracompact \hii\ regions.
Massive O-type YSOs are found in GMCs N\,159-E and N\,159-W, which are 
associated with ionized gas, i.e., where massive stars formed a few Myr ago.
The third GMC, N\,159-S, has neither
O-type YSOs nor evidence of previous massive star formation.  
This correlation between current and antecedent formation of massive stars 
suggests that energy feedback is relevant.
We present evidence that N\,159-W is forming YSOs spontaneously, while 
collapse in N\,159-E may be triggered.
Finally, we compare 
star formation rates determined from YSO counts with 
those from integrated 
\ha\ and 24 \um\ luminosities and expected from gas surface 
densities.
Detailed dissection of
extragalactic GMCs like the one presented here is key to revealing
the physics underlying commonly used star formation scaling laws.

\end{abstract}

\keywords{\hii\ regions -- infrared: stars -- ISM: individual (N\,159)
 -- Magellanic Clouds -- stars: formation -- stars: pre-main sequence}

\section{Introduction}

Despite significant progresses in understanding the physics 
of single star formation, we still have only crude ideas about 
why some giant molecular clouds (GMCs) form clusters, some form
distributed associations, and some form very few stars.
To make headway, key properties of GMCs displaying various
intensities of star formation must be explored observationally 
in detail.  After massive stars are formed, their energy feedback
through ionizing fluxes, fast stellar winds, and supernova explosions 
can alter the physical conditions of a GMC, subsequently enhancing 
or inhibiting further star formation.  If a knowledge of stellar 
content and energy feedback is obtained, it is then possible to 
assess the relative importance of self-propagating, triggered, 
and spontaneous star formation, and even further estimate the star 
formation efficiency of a GMC.  It is thus crucial to inventory
massive stars and young stellar objects (YSOs) associated with
GMCs.  The YSOs are particularly important as they provide the
most direct probes for causal relationship between the initial
condition (gas) and the end product (stars).

Such study is straightforward in concept, but difficult in practice
because in distant galaxies the stellar content is not resolved, and
in the Milky Way the distances and associations among stars and their
interstellar environments are uncertain. 
The Large and Small Magellanic Clouds (LMC and SMC) are the only
nearby star-forming galaxies in which stars are at common, known 
distances and can be resolved individually.
Recent high-sensitivity mid-IR {\it Spitzer Space Telescope (SST)} 
observations make it possible to study a large sample of YSOs in the 
MCs \citep[e.g.,][]{WBetal08,GC09,Setal07}.
The nearly face-on orientation and thin disk structure of the LMC
make it easy to identify physical associations between YSOs and 
their interstellar environments.  All objects in the LMC can be
considered to be at a uniform distance, with $<$10\% error, of 
50 kpc \citep{Fe99}.
In our recent study of massive YSOs in the LMC \hii\ complex 
N\,44 \citep[][hereafter Paper I]{CCetal09}, we examined three GMCs
with different star formation properties: the central and southern 
GMCs are associated with bright \hii\ regions and have O-type YSOs, 
while the northern GMC contains only a few faint, small \hii\ 
regions and has no O-type YSOs.
The close association of massive YSOs with \hii\ regions in GMCs
suggests that massive stars preferentially form in regions where
massive stars have been formed previously, within the last few Myr.
It is not clear whether this result is a special case for N\,44 or 
can be generalized.

\begin{figure}
\plotone{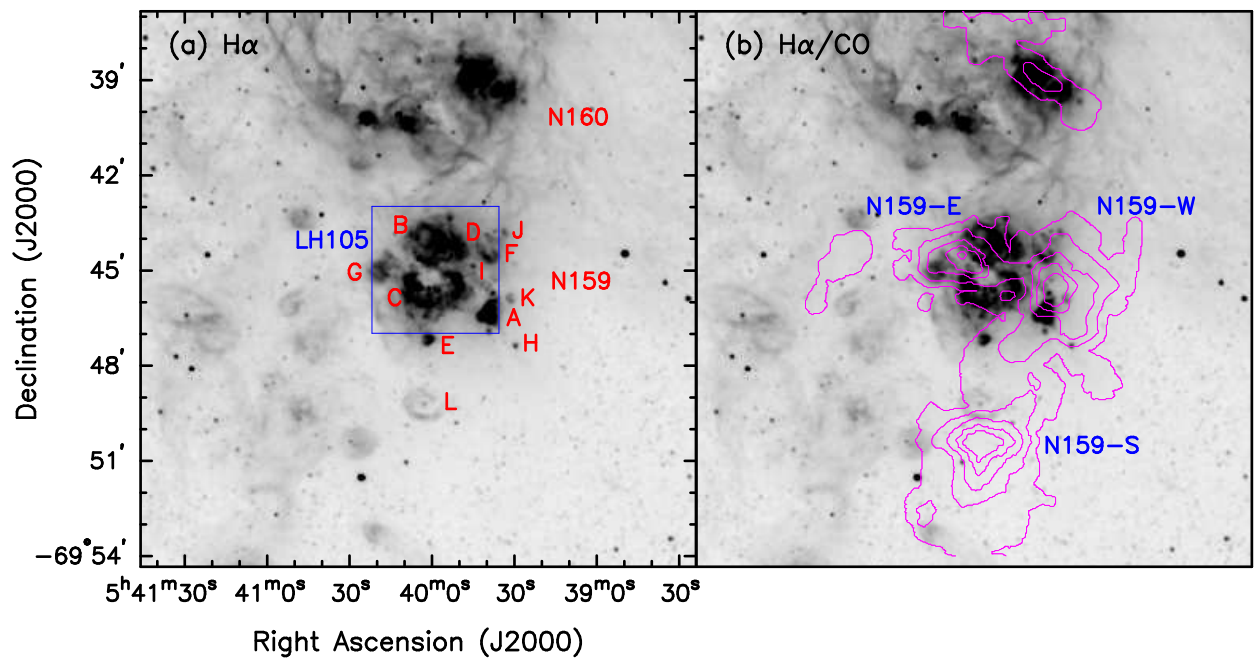}
\caption{MCELS \ha\ images of N\,159.
 (a) MCELS \ha\ image of N\,159, showing the nebular components A-L defined
 by \citet{He56} and the OB association LH105 cataloged by \citet{LH70}.
 (b) CO contours overlaid and three GMCs near CO peaks cataloged by 
 \citet{Joetal98} labeled on the MCELS \ha\ image.
 \label{fig:n159opt}}
\end{figure}

We have thus selected more LMC \hii\ complexes with different 
morphologies of ionized gas and molecular clouds to examine their 
YSO content and distribution.  
This study reports massive star formation in the \hii\ complex 
L\ha\ 120-N\,159 \citep[hereafter N159;][]{He56}.
N\,159 is located between two regions with contrasting activities:
to its north are a string of active star formation regions such as 
30 Doradus, N\,158, and N\,160, and to its south lies a large molecular
ridge with little star formation activity \citep[e.g.,][]{Inetal08}.
The ionized gas of N\,159 shows two prominent lobes in the central 
region, surrounded by smaller \hii\ regions (Figure~\ref{fig:n159opt}a).
The two prominent \hii\ lobes exhibit filamentary arcs resultant from
interactions with stellar winds and a supernova remnant (SNR) in the
northern lobe \citep{Chetal97,Wietal00}.  The northern and southern
lobes appear to be separated by dark clouds, but their independent
systems of circular filaments do not suggest a common origin for
their photoionization and dynamic shaping.
The stellar content of N\,159 has been identified loosely as
an OB association, i.e., LH105 \citep{LH70}.  Based on integrated 
colors and association with discrete \hii\ regions in N\,159, 
\citet{Bietal96} identified three young clusters of ages $<$ 10 Myr
in the A, F, and D components, and an OB association in the C 
component (see Figure~\ref{fig:n159opt}a for the identification of
\hii\ components).

Surveys of CO toward N\,159 (Figure~\ref{fig:n159opt}b) show three
molecular concentrations whose peaks correspond to three GMCs, 
N\,159-E, N\,159-W, and N\,159-S, respectively \citep{Joetal98}.
The similarity in their radial velocities, 
$V_{\rm lsr}=234.1-238.5$ km~s$^{-1}$, suggests that they are likely
physically related.
N\,159-E is associated with the central prominent \hii\ regions of
N\,159, the bulk of the OB association LH105 \citep{LH70}, which
include the young cluster in the D component.
N\,159-W, an immediate neighbor of N\,159-E, hosts a few small, 
bright \hii\ regions and the western part of LH105, which includes
the two young clusters in the A and F components.  
In sharp contrast, the GMC N\,159-S shows a much lower level of 
star formation with only a couple of very faint diffuse \hii\ 
regions and no OB associations, although it is similar in size 
and even larger in mass compared with each of its northern 
neighbors at $\sim 3$\arcmin\ away.
A recent study by \citet{Naetal05} identified candidate Herbig 
Ae/Be (HAeBe) stars in GMC N\,159-S and suggested that cluster 
formation had just begun in this GMC.  However, it is not clear 
whether the scarcity in massive stars seen in the optical and 
near-IR wavelengths in this GMC is caused by massive stars 
being still at infancy and deeply embedded in circumstellar
material, or not being formed actively at all.

To study the current massive star formation in N\,159, we have 
used archival {\it Spitzer} mid-IR observations and obtained 
complementary ground-based optical and near-IR observations.
These observations have been analyzed and the results are reported
in this paper.
Section~2 describes the observations and data reduction.
Section~3 reports the identification and classification of YSO candidates. 
In Section~4 we determine the physical properties of the YSOs by 
modeling their spectral shapes.
In Section~5 we discuss the massive star formation properties.
A summary is given in Section~6.

\section{Observations and Data Reduction}

We have used {\it Spitzer} mid-IR observations to diagnose YSOs.
To extend the spectral energy distribution (SED) and to improve the angular 
resolution, we have further obtained ground-based optical and near-IR 
imaging observations of N\,159.  
We have also retrieved available images in the {\it Hubble Space Telescope} 
({\it HST}) archive to examine the optical counterparts and environments 
of the YSOs.

\subsection{{\it Spitzer} IRAC and MIPS Observations}

Archival {\it Spitzer} observations from the Legacy Program Surveying 
the Agents of a Galaxy's Evolution \citep[SAGE;][]{MMetal06} and GTO 
Program 124 (PI: Gehrz) were used to study YSOs in N\,159.
These observations include images taken with the Infrared Array Camera
\citep[IRAC;][]{FGetal04} at 3.6, 4.5, 5.8, and 8.0 \um\ and the Multiband 
Imaging Photometer for {\it Spitzer} \citep[MIPS;][]{RGetal04} at 24, 70,
and 160 \um.
The observation summary is listed in Table~\ref{mirobs}, in which the 
program ID, principal investigator, and typical observation parameters 
are given.

We have adopted the {\it Spitzer} photometry of point sources from 
\citet{GC09}.
The photometric data are used to construct SEDs, and the images are 
used to examine the structure of cold and partially ionized 
interstellar matter.
To cover regions associated with the three GMCs around N\,159, the 
field we have analyzed is $12\arcmin\times$12\farcm5 and is shown in
Figure~\ref{fig:n159img} in the 3.6, 8.0, and 24 \um\ bands.
The 3.6 \um\ image is dominated by stellar emission, the 8.0 \um\
image shows the polycyclic aromatic hydrocarbon (PAH) emission,
and the 24 \um\ image is dominated by dust continuum emission 
\citep{LD01,LD02,DL07}.
To better illustrate the relative distribution of emission in
different bands, we have produced a color composite with 
3.6, 8.0, and 24 \um\ images mapped in blue, green, and red,
respectively.
In this color composite (Figure~\ref{fig:n159img}d), dust emission
appears red and diffuse, stars appear as blue point sources, red
supergiants appear yellow, and dust-shrouded YSOs and evolved 
stars appear red.

\subsection{CTIO 4~m ISPI Observations}

We obtained near-IR images in the $J$ and $K_s$ bands with the Infrared Side 
Port Imager \citep[ISPI;][]{vdBetal04} on the Blanco 4~m telescope at 
Cerro Tololo Inter-American Observatory (CTIO) on 2006 November 9.
The images were obtained with the $2{\rm K} \times 2$K HgCdTe HAWAII-2 
array, which had a pixel scale of 0\farcs3~pixel$^{-1}$ and a 
field-of-view of $10\farcm25 \times 10\farcm25$.
N\,159 was observed in one field with ten 30 s exposures in the $J$ 
band and twenty 30 s exposures in the $K_s$ band (each of the latter 
was coadded from two 15~s frames to avoid background saturation).
The observations were dithered to aid in the removal
of transients and chip defects.
The sky observations were made before and after each set of ten on-source
exposures.
All images were processed using the IRAF package \texttt{cirred} for 
dark and sky subtraction and flat-fielding.
The astrometry of individual processed images was solved with the 
routine \texttt{imwcs} in the package \texttt{wcstools}.
The astrometrically calibrated images are then coadded to produce a total 
exposure map for each filter.
The flux calibration was carried out using 2MASS photometry of 
isolated sources.

As some of the YSOs appear extended in the ISPI images, we have
used additional $J$ and $K_s$ images taken with the Nasmyth Adaptive Optics
System and Near-Infrared Imager and Spectrograph (NACO) on the 
8.2~m Very Large Telescope (VLT) at the European Southern Observatory 
(ESO) \citep{Teetal06,Teetal07}.
These images have a pixel size of 0\farcs027-0\farcs054 pixel$^{-1}$
and a superb resolution of 0\farcs11-0\farcs25, and
are thus used to resolve compact groups of YSOs.

\subsection{Archival {\it HST} Images}

We have searched the {\it HST} archive for Wide Field Planetary Camera 2 
(WFPC2) and Advanced Camera for Surveys (ACS) images in the field of N\,159.
These observations are listed in Table~\ref{hstobs}, in which the 
coordinate, program ID, PI, detector, filter, and total exposure time 
are given.
Each observation contains multiple exposures for the same pointing 
and filter, and these images have been pipeline-processed to remove 
cosmic rays and to produce a total exposure map, which is available
in the Hubble Legacy Archive\footnote{Based on observations made with 
the NASA/ESA Hubble Space Telescope, and obtained from the Hubble Legacy 
Archive, which is a collaboration between the Space Telescope Science 
Institute (STScI/NASA), the Space Telescope European Coordinating Facility 
(ST-ECF/ESA) and the Canadian Astronomy Data Centre (CADC/NRC/CSA).}.

Four fields have observations, but not all are useful; for example,
the wide $U$ band (F300W) images have very low S/N.
The most useful images are those taken with the \ha\ filter 
(WFPC2/F656N or ACS/F658N) or broad-band filters at longer wavelengths 
such as Str\"omgren $y$ (F547M) or WFPC2 $VI$ (F555W and F814W).
The former show ionized gas and the latter show stars at high resolution.

\subsection{Other Available Datasets}

To construct SEDs for sources in the {\it Spitzer} catalog from
\citet{GC09}, we have expanded the catalog by adding optical 
$UBVI$ photometry from the Magellanic Cloud Photometric Survey 
\citep[MCPS;][]{ZDetal04} and near-IR $JHK_s$ photometry from 
the Point Source Catalog of the Two Micron All Sky Survey 
\citep[2MASS;][]{SMetal06}.
As the 2MASS catalog is relatively shallow with a photometric limit
of $K_s \sim 14.5$, we have also used the point source catalog from 
the InfraRed Survey Facility \citep[IRSF;][]{Kaetal07}, which is 
$\sim$ 2 mag deeper than the 2MASS catalog, to match the IRAC sources 
that do not have 2MASS counterparts.
When merging the datasets, we allow a 1$''$ error margin for matching
{\it Spitzer} sources with optical or near-IR sources.
Thus, for each {\it Spitzer} source, the magnitudes from $U$ to 70 \um\ 
can be converted to flux densities using the corresponding zero-magnitude 
flux listed in Table~\ref{photpar} and then used to construct the SED.

We have used \ha\ images of N\,159 from the Magellanic Cloud 
Emission-Line Survey \citep[MCELS;][]{SRetal99} to examine the 
large-scale distribution of dense ionized gas and to compare with 
images at other wavelengths.
As the angular resolution of this survey is $\sim2''$, we have
used additional \ha\ images taken with the MOSAIC camera
on the CTIO Blanco 4~m telescope that cover the northern part
of our working field \citep{SFetal10}.
These images, with a pixel size of 0\farcs27 pixel$^{-1}$, are used 
to show the immediate environments of YSOs.
Finally, as stars are formed from molecular gas, we have also used 
the CO observations from the Magellanic Mopra Assessment 
\citep[MAGMA;][]{Otetal08,Huetal10} to examine the distribution
of molecular clouds in N\,159.

\section{Identification of Massive YSOs}

\subsection{Selection of Massive YSO Candidates}
YSOs are enshrouded in dust that absorbs stellar UV and optical
radiation and re-radiates in the IR.
Thus, YSOs can be distinguished from normal stars, i.e., 
main-sequence, giant, and supergiant stars, from their excess IR 
emission; they are positioned in redder parts of the color-color 
and color-magnitude diagrams (CMDs) than normal stars.
However, background galaxies and evolved stars such as 
asymptotic giant branch (AGB) stars can also be red sources, 
and these contaminants exist in non-negligible numbers.
In Paper I and \citet{GC09}, we have demonstrated that using 
two color-magnitude criteria $([4.5]-[8.0]) \ge 2.0$ and $[8.0] < 
14 - ([4.5]-[8.0])$, galaxies and evolved stars can be effectively 
excluded for the initial selection of massive YSOs.
We have thus applied the same criteria to select YSO candidates 
in N\,159. 
Note that although the full LMC catalogs of YSO candidates have 
been constructed using similar methods \citep{WBetal08,GC09}, 
as will be discussed in \S3.4, a detailed analysis on small regions 
together with new information such as deep near-IR photometry as 
we perform here will be more complete and also useful in diagnosing 
sources that might have been previously mis-identified.

\begin{figure}
\plotone{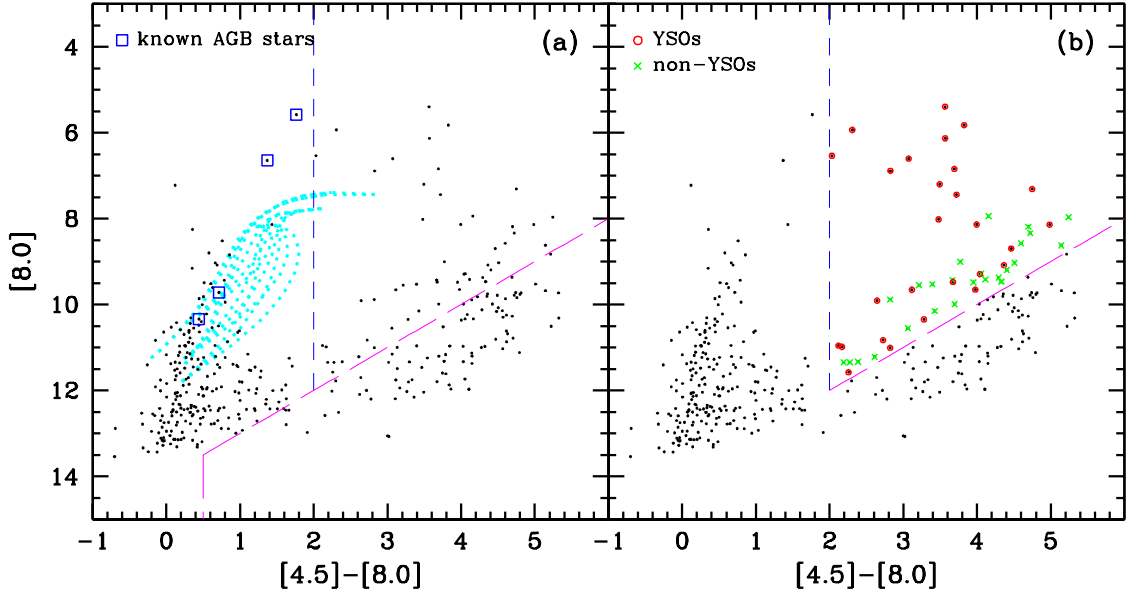}
\caption{(a) [8.0] vs.\ [4.5]$-$[8.0] CMD of all sources detected in N\,159. 
 Known AGB stars are marked with open blue squares and expected loci from AGB
 stellar models \citep{Gr06} with filled cyan squares.  The criterion to
 exclude normal and AGB stars is shown in short-dashed lines and that to
 exclude background galaxies in long-dashed lines. (b) 52 YSO candidates are
 found in the upper right wedge that has the minimum contamination from stars
 and background galaxies.  These candidates have been through detail
 examination using multi-wavelength images and SEDs.  Candidates that are
 most likely YSOs are marked with red open circles and non-YSOs with green
 crosses.
 \label{fig:cmds}}
\end{figure}

Figure~\ref{fig:cmds}a displays the [8.0] versus ([4.5]$-$[8.0]) 
CMD of all sources detected in N\,159.
To illustrate the locations of AGB stars in the CMD, models for
Galactic C- and O-rich AGB stars \citep{Gr06} are also plotted.
Although the LMC metallicity is only $1/3$ solar, as confirmed
in \citet{Bletal06}, these Galactic models should be good 
approximations for LMC objects since the chemistry of AGB atmospheres 
is dominated by nucleosynthesized material.
To avoid crowding, we plot only models for a stellar luminosity
of 3000 $L_\odot$ to illustrate the range of colors.
For a luminosity range $1\times10^3-6\times10^4 L_\odot$ 
\citep{Po93}, the expected loci of AGB stars in the CMD can
move vertically from 1.2 to $-3.3$ mag.
We have also searched the known evolved stars in N\,159 in 
the SIMBAD database and found 4 confirmed AGB stars 
at various evolutionary stages: 2 carbon stars and 2 OH/IR
stars.
These 4 objects are marked with open squares in the [8.0] 
versus ([4.5]$-$[8.0]) CMD (Figure~\ref{fig:cmds}a).
Their locations overlap with AGB stellar models and our
selection criteria indeed exclude them.

We obtain a list of 52 YSO candidates from the above selection 
criteria, however, they still include a significant number of small 
dust knots, obscured evolved stars, and bright background galaxies.
These contaminants need to be examined closely to assess their 
nature and to be excluded from the YSO list.
Following the procedures outlined in Paper I, to better resolve 
these IR sources and their environments, we examine the {\it HST}, 
CTIO 4m ISPI $JK_s$, and CTIO 4m MOSAIC \ha\ images.
We further use the multi-wavelength catalog described in \S2.4 to
construct SEDs from the $U$-band to 70 \um.
Using these images and SED data, we have assessed whether 
the candidates are truly YSOs.

\subsection{Identification of Contaminants}

Background galaxies can be identified from their morphologies 
if they are resolved.
Two of our CMD-selected YSO candidates, sources 054037.09$-$694521.5
and 054044.66$-$694550.9 show elongated, extended emission around 
the point sources in high-resolution MOSAIC and ISPI images.
The SEDs of these two sources are similar to those of late-type
galaxies, i.e., characterized by two broad humps, one from stellar 
emission over optical and near-IR wavelength range and the other 
from dust emission over mid- to far-IR range.
In addition, the interstellar environment around these two sources
are relatively dust free as they have little diffuse 8 \um\ emission
in the surroundings.
Based on these considerations, these two CMD-selected YSO candidates 
are reclassified as background galaxies.

Warm interstellar dust may show mid-IR SEDs indicative of PAH 
emission \citep[e.g.,][]{GVetal04}, similar to those of circumstellar 
dust in YSOs.
As the angular resolution of IRAC images is $\sim 2''$, corresponding 
to 0.5 pc for a LMC distance of 50 kpc, some small dust clumps may be 
identified as point sources and included in the YSO candidate list.
In addition, a main-sequence star projected near a dust clump or 
superposed on dust filaments may also make its way into our YSO 
candidate list.
To identify these two types of YSO imposters, we use our
high angular resolution (0\farcs3~pixel$^{-1}$) ISPI $JK_s$ images.
In the ISPI images, stars or YSOs appear unresolved and can
be further differentiated with the $J - K_s$ color as stars are 
brighter in $J$ and YSOs brighter in $K_s$.
Dust clumps may appear as extended emission or have no point-like 
counterparts in the ISPI $K_s$ image.
These dust clumps are unlikely to have highly embedded massive YSOs 
since none of them have bright 24 \um\ point sources as would have 
been expected for such YSOs; however, they may have YSOs with lower
masses. 
As a reference, Herbig Ae stars of A0-type at different evolutionary 
stages would have been detected, because AB Aur and Lk \ha\ 25
\citep{HLetal92} placed at the LMC distance would have $K_s = 16.9$
and 18.6, respectively, and the photometric limit of our $K_s$ 
images at 10-$\sigma$ level is $\sim 17.6$ mag and the 3-$\sigma$
detection limit is $\lesssim 19.0$ mag \citep{GC09}.
Aided by the ISPI images, we find that 16 of our 52 YSO candidates 
are interstellar dust clumps, and another 7 are stars projected
near dust clumps.

\subsection{Massive YSOs and Their Classification}
The results of our examination of 52 YSO candidates are given in 
Table~\ref{ysoclass}, which lists source name, ranking of the 
brightness at 8 \um, magnitudes in the 3.6, 4.5, 5.8, 8.0, 24, 
and 70 \um\ bands, source classification, and remarks.
Each source's magnitudes in the $U$, $B$, $V$, $I$, $J$, $H$, $K_s$ 
bands are given in Table~\ref{ysoclassb}.
Note that some of the YSOs appear as single sources in IRAC 
images but are resolved into multiple sources in high-resolution
ISPI and VLT/NACO images; for these sources, the photometric 
measurements in Table~\ref{ysoclassb} were made for the dominant 
YSO sources, i.e., the brightest $K_s$ sources with the reddest
$J-K_s$ colors.  
The ``Flag'' in Table~\ref{ysoclassb} gives the origin 
of data from which $JHK_s$ are measured.
The results of YSO identification are shown in the [8.0] versus 
([4.5]$-$[8.0]) CMD (Figure~\ref{fig:cmds}b).
Among the 52 YSO candidates, we identified 2 background galaxies, 
16 dust clumps, and 7 stars superposed on dust.
After excluding these sources, 27 YSO candidates remain.  
As these are most likely bona fide YSOs, we will simply call them 
YSOs in the rest of the paper. 

Following the classification scheme proposed in Paper I, we have 
further categorized these YSOs into Type I, II, and III by analyzing 
their properties and corresponding SED plots.
Type I YSOs have SEDs with a steep rise from the near-IR to 24
\um\ as the radiation is mostly from their circumstellar envelopes.
They are generally not visible in the optical or $J$-band, but are
visible in the $K_s$-band and bright at 24-70 \um.  Type I YSOs
are often found in or behind dark clouds.
Type II YSOs have SEDs with a low peak in the optical and a high
peak at 8-24 \um\ corresponding to stellar core and circumstellar
disk, respectively, after the envelope has dissipated.
Type II YSOs are faint in the optical, but bright in the $J$-band to
8 \um, and then faint again at the 24 \um. 
Type III YSOs have SEDs peaking in the optical with a modest
amount of dust emission in the near- to mid-IR as they are largely 
exposed but still possess remnant circumstellar material.
Their brightness fades in the longer wavelength and they are
often surrounded by \hii\ regions. 
The results of our classification of 27 YSOs are also given in 
Table~\ref{ysoclass}.
Multi-wavelength images and SEDs of these YSOs have been shown in 
\citet{GC09}, though differences exist in some SEDs between theirs 
and this study due to the inclusion of aforementioned deeper 
near-IR catalogs and that 24 \um\ upper limits were estimated 
for a few YSOs in order to better constrain model fitting to 
these SEDs as discussed later in \S 4.

\subsection{Comparison with YSO Samples Identified by Others}

N\,159 contains the first extragalactic protostar discovered by
\citet{Gaetal81} using ground-based near- to mid-IR $JHKL'$
photometry; this protostar corresponds to our YSO 053959.34$-$694526.3.
Using sensitive {\it Spitzer} observations, more YSOs in N\,159 
have been identified using various methods by \citet{JTetal05}, 
\citet{WBetal08}, and \citet{GC09}.
\citet{GC09} used the same {\it Spitzer} data and selection criteria,
and identified 25 YSOs.
Our assessments retain two more sources in the YSO sample, 
053952.60$-$694517.0 and 054050.85$-$695001.9, which were classified 
as a diffuse source and a background galaxy in their list.
The reclassification is aided by the ISPI $JK_s$ images and IRSF
$JHK_s$ photometry. 
For source 053952.60$-$694517.0 which was previously considered as 
(bright) stars superposed on a dust clump, the ISPI images show a 
faint red source ($K_s \sim 17.0$ and $J-K \sim 0.3$) resolved from 
nearby bright stars within the IRAC PSF.
For source 054050.85$-$695001.9 whose SED was once composed of only 
a nearly flat slope over 3.6--24 \um, the inclusion of 
IRSF $JHK_s$ photometry makes its SED (Figure~\ref{fig:fit}) more 
consistent with a YSO than a galaxy.

\citet{JTetal05} identified YSO candidates by comparing source 
locations in the ([3.6]$-$[4.5]) versus ([5.8]$-$[8.0]) 
color-color diagram with those  expected from low-mass YSO models
\citep{ALetal04} .
They studied a smaller region centered on the prominent central
\hii\ regions, within which we find 21 YSOs while they find only 
five YSO candidates.
In Figure~\ref{fig:ysocom}a, we mark the locations of their five YSO
candidates and our YSOs in the 8 \um\ image of N\,159, and in
Figure~\ref{fig:ysocom}b we compare their sources with ours in the
[8.0] versus ([4.5]$-$[8.0]) CMD.
All of their five YSO candidates fall within the YSO wedge bounded by 
[4.5]$-$[8.0] $\ge$ 2.0 and [8.0] $\ge 14.0-$([4.5]$-$[8.0]), and
form a subset of our YSO sample.
To investigate why the majority of our YSOs in N\,159 are missed 
in their study, we further compare our sources with their photometric
catalog \citep[Tables 1 and 2 in][]{JTetal05}.
We find that the discrepancy between our and their YSO lists is 
attributed to both different selection criteria for point sources  
and different criteria for YSOs.
Jones et al.\ used stringent parameters in the automated search 
for point sources, discarding any source that appear irregular
compared with the IRAC PSF; this omitted eight of our YSOs,
including three bright sources with [8.0] $< 8.0$.
Only 13 of our 21 YSOs were identified as point sources in
their catalog, and only five of these were also classified as YSOs
by Jones et al.
Of the remaining eight point sources, four were classified by
Jones et al.\ as \hii\ regions because their 8 \um\ images were 
slightly more extended than the PSF; three sources were not 
classified as they did not have photometric measurements in 
all four IRAC bands; and one was not selected as YSO because
its [5.8]$-$[8.0] color was $\sim 0.2$ mag bluer than those 
expected for Class I low-mass YSOs.
However, we find the SED of this last source, 053940.78$-$694632.0 in 
Figure~\ref{fig:fit}, to be consistent with those of bona fide YSOs. 

The comparison with the YSO sample of \citet{WBetal08} is not
straightforward as they used a complex set of criteria.  
Within N\,159, they found only 4 YSO candidates.
These four YSO candidates are also marked in Figure~\ref{fig:ysocom}.
The differences between our and their YSOs can be summarized as the 
following: \begin{enumerate}
\item  Within the YSO wedge bounded by [4.5]$-$[8.0] $\ge$ 2.0
and [8.0] $\ge 14.0-$([4.5]$-$[8.0]), we find 27 YSOs, while
\citet{WBetal08} find only 3 YSO candidates, a subset of our sample.
To investigate why the majority of the YSOs in N\,159 are missed in 
Whitney et al., we further compare our sources with their original catalog, 
i.e., the SAGE point source catalog.
We find that only one missed YSO results from different selection criteria,
and the rest from defining point sources and applying signal-to-noise (S/N) 
thresholds when making final lists in these two studies.
The SAGE catalog, similar to Jones et al., used stringent parameters 
for the automated search for point sources and Whitney et al.\ applied 
a high S/N threshold to select YSO candidates.
The former discards any irregular point sources compared to IRAC PSFs, 
resulting in only 14 of our 27 YSOs retaining in the SAGE catalog.
The latter tends to exclude sources near bright neighbors or over bright 
background, since varying backgrounds (Figure~\ref{fig:ysocom}a) raises 
photometric uncertainties and the SAGE pipeline reports conservative S/N 
ratios.
The S/N ratios are further reduced by applying additional 10\% and 30\% 
uncertainties respectively to 5.8 and 8.0 \um\ measurements 
\citep{WBetal08}, resulting in only 4 of the 14 sources being included 
in their list of YSO candidates.
One of these 4 sources does not match their set of  selection criteria
and hence is not included as a YSO in the \citet{WBetal08} catalog. 
However, the SED of this source, 054000.69$-$694713.4 in Figure~\ref{fig:fit}, 
is consistent with bona fide YSOs. 
\item One YSO candidate, W991 (SSTISAGE1CJ053945.37$-$694802.3), 
of the Whitney et al.\ sample was rejected by our more restrictive 
color-magnitude criteria for YSO selection that are intended to increase 
reliability though at the expense of missing some of the more evolved YSOs.  
\end{enumerate}

The photometric extraction method used for this work 
\citep[from][]{GC09} and Paper I has allowed the inclusion of 
marginally extended sources and sources amid a complex 
interstellar background in our initial photometric catalog.
Considering that $1''=$0.25 pc in the LMC, it is often inevitable
that within the IRAC PSF a YSO is blended with circumstellar
outflows or interstellar features.
Furthermore, YSOs in multiple systems or YSOs in ultracompact
or compact \hii\ regions (see \S4.2.2 for a detailed discussion)
are likely to be measured as a single source.
While it is necessary to relax the search criteria for point sources
in the LMC to obtain the most complete census of YSOs possible,
the relaxed search also requires that individual sources are examined
carefully in order to remove diffuse sources or local peaks in the 
dusty interstellar medium.

The color criteria commonly used for identifying low-mass YSOs,
as applied in \citet{JTetal05}, are too stringent for massive
YSOs, particularly when they are at early evolutionary stages.
As shown in models of massive YSOs, such emission is dominated 
by massive envelopes and hence SEDs peak toward wavelengths longer 
than 24 \um. 
These SEDs have a generally rising slope at shorter wavelengths 
such as IRAC bands of 3.6--8.0 \um, but show structures depending 
on, e.g., the detailed geometry of the dust distribution 
\citep{WBetal04b}.
The YSO selection method used in our study should make our YSO
list relatively complete for the region of the [8.0] vs.\ 
([4.5]$-$[8.0]) CMD analyzed.
It will however miss sources outside this region, i.e., the 
less embedded, more evolved, or less massive ($\lesssim 4 M_\odot$)
YSOs, as discussed in Paper I.
The \citet{WBetal08} analysis covers bluer and fainter
regions in the [8.0] vs.\ ([4.5]$-$[8.0]) CMD, and thus
may find YSOs that we have not analyzed, but confusion
from evolved stars and background galaxies in these regions
is more severe and careful consideration of complementary 
information is needed to distinguish between YSOs and contaminants.

\section{Determining YSO Properties from Model Fits of SEDs}
\subsection{Modeling the SEDs}
The observed SED of a YSO can be compared with model SEDs,
and the best-fit models selected by $\chi^2$ statistics can
be used to infer the probable ranges of physical parameters
for the YSO.  
We use a grid of dust radiative transfer models calculated with 
the HOCHUNK Monte-Carlo code \citep{BAW1}, each containing a point 
source photospheric emitter (pre-main-sequence or main-sequence spectrum), 
a flared circumstellar disk, a flattened rotating envelope, and 
a bipolar cavity \citep[see][for details]{RTetal06}.
The fitting can be carried out with the fitting tool Online SED 
Fitter\footnote{The Online SED Fitter is available at 
http://caravan.astro.wisc.edu/protostars/sedfitter.php.} 
\citep{RTetal07}.
The input parameters of the SED fitter include the fluxes of
a YSO and uncertainties of the fluxes.
The uncertainty of a flux has two origins: the measurement
itself and the absolute flux calibration. 
The fluxes and their measurement errors are given in 
Tables~\ref{ysoclass} and \ref{ysoclassb}.
The calibration errors are 10\% in $U$, 5\% in $B$, $V$, 
10\% in $I$, $J$, $H$, $K_s$, 5\% in  3.6, 4.5, 5.8, and 8.0 \um, 
10\% in 24 \um, and 20\% in 70 \um\ (Zaritsky et al.\ 2004; 
Skrutskie et al.\ 2006; Kato et al.\ 2007; IRAC Data Handbook; 
MIPS Data Handbook).
The total uncertainty of a flux is thus the quadratic sum 
of the measurement error and the calibration error.
We have used the above SED fitting tool to analyze the 27
YSOs in our sample.
22 of them appear single or are clearly the dominant source 
within the IRAC PSF, making comparisons with single-YSO models 
plausible.
The remaining five either have no dominant YSOs or contain 
luminous non-YSO sources such as background stars or dust clumps 
within the IRAC PSF, making single-YSO models suspect.
Although the total luminosities from well-reproduced SED fits 
to these five YSOs should still be good approximations, given 
the complicated nature, they are not discussed further in 
detailed comparisons with models.

The best-fit and acceptable SED models of the 27 YSOs are 
shown in Figure~\ref{fig:fit}, in which the 22 ``single'' YSOs 
are arranged in the order of Types I, I/II, II, II/III, and III 
from our empirical classification, and within each type by order
of increasing [8.0] magnitude, and the 5 ``multiple'' YSOs 
are arranged by order of increasing [8.0] magnitude. 
In each panel, data points are plotted in filled circles 
and upper limits are plotted in filled triangles; 
error bars are plotted but they are usually smaller than
the plot symbols.
The best-fit models, with minimum $\chi^2$ \citep[$\chi^2_{\rm min}$, 
see][for definitions of $\chi^2$]{RTetal07}, is shown in solid 
black line, and the radiation from the stellar core reddened by 
the best-fit foreground extinction $A_V$ is shown in dashed 
black line.
For most YSOs, their SEDs can be fitted similarly well by a 
range of models.
We have used a cutoff of $\chi^2 - \chi^2_{\rm min} \le 3$ 
for acceptable models and they are plotted in gray lines.

The results of the SED model fits are given in Table~\ref{sedfits},
where the YSOs are listed in the same order as Figure~\ref{fig:fit}.
The source name, [8.0] magnitude, and type from our empirical 
classification are listed on the left of the table, and selected 
physical parameters of the best-fit models are listed to the right,
i.e., the stellar parameters in mass, temperature, and radius 
($M_\star$, $T_\star$, and $R_\star$), envelope accretion rate 
($\dot{M}_{\rm env}$), disk mass ($M_{\rm disk}$), $A_V$, and
total luminosity ($L_{\rm tot}$).
In addition to the parameters of the best-fit model, we have also
used the acceptable models to show a possible range of stellar mass
and total luminosity.
We calculated weighted average $\bar{M}_\star$ and $\bar{L}_{\rm tot}$ 
using the $\chi^2$ values as weights, and defined the uncertainty to be 
the weighted standard deviation $\Delta M_\star$ and $\Delta L_{\rm tot}$ 
of the accepted models.
For each accepted model, we have also calculated the evolutionary 
stage using $\dot{M}_{\rm env}$/$M_\star$ and $M_{\rm disk}$/$M_\star$
ratios as defined in \citet{RTetal06}, i.e., 
$\dot{M}_{\rm env}$/$M_\star > 10^{-6}$~yr$^{-1}$ for Stage I sources,
$\dot{M}_{\rm env}$/$M_\star < 10^{-6}$~yr$^{-1}$ and 
$M_{\rm disk}$/$M_\star > 10^{-6}$ for Stage II sources, and 
$\dot{M}_{\rm env}$/$M_\star < 10^{-6}$~yr$^{-1}$ and 
$M_{\rm disk}$/$M_\star < 10^{-6}$ for Stage III sources.
We then calculated the range of evolutionary stage, Stage Range, 
with the same weighted scheme for average and standard deviation.
These ranges of stellar mass, total luminosity, and evolutionary 
stage are also given in Table~\ref{sedfits}.
Among these three quantities, total luminosity is most robust.
Mass depends on pre-main-sequence evolutionary tracks, which may introduce a
sysmtematic offset, and although the mass of warm circumstellar dust is
relatively robust, its translation into an accretion rate and evolutionary
state depends on the applicability of the single-YSO dust geometry
used in our radiative transfer models.

The SED fits of the 22 ``single'' YSOs, as displayed in Figure~\ref{fig:fit}, 
show different degrees of goodness-of-fit among our empirically
defined YSO types.
The 18 Type I, I/II, and II YSOs  show good agreement between
model and observed SEDs, though a number of them demonstrate 
that the observed fluxes are systematically lower than the modeled
ones at 4.5 \um.
Among the four Type III YSOs analyzed, two of them show reasonable
agreement in the SED fits, while the other two exhibit significant 
discrepancies between models and observed SEDs.
The above-mentioned discrepancies and possible causes are 
discussed below in \S4.2.

\subsection{Significant Discrepancies Between Model and Observed SEDs}

\subsubsection{PAH Emission in Massive YSOs}
A number of YSOs in N\,159 show a brightness dip 
at 4.5 \um\  in their SEDs (Figure~\ref{fig:fit}).
This is similar to YSOs in Paper I, and has been suggested that the 
apparent dip is not an absorption feature but caused by raised fluxes 
from PAH emission at 3.3, 6.2, 7.7, and 8.6 \um\ in the other three 
IRAC bands, and the difference between observed and modeled SEDs 
is due to models not including PAHs or small grains. 
The existence of PAH emission in massive YSOs in the LMC has been 
confirmed in a recent study using {\it Spitzer} InfraRed Spectrometer 
(IRS) observations \citep{Seetal09}.
Furthermore, PAH emission appears to almost ubiquitously present among 
massive YSOs, as it is detected in 87\% of 277 IRS spectra of massive 
YSOs.
We examined the IRS spectra of \citet{Seetal09} and found that 
28 of 29 YSOs in N\,44 (Paper I) and 11 of the 11 YSOs in N\,159 
show PAH emission in their IRS spectra, confirming our explanation 
of the 4.5 \um\ dip in SEDs as a result of PAH emission in the 
other three IRAC bands.

While including PAHs and small grains in YSO models is in progress 
\citep[e.g.,][]{Woetal08}, we can use IRS spectra to estimate the 
uncertainty in deriving physical parameters of YSOs with current 
models.
We have selected YSO 053935.99$-$694604.1 for illustration since 
this source is bright enough to ensure accurate photometric 
measurements and its SED shows an obvious dip at 4.5 \um.
Furthermore, this YSO has no silicate absorption at 10 \um\ in the 
IRS spectrum \citep{Seetal09} so that the contribution of PAH 
emission can be estimated by direct comparison between photometric 
data and the spectrum.
The photometric SED and IRS spectrum of YSO 053935.99$-$694604.1 
are shown in Figure~\ref{fig:pah}a, where PAH emission at 6.2, 7.7, 8.6
and 11.3 \um\ are marked.
Although 4.5 \um\ is outside of IRS's spectral range, this photometric 
datapoint appears to be consistent with the continuum level extending
from those around 10 and 15 \um.
We thus assumed that the difference between the photometric datapoints 
at 5.8 and 8.0 \um\ and the underlying continuum level is entirely
due to PAH emission and derived a factor of 1.8 for the 
(PAH+continuum)/continuum ratio.
We divided the observed fluxes at 3.6, 5.8, and 8.0 \um\ by this 
factor and modeled the PAH-corrected SED.
The model fits to the YSO's SED before and after the PAH correction
is shown in Figures~\ref{fig:pah}b and c; the PAH-corrected SED 
is well reproduced by YSO models.

The results of the fits are listed in Table~\ref{fitcom}, including
$\chi^2$-weighted average and standard deviation of stellar mass, 
total luminosity, envelope accretion rate, and Stage Range.
All four but one parameter inferred from fits to SEDs before and 
after PAH correction are in good agreement.
The exception is $\dot{M}_{\rm env}$, whose value is larger before
the PAH correction; however, this parameter is poorly constrained.
This result is understandable since the raised fluxes from PAH 
emission in the three IRAC bands is compensated by additional 
amount of larger dust grains incorporated in the models and hence
the larger $\dot{M}_{\rm env}$, and the ambiguous geometry of dust 
distribution (in disk or envelope) causes the large uncertainties
in $\dot{M}_{\rm env}$.
Nonetheless, the other three main parameters are not sensitive
to this contamination and thus the conclusions based on stellar
mass, total luminosity, and Stage Range derived from model
fits to observed SEDs without PAH corrections are not compromised.

\subsubsection{Discrepancies in Type III YSOs}

Among the four Type III YSOs, two show significant discrepancies 
between the best-fit models and the observed SEDs: 
053945.18$-$694450.4 and 054004.40$-$694437.6 (Figure~\ref{fig:fit}).
These two YSOs have SEDs agreeing with the stellar photospheric
emission in near-IR but deviant in optical bands.
To understand the cause of this discrepancy,
we examine available high-resolution optical images, i.e., the 4m 
MOSAIC \ha\ and archival {\it HST} images.
These images show that both YSOs have already formed compact
\hii\ regions.
YSO 053945.18$-$694450.4 has a small \hii\ region of diameter
$\sim 4\farcs8$ ($=$1.2 pc), resolved in the MOSAIC \ha\ image.
YSO 054004.40$-$694437.6 coincides with the compact \hii\ region 
the Papillon Nebula \citep{HMetal99} of diameters 
$\sim 4\farcs4\times 2\farcs8$ ($=1.1$pc $\times$ 0.7 pc), 
resolved in both MOSAIC and {\it HST} images.

For both YSOs, the optical $UBVI$ photometry were adopted from 
MCPS catalog.
This catalog used broad-band images taken with a CCD camera of 
a 0\farcs7 pixel$^{-1}$ scale and under seeings of 1\farcs2-1\farcs8 
\citep{ZDetal04}.
Thus, compact \hii\ regions of radii $\lesssim2''$ are unlikely to 
be resolved from the central stars. 
Bright nebular emission lines such as \hb\ would have significant 
contribution to $B$ and $V$ bands, and hence these fluxes deviate 
from stellar models, resulting in larger uncertainties in physical 
parameters inferred from SED fits.
Furthermore, the presence of \hii\ regions also suggests that
the dust emission can have an interstellar, instead of 
circumstellar, origin.
Thus, the inferred envelope accretion rate and disk mass are 
less reliable, and so is the Stage.
This is similar to the YSO associated with a small \hii\ region 
resolved by {\it HST} in N44, YSO\,052207.3$-$675819.9 (Paper I).
These results signify the importance of using high-resolution 
\ha\ images to reveal small \hii\ regions that are responsible
for the dust emission and indicate the last evolutionary stage
as a YSO approaches the main sequence stage.
On the other hand, the good agreement among mass estimates 
for YSO\,052207.3$-$675819.9 using different independent methods 
demonstrates that for a SED relatively well reproduced 
by models (and hence a good estimate for the total luminosity
of the system), the mass of the central star inferred from 
SED fits remains reliable.
This is likely the case for YSO 053945.18$-$694450.4.
As for YSO 054004.40$-$694437.6, the system is more complicated
and will be discussed in \S 4.3.

\subsection{Multiplicity Effect on SED Fits -- Modeling a Group of YSOs}

At the LMC's distance of 50 kpc, IRAC's $\sim$ 2\farcs0 angular resolution 
translates into 0.5 pc, which can hide a small group of YSOs.
Indeed, 10 of the YSOs are resolved by our ISPI images (with a 0\farcs3
pixel$^{-1}$ scale) into multiple sources or even small clusters within 
the IRAC PSF.
Another 10 YSOs appear more extended than the ISPI PSF; two of these
have NACO images available and both are resolved into multiple sources.
These results suggest that at least 44\% and up to 74\% of the 
IRAC-identified YSOs are multiple sources.
Similarly high fractions, 50--65\%, have been reported 
for YSOs in N44 (Paper~I) and for 82 LMC YSOs which have archival 
{\it HST} \ha\ images available \citep{Vaetal09}.
While the ISPI and NACO $JK_s$ images allow us to identify the
dominant YSO in a multiple system for accurate photometric 
measurements, the IRAC measurements correspond to the integrated 
light of all sources within the PSF.  
This mismatch is not problematic if one YSO dominates the emission
in IR bands; however, if two or more YSOs are present within the
IRAC PSF, the SED modeling can produce very uncertain results.
Below we use two examples to illustrate the extent of errors 
caused by multiplicity.

The YSO 053941.89$-$694612.0 in N\,159 appears extended in the ISPI 
$JK_s$ images and is resolved into four sources within 0\farcs6 in 
the NACO $J$ images.  
Among these four, the two brightest sources \citep[\#123 and \#121 in
N\,159-A7,][]{Teetal06}, have $K_s$ = 13.31 and 13.52 and $J-K_s$ = 
4.80 and 4.22, respectively; the extremely red colors 
originate from infrared excess unless
the foreground extinction was
exceedingly high, so these are most likely YSOs.
Therefore, 053941.89$-$694612.0 can be used to assess uncertainties 
in SED fitting for cases where two YSOs of comparable properties
exist within the IRAC PSF.
We construct SEDs for these two bright sources, 053941.89$-$694612.0a 
and b, using $JK_s$ from \citet{Teetal06} and IRAC/MIPS fluxes 
proportioned according to their $K_s$ mag; the latter assumption is 
reasonable since both YSOs have very red $J-K_s$ colors indicating 
early evolutionary stages.
Model fits to these two SEDs are shown in Figure~\ref{fig:multi}, 
along with fits to the SED of integrated fluxes of the multiple 
system.
All three SEDs have similarly good fits.

The results of the SED fits are listed in Table~\ref{fitcom}. 
The stellar mass inferred from the SED of integrated fluxes is
32 $M_\odot$, and those for the individual a and b components are 
27 and 25 $M_\odot$, respectively.
Component a makes up $\lesssim$ half of the total light in near-IR
and has an inferred mass $\sim 84$\% of that from the
SED of integrated fluxes.
Thus, for a source containing multiple YSOs at similar 
evolutionary stages such as 053941.89$-$694612.0, the 
inferred mass for individual (most massive) YSO could 
be overestimated by $\sim 20$\%, and the total YSO masses
may be underestimated by $\sim$40\%.

The second example of multiplicity in a YSO source is
\papillon\ in N\,159B, or the Papillon Nebula 
\citep{He56,HMetal99}.
{\it HST} images revealed a complex \hii\ region around \papillon.
\citet{HMetal99} used the integrated \hb\ flux of the \hii\ region
(assuming ionization-bounded) and determined that the spectral type 
of the ionizing source(s) of N\,159B is at least O8\,V or earlier,
while \citet{IJC04} reported an O6V ionizing star based on the 
integrated radio centimeter continuum flux of N\,159B.
The clumpy morphology seen in the {\it HST} images suggests 
that ionizing photons may leak from this region; thus the 
O8V spectral type is a lower limit.
As illustrated in Figure~\ref{fig:multi}d, our single-YSO 
model fails to simultaneously reproduce the mid-IR dust emission 
and the optical light.  
The latter, adopted from the relatively low-resolution MCPS
catalog, is almost certainly contaminated with nebular line 
emission from the ionized gas that is not included in the
YSO models.

To separate the central point source of \papillon\ from the 
surrounding nebulosity, we use photometric measurements made
with {\it HST} images in the optical and VLT/NACO adaptive 
optics images in $K_s$.  
At other IR bands, we use the IRSF $JH$ measurements as upper
limits, and IRAC and MIPS fluxes as they are.
The resultant SED and model fits are shown in 
Figure~\ref{fig:multi}e,  and the inferred YSO mass is
$\sim 21~M_\odot$ (Table~\ref{fitcom}).
The SED fitting is biased toward {\it HST} data points as they 
have smaller errors compared with those of {\it Spitzer} data points.
It is thus not surprising that the YSO mass is consistent 
with the B0/O9 spectral type implied by the {\it HST} $UbyI$
photometry.
In fact, \citet{HMetal99} and \citet{Teetal07} have already 
suggested that the point source detected in {\it HST} images,
which is also the brightest source in the VLT/NACO $K_s$ image 
within the IRAC PSF, does not account for the full luminosity 
or the centimeter radio flux of the nebula.

As the optically brightest star in \papillon\ may not be the YSO 
contributing to the observed mid-IR emission, we decouple the
mid-IR part of the SED from the rest and make SED fits to only 
the mid-IR segment.
This is essentially using the luminosity of dust emission as
a calorimeter to infer the effective luminosity of the main
energizing source(s) without specifically trying to separate 
those sources from the \hii\ region.
The result should be consistent with the spectral type derived 
from the total radio continuum emission from the \hii\ region.
Figure~\ref{fig:multi}f shows that the mid-IR part of the SED
can be reproduced by models reasonably well, and the inferred
mass, $\sim 41~M_\odot$ (Table~\ref{fitcom}), is indeed in 
agreement with the O6V spectral type derived from radio
observations \citep{IJC04}.
Therefore, the main ionizing sources of the Papillon Nebula 
are still hidden behind a large amount of dust.

In summary, when a YSO is in a multiple system, it is crucial
to use high-resolution optical and near-IR images to assess 
whether the multiple components are at different evolutionary
stages.  If the dominant sources are at similar YSO stages,
model fits of the integrated SEDs would overestimate the mass
of the most massive YSO but underestimate the total mass of 
the YSOs.
On the other hand, if multiple sources at different evolutionary
stages are present, the optical and near-IR emission may be
dominated by the exposed stars and the mid-IR emission dominated 
by more embedded YSOs.
Under these circumstances, model fits to the mid-IR portion of 
the SED can still provide useful information on the dominant
heating sources.
Lastly, we note that since only one case in each of these two 
types of system is investigated, our estimates on mulitiplicity
effect is thus approximate and a better quantification requires
a larger sample.
A cautionary corollary of our result is that the uncertainties 
would be even larger in mass estimates for YSOs of three or more 
comparably bright objects, such as the 5 multiple YSOs that are 
already resolved in the ISPI $JK_s$ images since each single 
source could be further resolved into multiples in even higher
resolution (e.g., the aforementioned NACO) images.

\subsection{Evolutionary Stage of YSOs}

\subsubsection{Comparisons among Empirical, Theoretical, and Spectral
 Classifications}

We use our analysis of 22 ``single'' YSOs to compare our empirical 
classification,
Type, with the theoretical classification, Stage \citep{RTetal06}.
As listed in Table~\ref{sedfits}, 
Types and Stages are only loosely correlated.
This lack of overwhelming correspondence is also seen in
Paper I and may be attributed to two reasons primarily related 
to the relatively large distance of the LMC.
In an unresolved cluster treated as a single source, details of the
circumstellar dust geometry may not correspond extremely well to our model
prescription, making the translation of fitted dust mass into an accretion rate
(quantitative Stage) less certain.  On the other hand, a Type determined from
environmental morphology on 
parsec scale may also not accurately
reflect the evolutionary state of a single YSO.  The methods are both useful,
neither definitive.
Second, limited angular resolution 
may cause
inclusion of extraneous dust emission from unresolved \hii\ regions.
This is particularly relevant to YSOs at a late evolutionary stage, 
such as the three Type III YSOs that show small \hii\ regions
in the MOSAIC \ha\ images; the mid-IR emission from unresolved
\hii\ regions would have been erroneously modeled as circumstellar
disks or envelopes.
Nevertheless, we expect the extreme Stage I YSOs to correspond to
our Type I YSOs, since they are deeply embedded in massive
envelopes.  
Indeed, Table~\ref{sedfits} shows that Type I YSOs are well fitted
by Stage I models.
We also expect Type III YSOs without \hii\ region confusions to
correspond to Stage III YSOs, and indeed this is the case for
054000.69$-$694713.4.

We also use the 11 YSOs with IRS observations to compare our empirical 
classification with the classification based on spectral features
\citep{Seetal09}.
Table~\ref{evo} lists their spectral classification and the presence
of silicate absorption.
Among the 11 YSOs, 10 have spectral type PE, whose spectra show PAH and 
fine-structure line emission, and 1 has spectral type P, whose spectrum 
shows PAH emission but not fine-structure line emission.
Our empirical classification for the 10 YSOs with PE-type spectra 
cover Types I, I/II, II, and III, indicating that the ionized gas
can be present early on.
Compared with PE-Type YSOs, the P-Type YSOs may have an earlier
evolutionary stage because of a higher accretion rate that quenches
the formation of an \hii\ region or have a stellar temperature
too low to provide ionizing flux \citep{Seetal09}.
The P-type YSO in N\,159, 053929.21$-$694719.0, is classified as 
Type I, and additional evidence is presented in \S4.4.2 to 
suggest that this P-type YSO is in an earlier evolutionary
stage than the PE-Type YSOs.
The strongest spectral feature that correlates with our empirical
classification is the silicate absorption.
The presence of silicate absorption is an indication of early
evolutionary stage \citep{Seetal09}.
The five YSOs in N\,159 that show silicate absorption are all 
of Types I and I/II.
They are also all of Stage I and have high $\dot{M}_{\rm env}$, 
$7.0\times10^{-4}-4.6\times10^{-3} M_\odot$/yr (Table~\ref{sedfits}), 
inferred from SED modeling, bolstering their evolutionary stage
being early.  
Thus, our classification using multiwavelength SEDs 
and images suggests an evolutionary sequence consistent with that
implied by classification based on spectra features.

\subsubsection{Evolutionary Stages of YSOs Associated with
 Masers and Ultracompact \hii\ Regions}
Massive stars inject energy into the surroundings even during 
their formation.
They undergo energetic mass ejection in the form of molecular 
outflows and produce maser phenomenon.
They can also ionize the circumstellar gas to form small, dense
\hii\ regions such as ultra-compact \hii\ regions (UCHIIs)
with diameters $\le10^{17}$ cm and densities 
$\ge10^4$ cm$^{-3}$ \citep{Fr00}.
In N\,159, a H$_2$O maser source and three UCHIIs have been identified
\citep{Lazetal02,IJC04}, and each of these four objects is
spatially coincident with a massive YSO (Figure~\ref{fig:jhkcmds}a).
The link between these objects and YSOs allows us to compare the 
YSOs' evolutionary stages inferred from SED models to the 
circumstellar conditions required to form masers or UCHIIs, 
providing an independent check for the feasibility of SED models.

The H$_2$O maser in N\,159 was first reported by  \citet{HW94} 
and confirmed with subarcsecond resolution and accurate 
position by \citet{Lazetal02}.
It coincides with our Type I YSO 053929.21$-$694719.0
within 1$''$.
As listed in Table~\ref{uch2}, this YSO has a high 
mass of 26.1$\pm$1.5 $M_\odot$ and a high $\dot{M}_{\rm env}$ 
of (5.1$\pm$1.5)$\times10^{-4} M_\odot$~yr$^{-1}$ derived
from model fits to its SED.
The high $\dot{M}_{\rm env}$ indicates it as a Stage I YSO.
Its IRS spectrum is classified as Type P (Table~\ref{evo}),
as it does not show any fine structure lines from ionized
gas \citep{Seetal09}.
The high mass of the stellar core implies that it does
emit ionizing radiation; however, the high envelope accretion
rate suggests that the circumstellar \hii\ region is most
likely too small to become detectable with our large beam.
The analysis of the YSO's SED leads to the conclusion that
YSO 053929.21$-$694719.0 is in an early evolutionary stage.
Indeed, it is commonly accepted that the water maser phase
precedes the UCHII phase \citep{Eletal07}.
Therefore, the production of the water maser and the YSO's SED models 
all point to a consistent picture that 053929.21$-$694719.0 is
at a very young evolutionary stage.

Three UCHIIs have been identified in N\,159: B0540$-$6946(1), 
B0540$-$6946(4), and B0540$-$6946(5) \citep{IJC04}.
All three UCHIIs are coincident with {\it Spitzer} sources within
1$''$, and these sources have been identified as Type I, II/III, 
and III YSOs (Table~\ref{uch2}).
As these sources are among the five most luminous YSOs in 
N\,159 at 8.0 \um, they are also bright in the 24 and even 70 
\um\ images from which reliable measurements are needed to  
constrain model fits to their SEDs.
The stellar luminosities determined from the best-fit SED models 
can be translated into spectral types, assuming the 
relationship for main sequence stars.
These spectral types can be compared with those implied by 
the ionizing fluxes determined from radio continuum observations 
\citep{IJC04}.
As seen in Table~\ref{uch2}, agreement exists between the spectral 
types independently determined with these two methods.  

The development of a UCHII depends on not only the ionizing 
flux provided by the central star, but also the opacity of the 
circumstellar medium.
For infalling rates higher than some critical value, 
$\dot{M}_{\rm crit}$, the circumstellar medium will have such high
opacities that the ionized region will be too small and 
too optically thick to be detectable \citep{CE02}.
For the respective spectral types of the central stars, the 
$\dot{M}_{\rm crit}$ of the three UCHIIs in N\,159 are computed to 
be $\sim 3-7\times10^{-5} M_\odot$~yr$^{-1}$ (Table~\ref{uch2}).
Compared to the $\dot{M}_{\rm env}$ determined from the best-fit 
models (Table~\ref{sedfits}) or weighted average $\dot{M}_{\rm env}$ 
determined from all accepted models (Table~\ref{uch2}) for these three 
YSOs, it is seen that all three YSOs have $\dot{M}_{\rm env}~\gg 
~\dot{M}_{\rm crit}$.
We have further examined whether any acceptable models of the 
three YSOs yield smaller envelope accretion rates.
We find that all the other acceptable models have $\dot{M}_{\rm env} 
\gg \dot{M}_{\rm crit}$.
This is similar to that found for 3 of the 4 UCHIIs 
associated with YSOs in N\,44 (Paper I).

There could be two causes for $\dot{M}_{\rm env} \gg \dot{M}_{\rm crit}$.
One is that most of the infalling envelope material is used in 
forming an accretion disk as modeled by \citet{YS02}, and the 
ionizing radiation escapes in the polar directions.
The other is that at the LMC distance, it is difficult to 
distinguish bound, circumstellar dust from more distant but
still heated interstellar dust.
The warm envelope of an UCHII contributes to the mid-IR 
emission modeled in the SED, and raises the derived 
accretion rate.
Thus, the high envelope accretion rates of YSOs associated 
with these UCHIIs and the H$_2$O maser mostly suggest that
they still have abundant dust/gas in the surroundings and 
have not significantly cleared their environments.
In terms of evolutionary sequence, we find that the maser is 
associated with Type I while UCHIIs are associated with Types 
I, I/II, and III; this is consistent with the general picture 
that the maser phase happens earlier than UCHII \citep{Eletal07}.
Although the SED of the maser source is probably also contaminated 
by hot dust in the environment that may not actually accrete onto 
the protostar.

\subsubsection{Evolutionary Stages of YSOs as Herbig Ae/Be stars}
HAeBe stars are intermediate-mass young stars that have evolved
from the most embedded phase and are now revealing their stellar 
components \citep[e.g.,][]{HLetal92}.
Since our YSO sample overlaps this mass range, we examine 
whether some of them might be previously classified as HAeBe stars.
Candidate HAeBe stars in N\,159 have been selected by 
\citet{Naetal05} using near-IR $JHK_s$ photometry.
Figure~\ref{fig:jhkcmds}a shows their locations marked on an 
8 \um\ image of N\,159, along with our YSOs marked in different 
colors to indicate their evolutionary stages.
Figure~\ref{fig:jhkcmds}b and c show color-color and color-magnitude
diagrams using $JHK_s$ photometry of sources in N\,159 that are 
detected in all three bands \citep[kindly provided by Yasushi
Nakajima;][]{Naetal05}.
As listed in Table~\ref{evo}, nine of these candidate HAeBe stars
are identified as YSOs and they have the following types of SEDs:
one Type I, one Type I/II, five Type II, and two Type III.
The majority (78\%) of these HAeBe candidates are Type II and III YSOs, 
consistent with the expectation that such objects are less embedded.
There are also two candidate HAeBe stars that we classify
as non-YSO sources: 053921.21$-$694409.4 is a star superposed on 
dust clumps, and 053938.80$-$694436.0 is the O7\,III stellar 
counterpart of the high-mass X-ray binary LMC X-1 \citep{Coetal95}.
The latter clearly is not a young star, and this explains why
its colors are bluer than the other HAeBe stars (see 
Figure~\ref{fig:jhkcmds}b and c).

\subsection{Masses of YSOs}

The mass estimates from the best and acceptable fits of 27 YSOs in 
N\,159 are listed in Table~\ref{sedfits}.
Among them, 22 YSOs have single or dominant sources within the IRAC 
PSF and their SEDs can be properly approximated by single-YSO 
models to obtain reliable mass estimates.
The other five have SEDs from multiple YSOs or even non-YSO
sources such as stars or nebulosities; their mass estimates 
have larger uncertainties and should be viewed as rough estimates.
Nevertheless, the results of SED fits show that twenty of the YSOs
in N\,159 have $\chi^2$-weighted average mass greater than 
$8~M_\odot$ and thus are most likely bona fide massive YSOs.
The remaining seven have $\chi^2$-weighted average $< 8~M_\odot$ 
and are likely intermediate-mass YSOs.

It has been suggested that the criterion [8.0] $\le 8.0$ may be 
used to select massive YSOs in the LMC \citep{GC09}.
We find that indeed the $\chi^2$-weighted average mass of the 
brightest YSOs, with [8.0] $\le 8.0$, are all $\ge 8~M_\odot$.
This criterion may be too conservative, as all (15 out of 15)
YSOs with [8.0] $\le 9.0$ still have masses $\ge 8~M_\odot$
The results are consistent with what we found for YSOs in
N\,44 (Paper I).

At the high-mass end, 
nine YSOs have $\chi^2$-weighted average 
mass $> 17~M_\odot$; these are most likely O-type YSOs.
Two YSOs have $\chi^2$-weighted average mass close to 17 $M_\odot$
and larger standard deviation that extending
beyond 17 $M_\odot$; these may or may not be O-type stars.
Therefore, there exist at least 
9 O-type YSOs in N\,159.
These most massive YSOs will be discussed further in \S5.1.

\section{Massive Star Formation in N\,159}

It is difficult to study the relationship between interstellar
conditions and the formation of massive stars because massive stars'
UV radiation fluxes and fast stellar winds quickly ionize and
disperse the ambient ISM.
Young massive YSOs that have not significantly altered the physical
conditions of their large-scale environments can be used to probe 
massive star formation.
The large number of massive YSOs found in N\,159 provides an 
excellent opportunity to investigate issues such as the relationship 
between star formation properties and interstellar conditions, 
progression of star formation, and evidence of triggered star 
formation.

\subsection{Interstellar Environments and Star Formation Properties}

\begin{figure}
\plotone{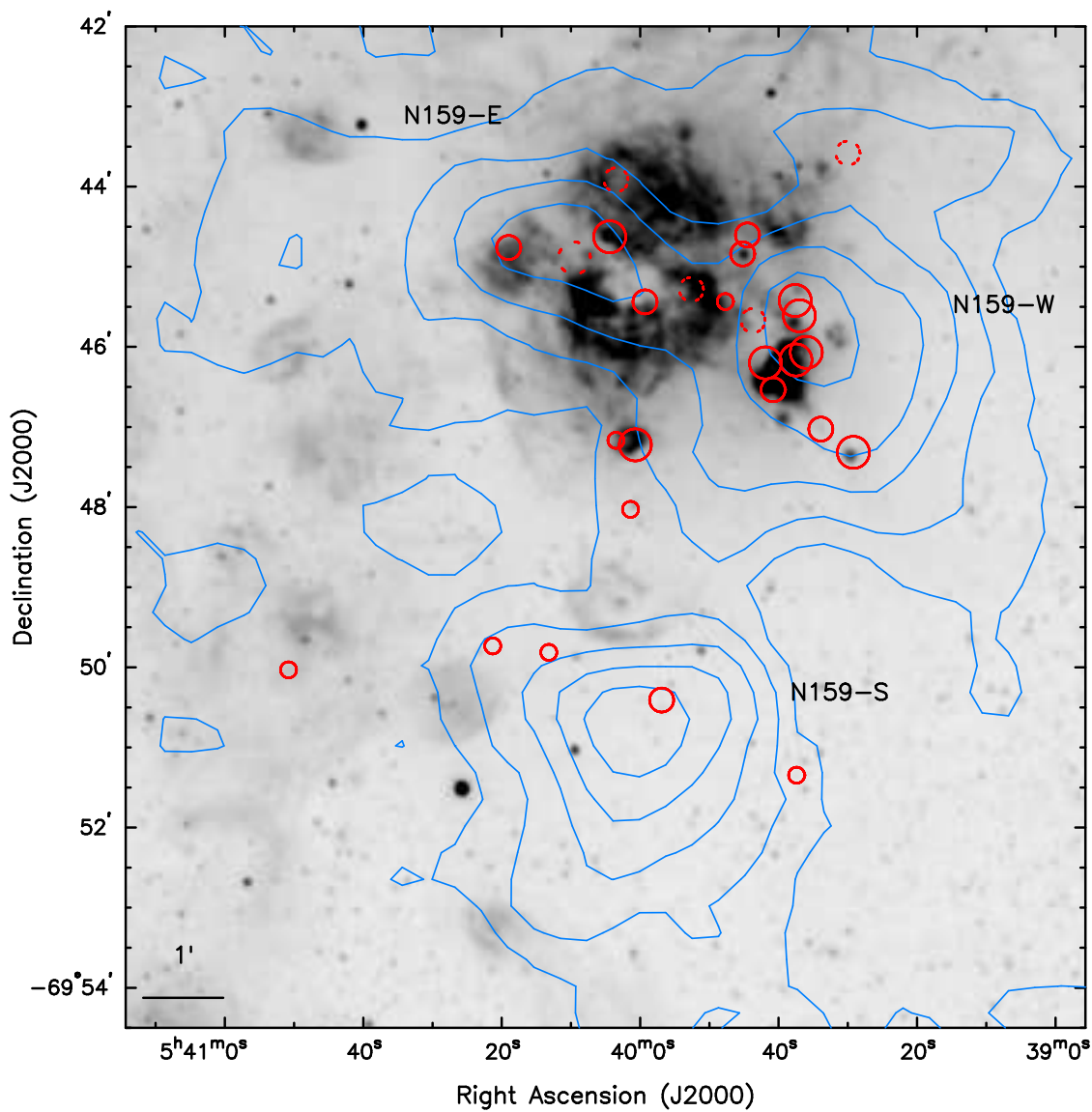}
\caption{Distribution of YSOs with respect to 
 interstellar environments of N\,159.  The \ha\ image of N\,159 is shown
 in grey scale, overlaid with MAGMA CO intensity contours (blue) in 
 n$^2$ K~km~s$^{-1}$ (n$=$1,2,3,...).
 YSOs with different mass estimates are marked with different symbols:
 O-type, i.e., ${\bar M}_\star \ge 17~M_\odot$, as large red open 
 circles; early-B type, i.e., $17~M_\odot > {\bar M}_\star \ge 8~M_\odot$, as 
 medium red open circles; and B-type, i.e., ${\bar M}_\star < 8~M_\odot$, 
 as small red open circles.  YSOs that appear single or are clearly the
 dominant source with the IRAC PSF are shown in solid-line circles, while
 YSOs that are multiple and have larger uncertainties in mass estimates
 are in dotted-line circles.
 \label{fig:yso_pos}}
\end{figure}

We examine the star formation properties of the molecular clouds in 
N\,159 as these clouds contain the bulk material to form stars. 
The SEST and MAGMA CO surveys of the LMC \citep{Joetal98,Otetal08}
show three large concentrations of molecular material in N\,159,
i.e., the eastern, western, and southern peaks (Figure~\ref{fig:yso_pos}).
These three concentrations exhibit different numbers of massive
stars formed in the last 10 Myr, as evidenced by their different 
amounts of ionized gas.
The eastern molecular peak is associated with bright \ha\ emission
from the prominent central \hii\ regions.
Star formation has been occurring in this $\sim 35$ pc $\times$ 50 pc 
region for an extended period of time, as it contains both evolved
and young massive stars.
For example, star \#54 of \citet{Faetal09} is a  B1-2 IV-II star
that is $\sim$ 10 Myr old, while their star \#62 is an O4-6 Vn star
that is $\lesssim 3$ Myr old \citep{SDetal93,SdK97}.
The western molecular peak has only several small, disjoint \hii\ regions,
but one of them, N\,159A, has the highest H$\alpha$ surface brightness in 
the entire N\,159.
The massive stars in N\,159A and its surroundings have been
imaged in $UBV$ and their colors and magnitudes indicate that
their spectral types range from B2\,V to O5--6\,V \citep{DC92}.
Given the absence of SNRs, these stars are most 
likely $\lesssim 5$ Myr or that star formation started only in 
the last few Myr.
The southern molecular peak only has a couple of small \hii\ regions, 
indicating that only modest massive star formation has taken place.

Figure~\ref{fig:yso_pos} shows that 96\% ($=26/27$) of YSOs 
in N\,159 are found within CO contours $\ge$ 1 K~km~s$^{-1}$.
This corresponds to an H$_2$ column density of 
$\ge 4\times10^{20}$~cm$^{-2}$, for a CO-to-H$_2$ conversion 
factor $X_{\rm CO} = 3.9\pm2.5 \times10^{20}$~cm$^{-2}$ 
(K km~s$^{-1}$)$^{-1}$ in the molecular ridge which includes 
N\,159 \citep{Pietal09}.
About 70\% of the YSOs are congregated toward the three molecular
peaks, similar to the 75\% found in N\,44 (Paper I).
The YSOs of the three molecular peaks show different characteristics
in their spatial distributions and interstellar environments.
The western molecular peak has the highest concentrations of YSOs, 
as 12 YSOs are found within $\sim$ 1\farcm5 ($=$ 22 pc)-radius of 
the CO contour peak and 6 of them cluster in a small region of 
$40''\times 80''$ ($=$10 pc $\times$ 20 pc) over the bright \hii\ 
region N\,159A and its north tip N\,159AN.
The eastern molecular peak has more widely distributed YSOs, and 
most of them are associated with \hii\ regions.
The southern molecular peak has 4 YSOs that are loosely distributed,
and they are not associated with any ionized gas.

The YSOs of the three molecular peaks in N\,159 also show differences
in their mass distributions.
In Figure~\ref{fig:yso_pos}, we have marked the YSOs with circles 
in three sizes that represent O-type stars with inferred $\bar{M}_\star 
\ge 17 M_\odot$, (early) B-type stars with $17 > \bar{M}_\star \ge 8 
M_\odot$, and intermediate-mass stars with $\bar{M}_\star < 8 M_\odot$.
We found $\sim 80$\% of the O- and B-type YSOs in N\,159 are in or 
adjacent to \hii\ regions, similar to those seen in N\,44 (Paper I).
The YSOs with intermediate masses or without mass estimates do not
show such preferred association with \hii\ regions, but this may be
caused by an observational bias against the detection of these
fainter intermediate-mass YSOs over the bright background dust 
emission in \hii\ regions.
The western and eastern molecular peaks, associated with bright \hii\
regions, have a respectable number of O- and B-type YSOs, while the 
southern molecular peak has one YSO with mass $\sim 8 M_\odot$ and 
no O-type YSOs at all.

Similar to the trend seen in N\,44 (Paper I), the characteristics 
of the current star formation in the three molecular peaks of N\,159 
appear to be dependent on the massive star formation that occurred 
in the recent past.
We have further examined whether the pattern of star formation 
is related to the physical properties of molecular clouds.
The eastern, western, and southern molecular concentrations 
contain GMCs N\,159-E, N\,159-W$+$N\,159-2, and N\,159-S$+$N\,159-3,
respectively \citep{Joetal98}.
Their physical properties adopted from \citet{Joetal98} are 
listed in Table~\ref{gmc}, including each GMC's local standard 
of rest velocity ($V_{\rm lsr}$), CO line-widths ($\Delta V$), 
virial mass ($M_{\rm vir}$), luminosity mass ($M_{\rm lum}$), 
and ratio of virial to luminosity mass.
Note that $M_{\rm lum}$ is converted from the CO luminosity 
$L_{\rm CO}$ using $M_{\rm lum} = 8.8\pm5.6 L_{\rm CO} 
M_\odot$, revised from Equation (6) in \citet{Boetal08} for 
an $X_{\rm CO}$ value suitable for N\,159 \citep{Pietal09}.
In addition, when there are two GMCs in a molecular concentration, 
we use averages of $V_{\rm lsr}$ and $\Delta V$ and sums of 
$M_{\rm vir}$ and $M_{\rm lum}$, respectively, to represent
the properties of a concentration.
For comparisons, in Table~\ref{gmc} we have also listed these 
quantities for GMCs in N\,44 \citep{MNetal01}, where the 
$M_{\rm lum}$ is converted from the CO luminosity $L_{\rm CO}$ 
using $M_{\rm lum} = 15.8\pm4.5 L_{\rm CO} M_\odot$ for 
$X_{\rm CO}=7\pm2 \times10^{20}$~cm$^{-2}$ (K km~s$^{-1}$)$^{-1}$ 
\citep{FYetal08}.

The  GMC N\,159-W has the smallest 
$\Delta V$ and $M_{\rm vir} / M_{\rm lum}$ ratio, which can be an 
indicator of lower gravitational support, and contains the 
largest number and most concentrated distribution of massive YSOs.
Similarly, the GMC N\,44-C has YSO properties like N\,159-W
and also has a smaller $M_{\rm vir} / M_{\rm lum}$ ratio among
the GMCs in N\,44.
The larger sample of 36 LMC GMCs in \citet{Inetal08} showed a similar trend 
of more star formation activity at lower  $M_{\rm vir} / M_{\rm lum}$.
This may reflect preferred physical conditions for GMCs to 
collapse and form a concentrated cluster, qualitatively consistent 
with the picture that a smaller turbulent kinetic energy results
in larger volumes exceeding the Jeans criterion for gravitational 
instability \citep[e.g.,][]{Kletal98}.
The similar values in $\Delta V$ and 
$M_{\rm vir} / M_{\rm lum}$ ratios found in the other two GMCs 
in N\,159 do not seem to scale with their different star formation 
activity,
although the uncertainties in the virial analysis are large compared to the
differences between the GMCs.
Detailed mapping of these three GMCs 
could reveal some of the physics underlying their different star 
formation characteristics.
In particular, observations of denser gas tracers such as HCO$^+$
and HCN reveal the virial ratios and physical conditions of the {\em clumps} that are actively
participating in star formation, whereas bright CO  emission can 
also come from more diffuse or photo-dissociation regions of the 
GMC.
High-resolution ($\sim 5''$) HCO$^+$ observations of N\,159-E 
and W have shown that dense clumps are spatially correlated 
with YSOs at early evolutionary stage (Chen et al, 2010, in preparation), 
and new observations of N\,159-S taken in October 2009 
will reveal whether dense clumps are present in this GMC as well.

\subsection{Massive Star Formation -- Triggered or Spontaneous?}

A number of observational studies have demonstrated that the current
star formation may be triggered by stellar energy feedback, such
as through the expansion of an \hii\ region 
\citep[e.g.,][]{ZAetal07,Poetal09}.
N\,159 experiences intense energy feedback evidenced with the 
filamentary structure in the \hii\ regions (Figure~\ref{fig:n159opt}) 
and the existence of SNR 0540$-$697 \citep{Chetal97,Wietal00}.
N\,159 also has abundant molecular material for continued star formation.
It is conceivable that its (main-sequence) massive stars photonionize 
the ambient gas to form \hii\ regions and the raised thermal 
pressure can compress the neighboring molecular cloud to 
trigger star formation.
In fact, as shown in Figure~\ref{fig:yso_pos}, about 2/3 of 
the YSOs are associated with \ha\ emission and almost all of 
them are in the eastern and western GMCs.
In contrast, none of the four YSOs in the southern GMC
are associated with \hii\ regions.
To investigate whether some of the current star formation in 
the eastern and western molecular concentrations may have been 
triggered, we compare the distributions of YSOs relative to
the massive stars, and use high-resolution MOSAIC \ha\ images to 
examine the immediate environment of YSOs.

The eastern GMC is associated with the 
prominent central \hii\ regions in N\,159.
This region has 17 spectroscopically identified massive 
stars \citep{Faetal09}.
As shown in the MOSAIC \ha\ image in Figure~\ref{fig:yso_mstar}a,
these massive stars are distributed in the bright northern and 
southern lobes of the central \hii\ region, while the YSOs are 
mostly found along the edges of an east-west oriented 
low-surface-brightness band between the two lobes.
As this band coincides with the peak of the GMC, 
the spatial distributions of YSOs appear to suggest that their
formation is likely triggered by the expansion of two \hii\ 
regions (lobes) into a central molecular cloud.
To test this hypothesis, we used two methods to search for massive 
stars in the central molecular cloud that were formed earlier 
before the external pressure was raised but are too obscured to
be detected at optical wavelengths.

We first use the 3~cm map to search for obscured O stars, since it
is nearly extinction-free signpost of gas ionized by massive stars.
Figure~\ref{fig:yso_mstar}a reveals that the 3~cm emission of 
N\,159 \citep{Dietal05,SFetal10} appears diffuse for a $22''$
resolution over (mature) massive stars in the bright \hii\ regions,
but shows four peaks in the east-west central band and the 
western molecular concentration.
Three of the four 3~cm peaks are centered at O-type YSOs 
associated with UCHIIs \citep{IJC04}.
The remaining 3~cm peak, i.e., the faintest peak in the east-west
band, is centered between two non-O-type YSOs and has no 
UCHIIs detected.
The absence of UCHIIs indicates that these YSOs have lower
masses than the detection limit of a B0 V star,
or that the \hii\ regions of optically obscured OB stars are
too extended and thus resolved out by the small synthesized 
beam \citep{IJC04}.

We next consider the candidate OB stars identified 
with $JHK_s$ photometry \citep{Naetal05}. 
Figure~\ref{fig:yso_mstar}b shows that some of them are 
distributed around, though not inside the east-west band.
To assess whether there might be OB stars hidden by extinction
even in the near-IR, we use the \ha\ and 3~cm maps to estimate 
the extinction.
Assuming that the 3~cm emission of N\,159 is completely 
thermal\footnote{This assumption is likely to result in an overestimate 
of the extinction as there is contamination of non-thermal emission 
from the nearby SNR 0540$-$697 \citep{SFetal10}.}, the extinction in 
the \ha\ emission, $A_{{\rm H}\alpha}$, of a 10,000 K \hii\ region 
can be estimated using
\begin{equation}
A_{{\rm H}\alpha} = -2.5{\rm log}~(\frac{F_{{\rm H}\alpha}}{S_{\rm 3cm}}
 / \frac{j_{{\rm H}\alpha}}{j_{\rm 3cm}}),
\end{equation}
where $j_{{\rm H}\alpha}/j_{\rm 3cm} = 8.86\times10^{-10}$ 
ergs~cm$^{-2}$~s$^{-1}$~Jy$^{-1}$ and then converted to the visual 
extinction using $A_V = 1.24 A_{{\rm H}\alpha}$
\citep{CD86}.
Figure~\ref{fig:yso_mstar}b shows the visual extinction map and its 
peak value is $\lesssim 5.0$ in the east-west band.
For a standard extinction curve of $A_J/A_V = 0.28$ and 
$A_K/A_V = 0.11$, an O8 V star would have $J=16.1$ and $K=15.1$ 
\citep{SK82,Ko83}, brighter than the catalog's 10-$\sigma$ limiting 
magnitudes of $J=18.8$ and $K=16.6$ \citep{Naetal05}, and would 
have been detected if the local variation in the extinction 
is within a factor of 3, i.e., $A_V \sim 15$.
As for massive YSOs which usually have [8.0] $\le 9.0$, a foreground 
extinction of $A_V = 5.0$ corresponds to $A_{8.0} = 0.24$ 
\citep{Inetal05} and imposes little effect on their detection.
The above considerations suggest that the peak of the eastern GMC
was not active in producing O-type stars a few Myr in the 
past and only began to form them currently in locations on the edges 
contacting \hii\ regions.
Thus, the formation of the YSOs in the GMC N\,159-E
is likely triggered.

It is interesting to note that YSOs 054003.49$-$694355.0 and 
054004.40$-$694437.6 in NGC\,159-E are 
projected within and at the edge of SNR 0540$-$697 
(Figure~\ref{fig:yso_mstar}a).
For this SNR to be responsible for the formation of these two YSOs, 
its age has to be longer than the ages of YSOs.
The age of this SNR, estimated from its size and expansion velocity, 
is $\sim 1-2\times10^4$ years \citep{Wietal00}.
The absolute ages of the YSOs are difficult to estimate.
However, the brighter and more massive of the two YSOs, 
054004.40$-$694437.6, has formed a small \hii\ region visible in 
the optical wavelength, i.e., the Papillon Nebula \citep{HMetal99}, 
and hence its age is likely to be more than several 10$^5$ years 
\citep{CE02}.
The fainter and less massive of the two YSOs has evolved beyond
the earliest evolutionary stage and thus unlikely to be younger 
than a few 10$^5$ years.
The age consideration suggests that the SNR is unlikely to be
responsible for the formation of these two YSOs.

The western GMC in N\,159 abuts the central \hii\ regions.
For the expanding \hii\ regions to trigger star formation in this
molecular concentration, the crossing time scale should be 
shorter than the ages of massive stars and YSOs.
Figure~\ref{fig:yso_mstar}a shows that the boundary of the 
\hii\ regions is $\gtrsim 40''$, or a projected distance of 10 pc 
from the massive stars and YSOs.
For this distance and a velocity dispersion $\sigma$ of 2.6 
km~s$^{-1}$ of the GMC N\,159-W \citep[from $\sigma = 0.43 
\Delta V$ and $\Delta V = 6$ km~s$^{-1}$,][]{Joetal98}, the traverse 
time is $\sim 4$ Myr.
This time scale is not shorter than the age of OB stars in 
the bright nebular component N\,159A, $\sim 3-4$ Myr,
indicated by the presence of a spectroscopically identified O8
star and a few photometrically identified mid- to late-O stars
\citep{DC92,SdK97,Faetal09}.
Thus it is not likely that the expansion of the central \hii\ 
regions is responsible for the formation of these massive stars.
Furthermore, YSOs in the western GMC are concentrated in the molecular 
core, in contrast to the YSOs in the eastern GMC that are spread out 
and distributed along the interface between the molecular cloud and 
\hii\ region.

It has been suggested that the star formation in N\,159 
started near the center of the diffuse radio emission where 
OB stars identified with $JHK$ photometry are located, 
and proceeded outwards to trigger next-generation star 
formation in the eastern and western GMCs 
\citep[e.g.,][]{JTetal05,Naetal05}.
Our detailed analysis finds a more complicated pattern of 
star formation.  The expansion of the central \hii\ region 
has triggered the star formation in the eastern GMC, but not 
the western GMC.  It is possible that the star formation in 
the western GMC was triggered by some force that is no longer
identifiable, such as an old SNR whose shock velocity had
slowed down to 20--45 km~s$^{-1}$ \citep{VC98}.  It is also 
known that in an active star forming region, the environment 
is usually too complex to pinpoint the triggering mechanism 
for star formation \citep{Deetal10}.  
Nevertheless, based on the distribution of YSOs concentrated 
near the core of the western GMC as well as its lowest velocity 
dispersion (kinetically cold) among GMCs in N\,159, 
we suggest that its star formation may have started spontaneously.  
This suggestion is further supported by the most massive stars 
formed in the western GMC, as shown below.

A trend is observed in YSOs in \hii\ regions in the Milky Way 
and the LMC that in triggered star formation, the second 
generation is less massive than the first  
\citep[e.g.,][]{Poetal09,Fletal10}.
In N\,159, the most massive YSO formed in the eastern GMC, 
054004.40$-$694437.6, is O6 V \citep{IJC04}, not as massive as 
the earliest type O4-6 Vn found in the central \hii\ region 
\citep[\#54 in][]{Faetal09}.
The most massive YSO formed in the western GMC, 053937.53$-$694609.8, 
is O5.5V \citep{IJC04}, more massive than the only spectroscopically 
classified O8 star\footnote{\citet{DC92} suggested another star as 
mid- to late-O type based on its $UBV$ colors and magnitudes, however 
such conversion is known to have large uncertainties \citep[e.g.,][]{Ma85}.} 
in the bright \hii\ region N\,159A \citep{Faetal09}.
The spectral types derived from radio continuum by \citet{IJC04} 
may be underestimates if significant ionizing flux is leaking from 
the UCHII, and there is a systematic uncertainty in converting 
from ionizing flux to spectral type due to differences in massive 
stellar atmosphere models. 
However, the comparison of most massive YSOs and main-sequence 
stars in these two GMCs suggests that the eastern GMC is more 
consistent with the aforementioned trend than the western GMC.
In cases of spontaneous star formation, the masses of YSOs 
do not depend on the masses of stars that were formed earlier 
in the molecular cloud.  
It is thus the cloud properties that determine the types of 
stars formed, though once massive stars are formed, their 
energy feedback can trigger further massive star formation 
as demonstrated in the eastern GMC.
These two factors can naturally result in massive YSOs 
preferentially found in GMCs associated with \hii\ regions 
as the GMCs likely already have conditions to form massive 
stars and can form even more with the aide of stellar 
energy feedback.

\subsection{Star Formation Efficiency and Rate}
The correlation or lack thereof between GMC properties in 
galaxies and their star formation activity is of critical 
importance to understand galaxy evolution and star formation 
in general.  
In particular, we need to understand the physics underlying 
the empirically observed correlation between star formation 
rate and gas surface density, the ``Schmidt-Kennicutt''
(S-K) relation \citep[e.g.,][]{Ke89,Caetal07,Keetal09}. 
This correlation is tight when properties are averaged over 
a kpc or more, but it is difficult to justify why it should 
hold on sub-kpc scales.  
Only recently have datasets been rich enough to explore the
smaller scales.
The THINGS project \citep{Waetal08} found that whether molecular 
or total (= molecular + atomic) gas surface density is a better 
predictor of star formation depends on which is the dominant phase in
that part of the galaxy \citep{Leetal08}, and in molecular regions, 
star formation scales linearly with H$_2$ surface density 
\citep{Bietal08}.  

Large-scale star formation in a galaxy depends on the formation
rate of GMCs and how GMCs collapse to form stars.
Theoretical models assuming that GMCs have a constant star 
formation efficiency within a free-fall timescale find that 
GMC formation is the dominant driver in the star formation rate 
(SFR) in galaxies, and that the S-K relation can be reproduced
\citep{KMT09}.
It is of fundamental importance to determine how and when GMCs
collapse to form stars.
Only a small number of detailed extragalactic studies have 
been made.  
In N\,44 (Paper I), N\,159 (this study), and the LMC's quiescent 
molecular ridge \citep{Inetal08}, we find that although most 
massive young stars are found near peaks in $^{12}$CO emission, 
there is not always a clear correlation between $^{12}$CO properties 
and YSO properties. 
In N\,44 and N\,159, massive YSOs are mostly found in GMCs 
associated with ionized gas.
In N\,44, the most massive GMC N\,44-S has $\sim 5$-times 
the cloud mass but only half the YSO content of N\,44-C; in N\,159, 
the most massive GMC N\,159-S formed only a few YSOs and 
none of them are massive (Table~\ref{gmc}).
The current star formation activity appears to correlate more 
with the energy feedback from antecedent massive stars than 
the mass of a GMC.
\citet{Inetal08} found that in the molecular ridge, the SFR measured 
from either integrated mid-IR light or detailed analysis of the 
YSOs is correlated with the $M_{\rm vir}/M_{\rm lum}$ ratio, 
an indicator of gravitational support.
However, the total SFR measured from the integrated light is far below 
that expected from the S-K relation for the given GMC surface density.
The molecular ridge is known for its anomalously low star formation
activity, but the GMCs in N\,44 and N\,159 host a range of star 
formation activities and may be used to investigate whether the S-K 
relation applies on scales of individual GMCs.

The known massive YSO content allows us to assess the instantaneous
star formation efficiency, SFE$_{\rm YSO}$, of a GMC.
We use the number counts of YSOs within different mass bins and the
assumption of Salpeter's initial mass function (IMF) to estimate the 
total mass of the current star formation, $M^{\rm total}_{\rm YSO}$.
The lowest mass bin used for counting depends on
the photometric completeness in a region.
For regions with bright diffuse background, such as N\,159-W and 
N\,44-C, only higher-mass bins are used to avoid problems from 
photometric incompleteness at $< 9.0$ mag, though at the expense 
of small number statistics.
The brightness limit [8.0]$=9.0$ corresponds to $\sim 8 M_\odot$ 
(recall masses of YSOs inferred from SED fits in \S4.5).
The total mass,  $M^{\rm total}_{\rm YSO}$, is calculated by integrating
the initial mass function from $M_{\rm l}$ = 1 $M_\odot$ to $M_{\rm u}$, 
the highest mass of YSO observed. 
We adopt the commonly used lower mass limit of 1 $M_\odot$ to facilitate
comparisons with other work, although the formation time scale of a 1 $M_\odot$
star is much longer than that of the observed massive YSOs.
The $M^{\rm total}_{\rm YSO}$ is then divided by the virial mass 
of the GMC ($M_{\rm vir}$) to obtain SFE$_{\rm YSO}$, and all
these three quantities are listed in Table~\ref{gmc}. 
Uncertainties in $M^{\rm total}_{\rm YSO}$ are also listed; they are 
directly related to the uncertainty in mass inferred for individual 
massive YSOs used in number counts and are thus estimated from 
the largest and small mass ranges covered by these massive YSOs.

To determine the SFR from the total mass of current star formation 
requires a timescale.  
Although we can constrain the age of each massive YSO from its 
evolutionary stage, statistics of solar-mass YSOs in Class I, 
II, III, and models of high-mass protostellar accretion, such 
ages are quite uncertain aside from being $\lesssim$ 1 Myr.
On the other hand, since we consider all YSOs with high and 
intermediate masses within each GMC that are currently forming 
in a burst, this seems more relevant to the formation timescale 
of a cluster or association.
Proceeding with this assumption and assigning a cluster formation 
time ($t_{\rm cluster}$) of $\sim 1$ Myr \citep[e.g.,][]{BBV03},
we derive the current star formation rate, SFR$_{\rm YSO}$.
We have also estimated SFE per free-fall time ($t_{\rm ff}$),
\begin{equation}
\epsilon_{\rm ff} = \frac{t_{\rm ff}}{t_{\rm cluster}}
 \frac{M^{\rm total}_{\rm YSO}}{M_{\rm vir}}
\end{equation}
\citep[e.g.,][]{HS07} which has gained popularity recently 
as a normalized measure of star formation activity.
Note that the angular resolution for CO data of N\,44 and N\,159
are different: N\,44 was observed at 145$''$ angular 
resolution while N\,159 at 45$''$ resolution.
Thus, in N\,44, multiple GMCs may not be resolved, resulting in 
much larger GMC size and longer $t_{\rm ff}$.
Table~\ref{gmc} shows that the SFE$_{\rm YSO}$ for GMCs in N\,44 
and N\,159 are low, $\sim 0.0002-0.009$.
As a comparison, the molecular cloud associated with the Pipe 
Nebula, a very low-level Galactic star formation region, has 
SFE$_{\rm YSO} \sim 0.0006$ for YSOs $\ge 0.3 M_\odot$ \citep{Foetal09};
the SFE$_{\rm YSO}$ found in N\,44 and N\,159, after multiplied by 
a factor of 1.7-2 to account for a lower mass limit extending to 
$0.3M_\odot$ under the Salpeter's IMF assumption, are $\sim 1-25$ 
times that in the Pipe Nebula.
The $\epsilon_{\rm ff}$ is also low, mostly $\sim 10^{-4}$ to 
10$^{-3}$.

For comparison, we have also estimated the SFR associated with 
each GMC using integrated \ha\ and 24 \um\ fluxes with the 
prescription of \citet{Caetal07},
\begin{equation}
{\rm SFR}_{{\rm H}\alpha+24}(M_\odot~{\rm yr}^{-1}) = 
5.3\times10^{-42}[L(H\alpha )_{\rm obs}
+(0.031\pm0.006)L(24~\mu m)],
\end{equation}
where $L$(\ha ) and $L$(24 \um ) are \ha\ and 24 \um\ luminosities in
ergs~s$^{-1}$, respectively.
To measure the \ha\ and 24 \um\ luminosities, we use an aperture size
to include the bulk of a GMC; the largest uncertainties come from flux 
calibration, i.e., 5-10\% in \ha\ and 10\% in 24 \um\ (Sean Points, 
private communication; MIPS Data Handbook).
The aperture size, \ha\ and 24 \um\ luminosities, SFR$_{{\rm H}\alpha+24}$
in $M_\odot$~yr$^{-1}$, and normalized SFR$_{{\rm H}\alpha+24}$
in $M_\odot~{\rm yr}^{-1}$~kpc$^{-2}$ are all given in Table~\ref{gmc}.
Among the 6 GMCs, three have SFR$_{\rm YSO}$ higher than
SFR$_{{\rm H}\alpha+24}$.
Two of these three GMCs, N\,44-N and N\,159-S, have low star formation 
activities in the optical and near-IR wavelengths, similar to that 
seen in the molecular ridge (N\,159-S is at the north tip of the 
molecular ridge).
As suggested in the study of molecular ridge \citep{Inetal08}, 
such difference likely results from the lower luminosity to
mass ratios in the regions studied.
In N\,44-N and N\,159-S, star formation occurred mostly in lower-mass 
or less rich clusters.
As these clusters do not fully sample the high-mass end of the
stellar initial mass function, they have a lower luminosity 
to mass ratio than the rich clusters, which are analyzed in \citet{Caetal07}.
In GMCs like N\,44-N and N\,159-S, SFR$_{H\alpha +24}$ would severely
underestimate the star formation rate by almost an order of magnitude.
The third GMC with SFR$_{H\alpha +24} <$ SFR$_{\rm YSO}$, N\,159-W, 
has massive main-sequence stars, and it is still actively forming
new massive stars (\S 5.2), similar to the three GMCs with
SFR$_{H\alpha +24} >$ SFR$_{\rm YSO}$.  
The ratios of current-to-past massive star formation rates,
SFR$_{\rm YSO}$/SFR$_{H\alpha +24}$, of these four GMCs range
from 0.4 to 2.1.
These ratios suggest that the current star formation rate 
may be lower or even higher than the average star formation rate
in the last few Myr, not constant over time.
In active star forming GMCs, SFR$_{H\alpha +24}$ would miss
the current star formation, which amounts to 30--70\% of the total
star formation rate (calculated from the above four GMCs).

\begin{figure}
\plotone{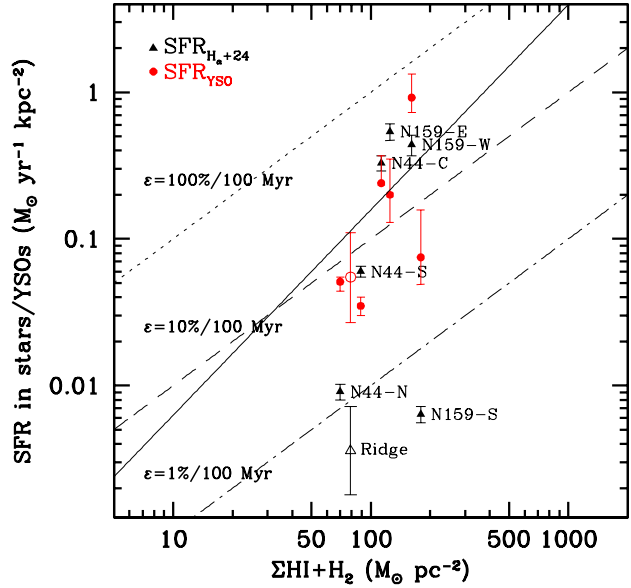}
\caption{Relation between the SFR per unit area and gas density for
 six GMCs in N\,44 and N\,159.  Each GMC has two estimated SFRs using 
 different tracers, SFR$_{{\rm H}\alpha+24}$ (filled triangles and GMC 
 name labeled) and SFR$_{\rm YSO}$ (filled circles).  For comparison, 
 SFRs estimated in the molecular ridge \citep{Inetal08} are also plotted,
 i.e., SFR$_{{\rm H}\alpha+24}$ in open triangle and SFR$_{\rm YSO}$ in
 open circle.  The solid line is the S-K relation, and the dotted, dashed, 
 dashed-dotted lines correspond to constant SFRs per unit gas mass, in 
 units of 1\%, 10\%, and 100\% per 100 Myr, commonly used in estimating 
 SFRs of a galaxy \citep{Ke98}. GMCs without prominent \hii\
 regions, i.e.,  N\,44-N,  N\,159-S, and the molecular ridge, have
 SFR$_{{\rm H}\alpha+24}$ 11--56 times smaller than expected from the S-K
 relation, but their SFR$_{\rm YSO}$ are in better agreement with the S-K
 relation. \label{fig:sfr}}
\end{figure}

Finally, we use the star formation properties of the GMCs to 
evaluate the S-K relation.
The normalized SFR expected from the S-K relation for a region is:
\begin{equation}
{\rm SFR}(M_\odot~{\rm yr}^{-1}~{\rm kpc}^{-2}) =
 2.5\times10^{-4}(\frac{\Sigma}{M_\odot~{\rm pc}^{-2}})^{1.4}
\end{equation}
where $\Sigma$ is the sum of molecular and atomic gas surface densities 
\citep{Ke98}.
The average molecular surface densities are measured from CO maps,
adopting $X_{\rm CO}=7\times10^{20}$ and $3.9\times10^{20}$~cm$^{-2}$ 
(K km~s$^{-1}$)$^{-1}$ for N\,44 and N\,159 from \citet{FYetal08} and 
\citet{Pietal09}, respectively.
The atomic surface densities are measured from HI maps \citep{KSetal03}. 
Then the expected SFRs from the total gas surface densities, 
SFR$_{\Sigma}$, are given in Table~\ref{gmc}.
As shown in Figure~\ref{fig:sfr}, among the six GMCs, four of them 
have SFR$_{\Sigma}$ and SFR$_{H\alpha +24}$ in agreement 
within a factor of 3.
The remaining two GMCs, N\,44-N and N\,159-S, have 
SFR$_{\Sigma} \sim 11$ and 56 times SFR$_{H\alpha +24}$,
respectively.
When compared with SFR$_{\rm YSO}$, which is a better
estimate of SFRs in these two low-luminosity regions, 
N\,44-N is within a factor of 2 of SFR$_{\Sigma}$, while 
N\,159-S is still only $1/5$ SFR$_{\Sigma}$.
The GMC N\,159-S appears very different from the other 
five GMCs as its star formation rate, estimated 
with a comprehensive YSO inventory, is still much lower 
than that expected from the S-K relation.
It is known that the S-K relation does not apply to small 
regions; thus is is not surprising that these GMCs with 
region sizes ranging from 45 to 135 pc do not follow the
S-K relation in star formation. 
It is interesting that the SFR averaged over all GMCs 
within the entire N44 complex or the entire N159 complex
is very close to that expected from the S-K relation.

\section{Summary}

We have studied the \hii\ complex N\,159 in the LMC with 
{\it Spitzer} IRAC and MIPS data at 3.6, 4.5, 5.8, 8.0, 24, 70, 
and 160 \um\ and CTIO Blanco 4~m ISPI in the $JK_s$ and MOSAIC in
the \ha\ bands.
Following procedures outlined in Paper I, we first identified YSOs
using the criteria [4.5]$-$[8.0] $\ge 2.0$ to exclude normal and 
evolved stars and [8.0] $<$ 14.0$-$([4.5]$-$[8.0]) to exclude 
background galaxies.
A total of 52 YSO candidates were identified.
We then inspected the SED and close-up images of each YSO candidate
in \ha, $VRIJK_s$, IRAC, and MIPS bands simultaneously to further
identify evolved stars, galaxies, and dust clumps, resulting a 
sample of 27 YSO candidates that are most likely bona fide YSOs
of high and intermediate masses.   
We classified these YSOs into Type I, II, and III according to 
their SED shapes.

In our sample of 27 YSOs, $\sim 74$\% of them are resolved 
into multiple components or extended sources in high-resolution 
$JK_s$ images.  
To assess the physical properties of the YSOs, we have used 
the Online SED Model Fitter \citep{RTetal07} to model SEDs of 
these YSOs and further analyzed in detail 22 YSOs that appear 
single or dominant within a group.
We find good fits for Types I and II YSOs, though they show 
modest deviations between their observed SEDs and the models at 5.8 and 8.0\um, 
because the models do not include PAH emission.  
We have used a YSO with a {\it Spitzer} IRS observation to estimate
the fraction of PAH emission in IRAC 3.6, 5.8, and 8.0 \um\ bands and 
modeled the SED with aromatic emission removed.
That SED is well fit by the models, and comparisons between SEDs
before and after PAH correction show that mass and total luminosity
of the YSO is not significantly affected by including PAHs in models.

Some of the Type III YSOs show large deviations from the models at optical 
wavelengths.
This is due to the modest angular resolution of the  
MCPS $UBVI$ catalog since the nebular line emission from small 
\hii\ regions surrounding massive YSOs is not resolved from the 
point source and contributes to the broad-band photometry.
We have also examined the effect of multiplicity (due to inadequate 
angular resolution) on inferring parameters of YSOs from SED fits 
by modeling a group of YSOs resolved in the VLT/NACO adaptive 
optics $JK_s$ images.  
We find that for a YSO of multiple sources at similar evolutionary
stages, the integrated fluxes may result in up to 20\% over-estimate 
in mass of the most massive component, but up to $\sim$ 40\% 
under-estimate in the total mass of the system.
For a YSO of sources at mixed evolutionary stages, the mid-IR 
luminosity is a good estimate for the system's total luminosity, 
but high-resolution optical and near-IR images are needed to 
separate main-sequence components from the YSO components.

YSO counterparts are found in one maser, three UCHIIs, and nine 
candidate HAeBe stars.  
The maser is associated with a Type I YSO and the majority of the 
HAeBe stars correspond to Type II and III YSOs, supporting the 
evolutionary sequence of our empirical classification.
The maser has IRS spectral features similar to the ``P'' group 
classification used in \citet{Seetal09}, while the three UCHIIs are 
similar to the ``PE'' group.
Masers are typically associated with YSOs less evolved than UCHIIs, 
supporting the proposed spectral evolutionary sequence that YSOs with the ``PE''
spectral type are more evolved than those with the ``P'' spectral
type.
We have further found that for YSOs associated with ultracompact \hii\ regions,
(proto) stellar masses determined from SED model fits agree well with those 
estimated from the ionization requirements of the \hii\ regions.
Using the SED model fits, we find at least 9 O-type YSOs in N\,159.

N\,159 encompasses three molecular concentrations with different
star formation histories and intensities.
O-type YSOs are found in the two GMCs that are associated with 
ionized gas, i.e., where massive stars have formed a few Myr ago, 
while no O-type YSOs are found in the third GMC that shows no 
signs of massive star formation in the optical and near-IR 
wavelengths.  
This result is similar to that seen in YSOs in the three GMCs in 
N\,44 (Paper I), indicating that energy feedback plays a role in 
the formation of massive YSOs.
However it remains unclear whether the less active GMCs will form 
massive stars later.  
Although the uncertainties in $\Delta V$ are large, 
the GMC N\,159-W does have the smallest $\Delta V$ and 
$M_{\rm vir}/M_{\rm lum}$ ratio and also contains the largest number 
and most concentrated distribution of YSOs, consistent with the
hypothesis that concentrated clusters are formed in a less
gravitationally stable region.
A smaller $M_{\rm vir}/M_{\rm lum}$ ratio is also found in the 
GMC N\,44-C which has the highest number and the most concentrated 
distribution of YSOs.

To investigate whether star formation may be triggered or 
spontaneous in GMCs in N\,159, we have performed a detailed
comparison between the mid-IR YSO distribution and 
the distribution of already formed massive 
stars and their associated ionized gas.
We find that the current star formation in the GMC N\,159-E
is likely to be triggered by \hii\ regions expanding into 
the molecular cloud, while the massive YSOs in GMC N\,159-W are 
more likely forming spontaneously.

Finally, we estimated star formation efficiencies and rates by counting the 
massive YSOs, and compared these rates to those inferred from integrated 
\ha\ and 24 \um\ luminosities. 
We also compared with expected rates calculated from gas surface densities
and using the Schmidt-Kennicutt extragalactic scaling relation.
In GMCs with relatively high levels of activity, such as N159\,-E and -W, 
the various measures are consistent. 
However, in more quiescent regions such as N\,159-S and the giant molecular 
ridge which continues south from there, we find that star formation is 
distributed in relatively low luminosity regions 
and dominated by current (mid-IR detected) star formation, with a lack of 
massive main-sequence stars (representing few-Myr old activity).
Star formation rates derived from YSO counting are higher than those
predicted by total \ha\ and 24 \um\ luminosities, and both are lower
than would be implied by the total gas surface density.  This
discrepancy can be explained if the star formation sites are poor
clusters which do not fully sample the stellar mass function (so their
mass-to-light ratios are elevated).  Alternatively, the youngest and
most embedded still accreting massive protostars have elevated
mass-to-light ratios due to late spectral type pre-main-sequence
photospheres and possible having not yet accreted their final mass.
Either possibility indicates unusual conditions in N\,159-S and a
breakdown of the scaling laws determined by averaging over kiloparsec
scales in galaxies.  Detailed studies such as this of archival {\it
Spitzer} data, soon to be complemented by detailed molecular gas
properties obtained with ALMA, are revealing the full complexity of
extragalactic star formation.

\acknowledgments
We thank the anonymous referee for the careful reading of 
the manuscript to improve the paper.
We also thank Tony Wong for providing the MAGMA CO map.
This work is supported through NASA grants JPL 1282653 
and 1288328.  R.~A.~G. and J.~P.~S. acknowledge supports
from JPL grant 1316421 and NSF grant AST 08-07323.
M.~M. acknowledges support in part from {\it Spitzer} 
grant 1275598 and NASA grant NAG5-12595. 
This study made use of data products of the Two Micron 
All Sky Survey, which is a joint project of the University 
of Massachusetts and the Infrared Processing and Analysis 
Center/California Institute of Technology, funded by the 
National Aeronautics and the Space Administration and the 
National Science Foundation.


\begin{thebibliography}{85}
\expandafter\ifx\csname natexlab\endcsname\relax\def\natexlab#1{#1}\fi

\bibitem[{{Allen} {et~al.}(2004)}]{ALetal04}
{Allen}, L.~E., {et~al.} 2004, \apjs, 154, 363

\bibitem[{{Bica} {et~al.}(1996){Bica}, {Claria}, {Dottori}, {Santos}, \&
  {Piatti}}]{Bietal96}
{Bica}, E., {Claria}, J.~J., {Dottori}, H., {Santos}, Jr., J.~F.~C., \&
  {Piatti}, A.~E. 1996, \apjs, 102, 57

\bibitem[{{Bigiel} {et~al.}(2008)}]{Bietal08}
{Bigiel}, F., {et~al.} 2008, \aj, 136, 2846

\bibitem[{{Blum} {et~al.}(2006)}]{Bletal06}
{Blum}, R.~D., {et~al.} 2006, \aj, 132, 2034

\bibitem[{{Bolatto} {et~al.}(2008){Bolatto}, {Leroy}, {Rosolowsky}, {Walter},
  \& {Blitz}}]{Boetal08}
{Bolatto}, A.~D., {Leroy}, A.~K., {Rosolowsky}, E., {Walter}, F., \& {Blitz},
  L. 2008, \apj, 686, 948

\bibitem[{{Bonnell} {et~al.}(2003){Bonnell}, {Bate}, \& {Vine}}]{BBV03}
{Bonnell}, I.~A., {Bate}, M.~R., \& {Vine}, S.~G. 2003, \mnras, 343, 413

\bibitem[{{Calzetti} {et~al.}(2007)}]{Caetal07}
{Calzetti}, D., {et~al.} 2007, \apj, 666, 870

\bibitem[{{Caplan} \& {Deharveng}(1986)}]{CD86}
{Caplan}, J., \& {Deharveng}, L. 1986, \aap, 155, 297

\bibitem[{{Chen} {et~al.}(2009){Chen}, {Chu}, {Gruendl}, {Gordon}, \&
  {Heitsch}}]{CCetal09}
{Chen}, C.-H.~R., {Chu}, Y.-H., {Gruendl}, R.~A., {Gordon}, K.~D., \&
  {Heitsch}, F. 2009, \apj, 695, 511

\bibitem[{{Chu} {et~al.}(1997){Chu}, {Kennicutt}, {Snowden}, {Smith},
  {Williams}, \& {Bomans}}]{Chetal97}
{Chu}, Y.-H., {Kennicutt}, R.~C., {Snowden}, S.~L., {Smith}, R.~C., {Williams},
  R.~M., \& {Bomans}, D.~J. 1997, \pasp, 109, 554

\bibitem[{{Churchwell}(2002)}]{CE02}
{Churchwell}, E. 2002, \araa, 40, 27

\bibitem[{{Cowley} {et~al.}(1995){Cowley}, {Schmidtke}, {Anderson}, \&
  {McGrath}}]{Coetal95}
{Cowley}, A.~P., {Schmidtke}, P.~C., {Anderson}, A.~L., \& {McGrath}, T.~K.
  1995, \pasp, 107, 145

\bibitem[{{Deharveng} \& {Caplan}(1992)}]{DC92}
{Deharveng}, L., \& {Caplan}, J. 1992, \aap, 259, 480

\bibitem[{{Desai} {et~al.}(2010)}]{Deetal10}
{Desai}, K.~M., {et~al.} 2010, \apj, in press (ArXiv e-prints1006.3344)

\bibitem[{{Dickel} {et~al.}(2005){Dickel}, {McIntyre}, {Gruendl}, \&
  {Milne}}]{Dietal05}
{Dickel}, J.~R., {McIntyre}, V.~J., {Gruendl}, R.~A., \& {Milne}, D.~K. 2005,
  \aj, 129, 790

\bibitem[{{Draine} \& {Li}(2007)}]{DL07}
{Draine}, B.~T., \& {Li}, A. 2007, \apj, 657, 810

\bibitem[{{Ellingsen} {et~al.}(2007){Ellingsen}, {Voronkov}, {Cragg},
  {Sobolev}, {Breen}, \& {Godfrey}}]{Eletal07}
{Ellingsen}, S.~P., {Voronkov}, M.~A., {Cragg}, D.~M., {Sobolev}, A.~M.,
  {Breen}, S.~L., \& {Godfrey}, P.~D. 2007, in IAU Symposium, Vol. 242, IAU
  Symposium, ed. {J.~M.~Chapman \& W.~A.~Baan}, 213--217

\bibitem[{{Fari{\~n}a} {et~al.}(2009){Fari{\~n}a}, {Bosch}, {Morrell},
  {Barb{\'a}}, \& {Walborn}}]{Faetal09}
{Fari{\~n}a}, C., {Bosch}, G.~L., {Morrell}, N.~I., {Barb{\'a}}, R.~H., \&
  {Walborn}, N.~R. 2009, \aj, 138, 510

\bibitem[{{Fazio} {et~al.}(2004)}]{FGetal04}
{Fazio}, G.~G., {et~al.} 2004, \apjs, 154, 10

\bibitem[{{Feast}(1999)}]{Fe99}
{Feast}, M. 1999, in IAU Symp. 190: New Views of the Magellanic Clouds, ed.
  Y.-H. {Chu}, N.~{Suntzeff}, J.~{Hesser}, \& D.~{Bohlender}, 542

\bibitem[{{Fleener} {et~al.}(2010){Fleener}, {Payne}, {Chu}, {Chen}, \&
  {Gruendl}}]{Fletal10}
{Fleener}, C.~E., {Payne}, J.~T., {Chu}, Y., {Chen}, C., \& {Gruendl}, R.~A.
  2010, \aj, 139, 158

\bibitem[{{Forbrich} {et~al.}(2009){Forbrich}, {Lada}, {Muench}, {Alves}, \&
  {Lombardi}}]{Foetal09}
{Forbrich}, J., {Lada}, C.~J., {Muench}, A.~A., {Alves}, J., \& {Lombardi}, M.
  2009, \apj, 704, 292

\bibitem[{{Franco} {et~al.}(2000){Franco}, {Kurtz}, {Garc{\'{\i}}a-Segura}, \&
  {Hofner}}]{Fr00}
{Franco}, J., {Kurtz}, S.~E., {Garc{\'{\i}}a-Segura}, G., \& {Hofner}, P. 2000,
  \apss, 272, 169

\bibitem[{{Fukui} {et~al.}(2008)}]{FYetal08}
{Fukui}, Y., {et~al.} 2008, \apjs, 178, 56

\bibitem[{{Gatley} {et~al.}(1981){Gatley}, {Becklin}, {Hyland}, \&
  {Jones}}]{Gaetal81}
{Gatley}, I., {Becklin}, E.~E., {Hyland}, A.~R., \& {Jones}, T.~J. 1981,
  \mnras, 197, 17P

\bibitem[{{Gorjian} {et~al.}(2004)}]{GVetal04}
{Gorjian}, V., {et~al.} 2004, \apjs, 154, 275

\bibitem[{{Groenewegen}(2006)}]{Gr06}
{Groenewegen}, M.~A.~T. 2006, \aap, 448, 181

\bibitem[{{Gruendl} \& {Chu}(2009)}]{GC09}
{Gruendl}, R.~A., \& {Chu}, Y. 2009, \apjs, 184, 172

\bibitem[{{Henize}(1956)}]{He56}
{Henize}, K.~G. 1956, \apjs, 2, 315

\bibitem[{{Heydari-Malayeri} {et~al.}(1999){Heydari-Malayeri}, {Rosa},
  {Charmandaris}, {Deharveng}, \& {Zinnecker}}]{HMetal99}
{Heydari-Malayeri}, M., {Rosa}, M.~R., {Charmandaris}, V., {Deharveng}, L., \&
  {Zinnecker}, H. 1999, \aap, 352, 665

\bibitem[{{Hillenbrand} {et~al.}(1992){Hillenbrand}, {Strom}, {Vrba}, \&
  {Keene}}]{HLetal92}
{Hillenbrand}, L.~A., {Strom}, S.~E., {Vrba}, F.~J., \& {Keene}, J. 1992, \apj,
  397, 613

\bibitem[{{Huff} \& {Stahler}(2007)}]{HS07}
{Huff}, E.~M., \& {Stahler}, S.~W. 2007, \apj, 666, 281

\bibitem[{{Hughes} {et~al.}(2010)}]{Huetal10}
{Hughes}, A., {et~al.} 2010, \mnras, 873

\bibitem[{{Hunt} \& {Whiteoak}(1994)}]{HW94}
{Hunt}, M.~R., \& {Whiteoak}, J.~B. 1994, Proceedings of the Astronomical
  Society of Australia, 11, 68

\bibitem[{{Indebetouw} {et~al.}(2004){Indebetouw}, {Johnson}, \&
  {Conti}}]{IJC04}
{Indebetouw}, R., {Johnson}, K.~E., \& {Conti}, P. 2004, \aj, 128, 2206

\bibitem[{{Indebetouw} {et~al.}(2005)}]{Inetal05}
{Indebetouw}, R., {et~al.} 2005, \apj, 619, 931

\bibitem[{{Indebetouw} {et~al.}(2008)}]{Inetal08}
---. 2008, \aj, 136, 1442

\bibitem[{{Johansson} {et~al.}(1998)}]{Joetal98}
{Johansson}, L.~E.~B., {et~al.} 1998, \aap, 331, 857

\bibitem[{{Jones} {et~al.}(2005){Jones}, {Woodward}, {Boyer}, {Gehrz}, \&
  {Polomski}}]{JTetal05}
{Jones}, T.~J., {Woodward}, C.~E., {Boyer}, M.~L., {Gehrz}, R.~D., \&
  {Polomski}, E. 2005, \apj, 620, 731

\bibitem[{{Kato} {et~al.}(2007)}]{Kaetal07}
{Kato}, D., {et~al.} 2007, \pasj, 59, 615

\bibitem[{{Kennicutt} {et~al.}(2009)}]{Keetal09}
{Kennicutt}, R.~C., {et~al.} 2009, \apj, 703, 1672

\bibitem[{{Kennicutt}(1989)}]{Ke89}
{Kennicutt}, Jr., R.~C. 1989, \apj, 344, 685

\bibitem[{{Kennicutt}(1998)}]{Ke98}
---. 1998, \apj, 498, 541

\bibitem[{{Kim} {et~al.}(2003){Kim}, {Staveley-Smith}, {Dopita}, {Sault},
  {Freeman}, {Lee}, \& {Chu}}]{KSetal03}
{Kim}, S., {Staveley-Smith}, L., {Dopita}, M.~A., {Sault}, R.~J., {Freeman},
  K.~C., {Lee}, Y., \& {Chu}, Y. 2003, \apjs, 148, 473

\bibitem[{{Klessen} {et~al.}(1998){Klessen}, {Burkert}, \& {Bate}}]{Kletal98}
{Klessen}, R.~S., {Burkert}, A., \& {Bate}, M.~R. 1998, \apjl, 501, L205

\bibitem[{{Koornneef}(1983)}]{Ko83}
{Koornneef}, J. 1983, \aap, 128, 84

\bibitem[{{Krumholz} {et~al.}(2009){Krumholz}, {McKee}, \& {Tumlinson}}]{KMT09}
{Krumholz}, M.~R., {McKee}, C.~F., \& {Tumlinson}, J. 2009, \apj, 699, 850

\bibitem[{{Lazendic} {et~al.}(2002){Lazendic}, {Whiteoak}, {Klamer},
  {Harbison}, \& {Kuiper}}]{Lazetal02}
{Lazendic}, J.~S., {Whiteoak}, J.~B., {Klamer}, I., {Harbison}, P.~D., \&
  {Kuiper}, T.~B.~H. 2002, \mnras, 331, 969

\bibitem[{{Leroy} {et~al.}(2008)}]{Leetal08}
{Leroy}, A.~K., {et~al.} 2008, \aj, 136, 2782

\bibitem[{{Li} \& {Draine}(2001)}]{LD01}
{Li}, A., \& {Draine}, B.~T. 2001, \apj, 554, 778

\bibitem[{{Li} \& {Draine}(2002)}]{LD02}
---. 2002, \apj, 576, 762

\bibitem[{{Lucke} \& {Hodge}(1970)}]{LH70}
{Lucke}, P.~B., \& {Hodge}, P.~W. 1970, \aj, 75, 171

\bibitem[{{Massey}(1985)}]{Ma85}
{Massey}, P. 1985, \pasp, 97, 5

\bibitem[{{Meixner} {et~al.}(2006)}]{MMetal06}
{Meixner}, M., {et~al.} 2006, \aj, 132, 2268

\bibitem[{{Mizuno} {et~al.}(2001)}]{MNetal01}
{Mizuno}, N., {et~al.} 2001, \pasj, 53, 971

\bibitem[{{Nakajima} {et~al.}(2005)}]{Naetal05}
{Nakajima}, Y., {et~al.} 2005, \aj, 129, 776

\bibitem[{{Ott} {et~al.}(2008)}]{Otetal08}
{Ott}, J., {et~al.} 2008, \pasa, 25, 129

\bibitem[{{Pineda} {et~al.}(2009){Pineda}, {Ott}, {Klein}, {Wong}, {Muller}, \&
  {Hughes}}]{Pietal09}
{Pineda}, J.~L., {Ott}, J., {Klein}, U., {Wong}, T., {Muller}, E., \& {Hughes},
  A. 2009, \apj, 703, 736

\bibitem[{{Pomar{\`e}s} {et~al.}(2009)}]{Poetal09}
{Pomar{\`e}s}, M., {et~al.} 2009, \aap, 494, 987

\bibitem[{{Pottasch}(1993)}]{Po93}
{Pottasch}, S.~R. 1993, in Infrared Astronomy, ed. A.~{Mampaso}, M.~{Prieto},
  \& F.~{Sanchez}, 63

\bibitem[{{Rieke} {et~al.}(2004)}]{RGetal04}
{Rieke}, G.~H., {et~al.} 2004, \apjs, 154, 25

\bibitem[{{Robitaille} {et~al.}(2007){Robitaille}, {Whitney}, {Indebetouw}, \&
  {Wood}}]{RTetal07}
{Robitaille}, T.~P., {Whitney}, B.~A., {Indebetouw}, R., \& {Wood}, K. 2007,
  \apjs, 169, 328

\bibitem[{{Robitaille} {et~al.}(2006){Robitaille}, {Whitney}, {Indebetouw},
  {Wood}, \& {Denzmore}}]{RTetal06}
{Robitaille}, T.~P., {Whitney}, B.~A., {Indebetouw}, R., {Wood}, K., \&
  {Denzmore}, P. 2006, \apjs, 167, 256

\bibitem[{{Schaerer} \& {de Koter}(1997)}]{SdK97}
{Schaerer}, D., \& {de Koter}, A. 1997, \aap, 322, 598

\bibitem[{{Schaerer} {et~al.}(1993){Schaerer}, {Meynet}, {Maeder}, \&
  {Schaller}}]{SDetal93}
{Schaerer}, D., {Meynet}, G., {Maeder}, A., \& {Schaller}, G. 1993, \aaps, 98,
  523

\bibitem[{{Schmidt-Kaler}(1982)}]{SK82}
{Schmidt-Kaler}, T. 1982, {in Landolt-B{\"o}rnstein New Series, Group 6, Volume
  2b, Stars and Star Clusters}, ed. K.~{Schaifers} \& H.~H. {Voigt} (Berlin:
  Springer)

\bibitem[{{Seale} {et~al.}(2009)}]{Seetal09}
{Seale}, J.~P., {et~al.} 2009, \apj, 699, 150

\bibitem[{{Seward} {et~al.}(2010){Seward}, {Williams}, {Chu}, {Gruendl}, \&
  {Dickel}}]{SFetal10}
{Seward}, F.~D., {Williams}, R.~M., {Chu}, Y., {Gruendl}, R.~A., \& {Dickel},
  J.~R. 2010, \aj, 140, 177

\bibitem[{{Simon} {et~al.}(2007)}]{Setal07}
{Simon}, J.~D., {et~al.} 2007, \apj, 669, 327

\bibitem[{{Skrutskie} {et~al.}(2006)}]{SMetal06}
{Skrutskie}, M.~F., {et~al.} 2006, \aj, 131, 1163

\bibitem[{{Smith} \& {The MCELS Team}(1999)}]{SRetal99}
{Smith}, R.~C., \& {The MCELS Team}. 1999, in IAU Symp. 190: New Views of the
  Magellanic Clouds, ed. Y.-H. {Chu}, N.~{Suntzeff}, J.~{Hesser}, \&
  D.~{Bohlender}, 28

\bibitem[{{Testor} {et~al.}(2006){Testor}, {Lemaire}, {Field}, \&
  {Diana}}]{Teetal06}
{Testor}, G., {Lemaire}, J.~L., {Field}, D., \& {Diana}, S. 2006, \aap, 453,
  517

\bibitem[{{Testor} {et~al.}(2007){Testor}, {Lemaire}, {Kristensen}, {Field}, \&
  {Diana}}]{Teetal07}
{Testor}, G., {Lemaire}, J.~L., {Kristensen}, L.~E., {Field}, D., \& {Diana},
  S. 2007, \aap, 469, 459

\bibitem[{{Vaidya} {et~al.}(2009){Vaidya}, {Chu}, {Gruendl}, {Chen}, \&
  {Looney}}]{Vaetal09}
{Vaidya}, K., {Chu}, Y., {Gruendl}, R.~A., {Chen}, C., \& {Looney}, L.~W. 2009,
  \apj, 707, 1417

\bibitem[{{van der Bliek} {et~al.}(2004)}]{vdBetal04}
 van der Bliek, N.~S., et al.\ 2004, \procspie, 5492, 1582 

\bibitem[{{Vanhala} \& {Cameron}(1998)}]{VC98}
{Vanhala}, H.~A.~T., \& {Cameron}, A.~G.~W. 1998, \apj, 508, 291

\bibitem[{{Walter} {et~al.}(2008)}]{Waetal08}
{Walter}, F., {et~al.} 2008, \aj, 136, 2563

\bibitem[{{Whitney} {et~al.}(2004){Whitney}, {Indebetouw}, {Bjorkman}, \&
  {Wood}}]{WBetal04b}
{Whitney}, B.~A., {Indebetouw}, R., {Bjorkman}, J.~E., \& {Wood}, K. 2004,
  \apj, 617, 1177

\bibitem[{{Whitney} {et~al.}(2003){Whitney}, {Wood}, {Bjorkman}, \&
  {Wolff}}]{BAW1}
{Whitney}, B.~A., {Wood}, K., {Bjorkman}, J.~E., \& {Wolff}, M.~J. 2003, \apj,
  591, 1049

\bibitem[{{Whitney} {et~al.}(2008)}]{WBetal08}
{Whitney}, B.~A., {et~al.} 2008, \aj, 136, 18

\bibitem[{{Williams} {et~al.}(2000){Williams}, {Petre}, {Chu}, \&
  {Chen}}]{Wietal00}
{Williams}, R.~M., {Petre}, R., {Chu}, Y.-H., \& {Chen}, C.-H.~R. 2000, \apjl,
  536, L27

\bibitem[{{Wood} {et~al.}(2008){Wood}, {Whitney}, {Robitaille}, \&
  {Draine}}]{Woetal08}
{Wood}, K., {Whitney}, B.~A., {Robitaille}, T., \& {Draine}, B.~T. 2008, \apj,
  688, 1118

\bibitem[{{Yorke} \& {Sonnhalter}(2002)}]{YS02}
{Yorke}, H.~W., \& {Sonnhalter}, C. 2002, \apj, 569, 846

\bibitem[{{Zaritsky} {et~al.}(2004){Zaritsky}, {Harris}, {Thompson}, \&
  {Grebel}}]{ZDetal04}
{Zaritsky}, D., {Harris}, J., {Thompson}, I.~B., \& {Grebel}, E.~K. 2004, \aj,
  128, 1606

\bibitem[{{Zavagno} {et~al.}(2007){Zavagno}, {Pomar{\`e}s}, {Deharveng},
  {Hosokawa}, {Russeil}, \& {Caplan}}]{ZAetal07}
{Zavagno}, A., {Pomar{\`e}s}, M., {Deharveng}, L., {Hosokawa}, T., {Russeil},
  D., \& {Caplan}, J. 2007, \aap, 472, 835

\end{thebibliography}

\clearpage
\newpage

\begin{figure}
\epsscale{0.9}
\caption{IRAC and MIPS images of N\,159.
 (a) 3.6 \um\ image showing stars and modest PAH emission, with OB
 association LH105 labeled; (b) 8.0 \um\ image showing PAH and dust
 emissions, (c) 24 \um\ image showing dust emission and overlaid with 
 CO contours from \citet{Joetal98} , and (d) color composite of 3.6, 8.0,
 and 24 \um\ images.  Dust shrouded objects, e.g., YSOs and AGB stars,
 appear red. 
 \label{fig:n159img}}
\end{figure} 

\begin{figure}
\caption{(a) 8 \um\ image of N\,159 overlaid with MAGMA CO intensity
 contours (blue) in n$^2$ K~km~s$^{-1}$ (n$=$1,2,3,...) and marked 
 with YSOs from three studies.  27 YSOs from this study are marked in
 the red circles, 5 YSOs from \citet{JTetal05} are marked in yellow
 triangles and labeled with numbers 1-5, and 4 YSOs from \citet{WBetal08}
 are marked in cyan boxes and labeled with numbers 1-4.  The FOV of the
 \citet{JTetal05} study is smaller and marked with dashed yellow lines.
 (b) YSOs in N\,159 from this study, \citet[][abbreviated as 
 J05 in the figure]{JTetal05}, and \citet[][abbreviated as 
 W08 in the figure]{WBetal08} marked on a [8.0] vs. ([4.5]$-$[8.0]) CMD.  
 Symbols are the same as in (a) and numbers are attached with an extra
 letter of J or W to separate samples from the \citet{JTetal05} and 
 \citet{WBetal08} studies.
 \label{fig:ysocom}}
\end{figure} 

\clearpage
\begin{figure}
\epsscale{0.8}
\plotone{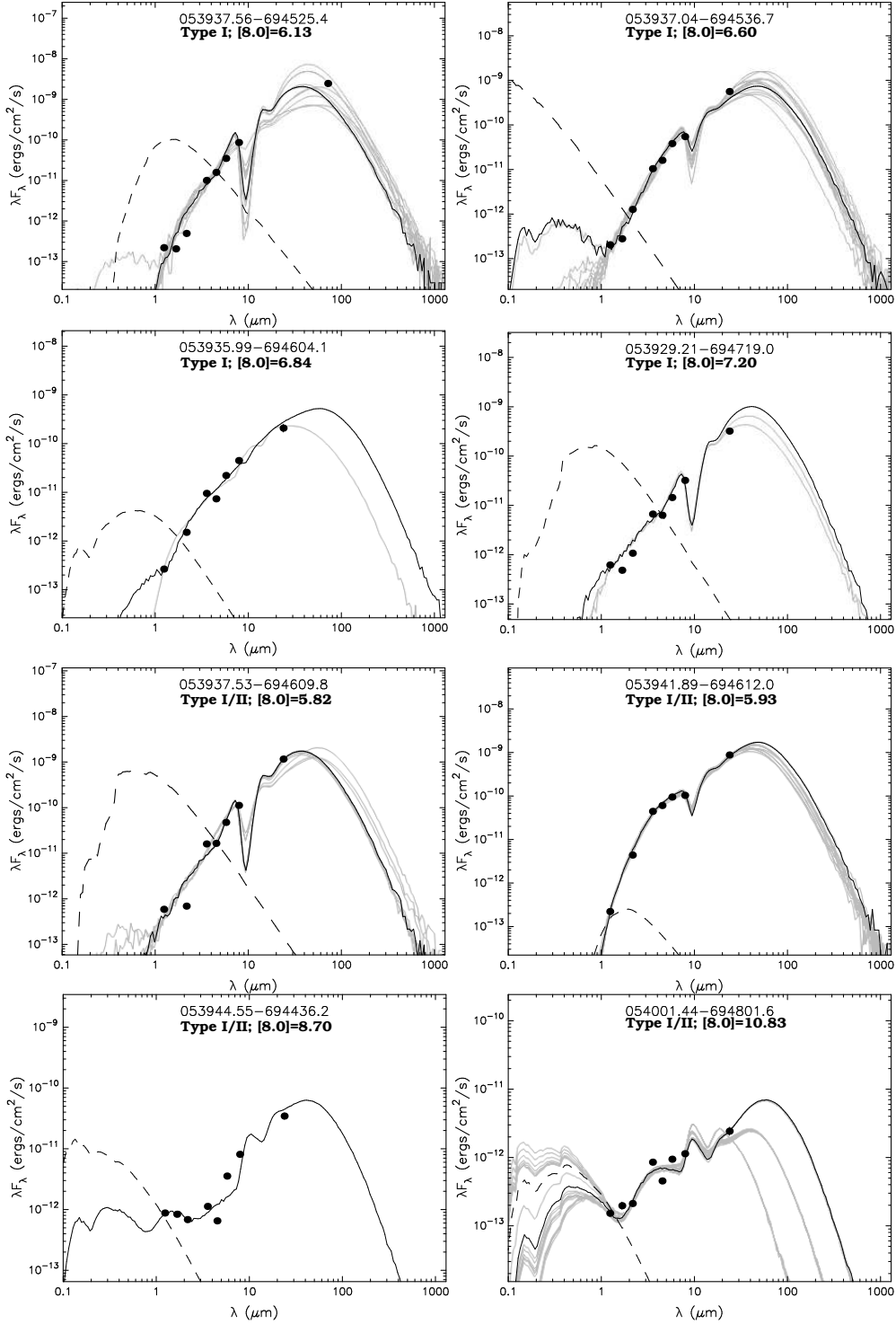}
\caption{SEDs of 27 YSOs analyzed in this study.  
 Filled circles are the flux values converted from magnitudes listed in 
 Table~\ref{ysoclass}.  The source name, Type from our empirical 
 classification, and [8.0] mag are labeled at the top of the plot.
 Triangles are upper limits.  Error bars are shown if larger than the data
 points. The solid black line shows the best-fit model, and the dashed black
 line illustrates the radiation from the central star reddened by the best-fit
 $A_V$.  The gray lines  show all acceptable models.  \label{fig:fit}}.
\end{figure} 

\clearpage
{\plotone{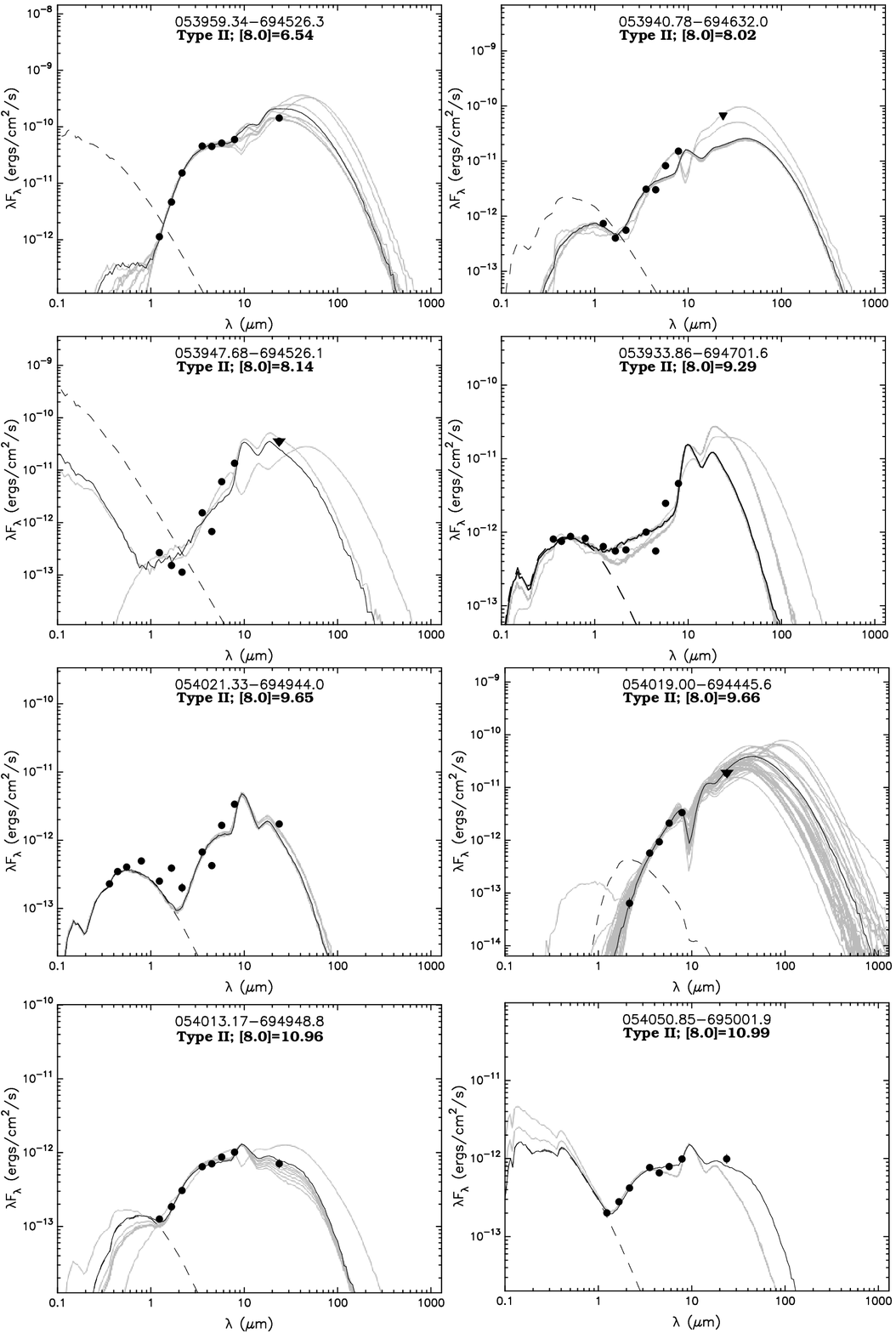}}\\[5mm]
\centerline{Figure~\ref{fig:fit} --- Continued.}

\clearpage
{\plotone{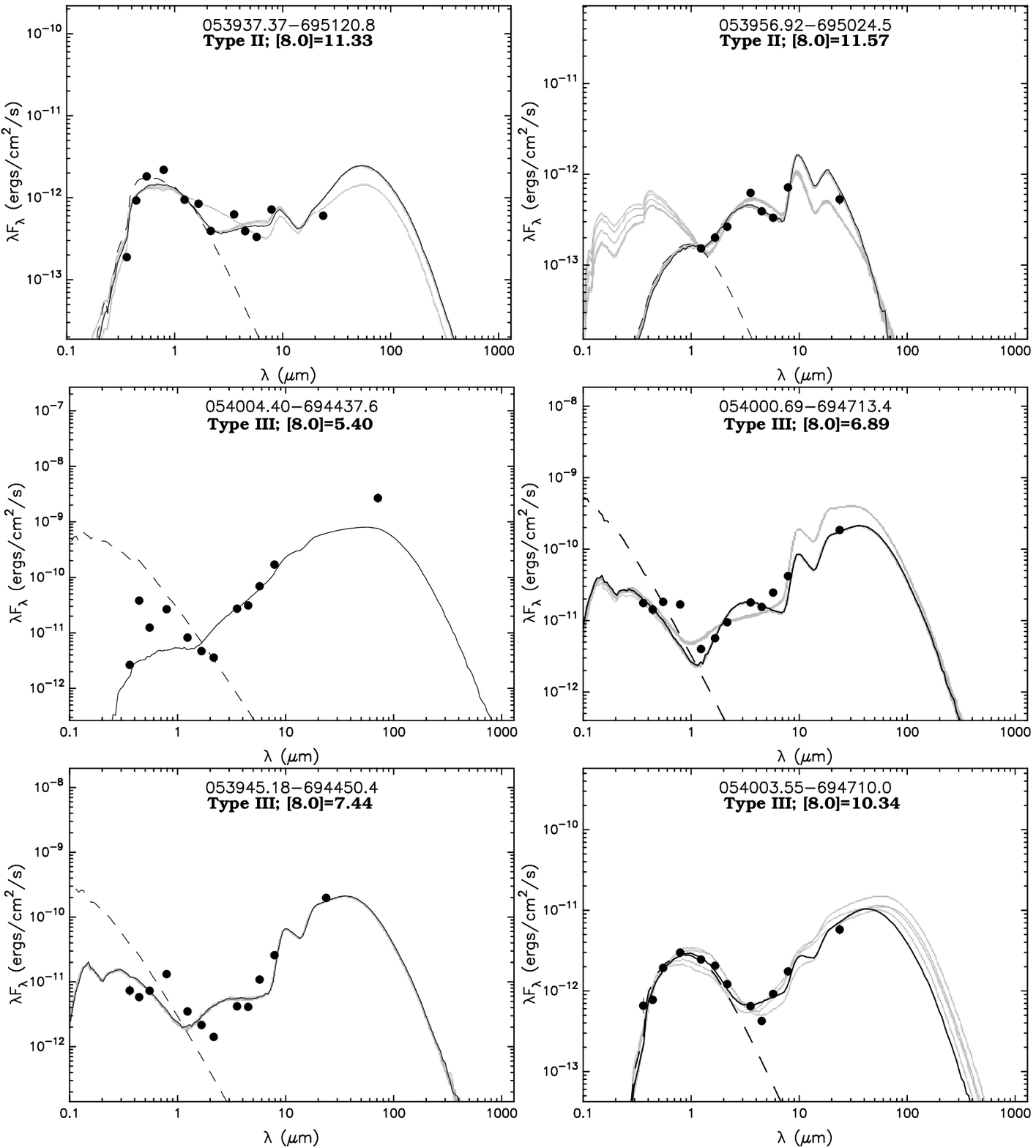}}\\[5mm]
\centerline{Figure~\ref{fig:fit} --- Continued.}

\clearpage
{\plotone{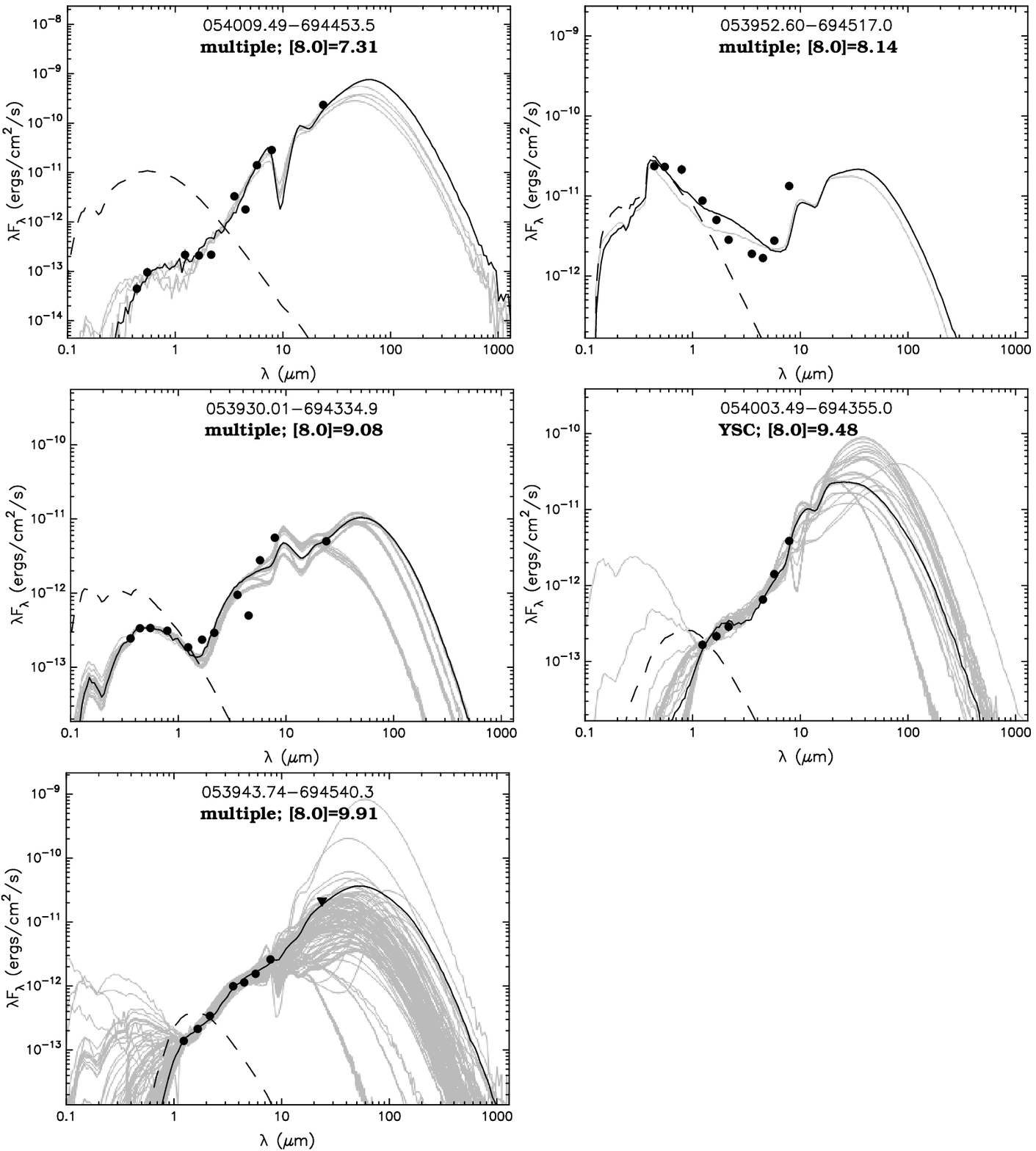}}\\[5mm]
\centerline{Figure~\ref{fig:fit} --- Continued.}

\epsscale{1.0}
\begin{figure}
\plotone{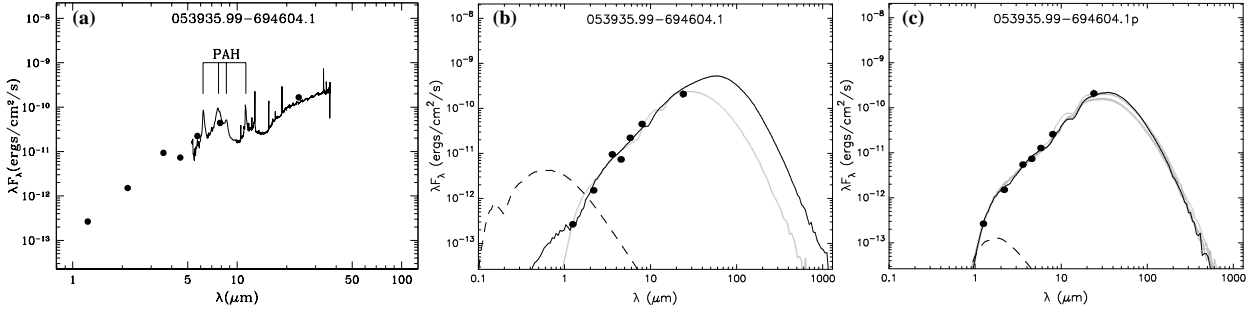}
\caption{(a)Multi-wavelength SED (filled circles, this study) and IRS spectrum
 \citep[solid line,][]{Seetal09} of YSO 053935.99$-$694604.1.  PAH features
 at 6.2, 7.7, 8.6, and 11.3 \um\ are labeled in the IRS spectrum.
 (b) and (c) Model fits to the SED before and after correcting contribution
 of PAH emission.  Symbols in (b) and (c) are the same as in
 Figure~\ref{fig:fit}.
 \label{fig:pah}}
\end{figure} 

\begin{figure}
\plotone{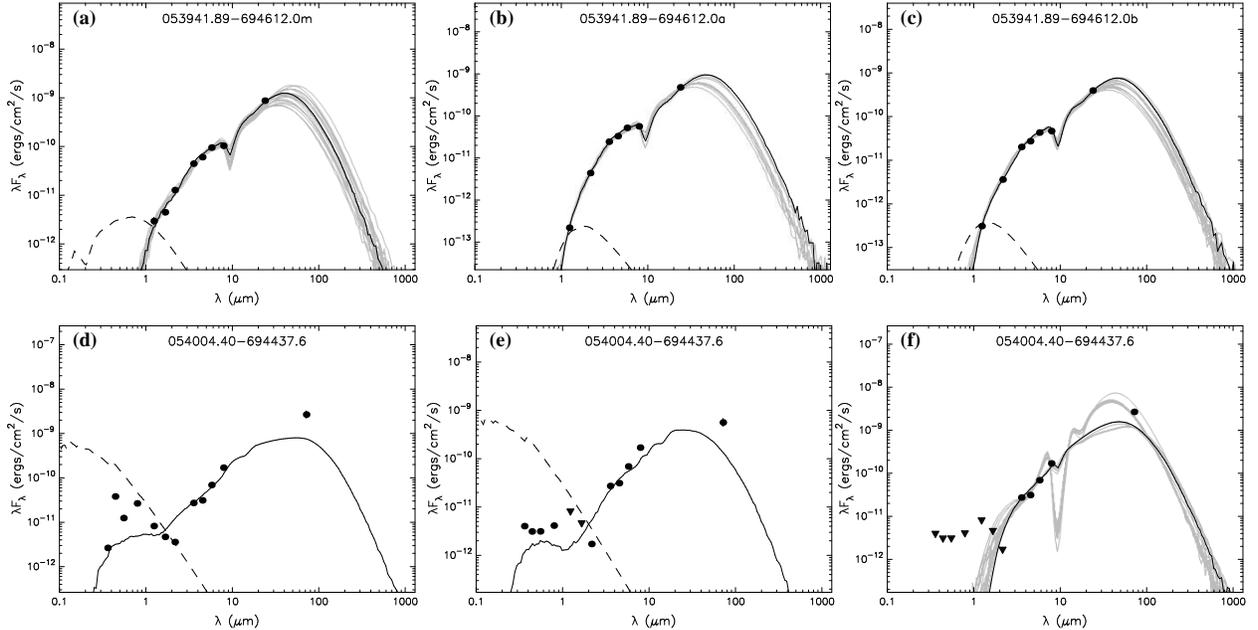}
\caption{Upper panel: model fits to a multiple system with sources at similar
 evolutionary stage.  (a) SED of 053941.89$-$694612.0 from integrated fluxes
 of the system. (b,c) SEDs of the two brightest YSOs, labeled as a and b
 respectively in the figure, from their NACO $K_s$ luminosities and
 proportioned IRAC and MIPS fluxes (according to the $K_s$ luminosity).
 Lower panel: model fits to a multiple system with sources at different
 different evolutionary stages.  Optical and near-IR segments of SEDs of
 YSO 054004.40$-$694437.6 are constructed from (d) $UBVI$ from MCPS and
 $JHK_s$ from IRSF, (e) high-resolution $UbyI$ from {\it HST} and $K_s$ from
 VLT/NACO, and (f) optical and near-IR fluxes used only as upper limits. 
 See text for detail. Symbols are the same as in Figure~\ref{fig:fit}.
 \label{fig:multi}}
\end{figure} 

\epsscale{0.9}
\begin{figure}
\caption{(a) 8 \um\ image of N\,159 overlaid with MAGMA CO intensity
 contours (blue) in n$^2$ K~km~s$^{-1}$ (n$=$1,2,3,...).  YSOs at 
 different evolutionary stages are marked as follows: red circle --
 Type I and I/II, green circle -- Type II and II/III, and blue circle
 -- Type III.  Known maser and UCHIIs are marked with additional triangle
 and boxes, respectively \citep{Lazetal02,IJC04}. Candidate HAeBe stars
 from  \citet{Naetal05} are marked in orange dots.  Candidate HAeBe stars in 
 common with YSOs are labeled with positive numbers, and those in common
 with non-YSO red sources (yellow pluses) are labeled with negative numbers.
 (b) $J-H$ vs.\ $H-K_s$ CMD of all sources detected in all three $JHK_s$
 bands in N\,159.  YSOs and non-YSO red sources in common with candidate
 HAeBe stars are labeled with numbers as in (a) and marked with
 additional red circles and yellow crosses, respectively.    
 Dashed magenta lines indicate the criterion used to select candidate
 HAeBe stars in the \citet{Naetal05} study. (c) same as in (b) for the
 $K$ vs.\ $J-K_s$ CMD.
 \label{fig:jhkcmds}}
\end{figure} 

\epsscale{.9}

\epsscale{1.0}
\begin{figure}
\caption{(a) MOSAIC \ha\ image of N\,159 superbubble overlaid with 3~cm
 contours (yellow).  The position and size of the SNR 0540$-$697 are marked
 in dashed cyan line.  YSOs are marked in the same symbols as
 Figure~\ref{fig:yso_pos}.  Spectroscopically identified massive stars 
 \citep{Faetal09} are marked with triangles in three sizes to show
 different evolutionary stages: large -- young phase including O4-5 V-I
 and O Vz, medium -- middle-aged phase including O6-9 V-I and B0 I,
 and small -- evolved phase including B0-2 V-III and B1-8 I. 
 (b) Extinction map overlaid with MAGMA CO intensity contours (magenta).
 YSOs and spectroscopically identified massive stars are marked in the
 same symbols as (a).  Candidate OB stars from \citet{Naetal05} are marked
 in green squares. \label{fig:yso_mstar}}
\end{figure}

\epsscale{0.5}

\clearpage
\newpage

\begin{deluxetable}{rlrc}
\tablecolumns{4}
\tablecaption{Archival {\it Spitzer} Observations of N\,159
 \label{mirobs}} 
\tablewidth{0pc}
\tablehead{
\colhead{Program} & \colhead{Principal} & 
\multicolumn{2}{c}{Observation Parameters}\\
\colhead{ID} & \colhead{Investigator} & \multicolumn{1}{c}{IRAC} & \multicolumn{1}{c}{MIPS} 
}

\startdata

124   & Gehrz   & $10\times12$\tablenotemark{a} s & \\
20203 & Meixner & $4\times12$\tablenotemark{a} s  & $2\times$ Fast Scan \\
\enddata
\tablenotetext{a}{Observations used high dynamic range mode and had
 complementary exposures.}
\end{deluxetable}

\begin{deluxetable}{rrllr}
\tablecolumns{5}
\tablecaption{Archival {\it HST} WFPC2/ACS Observations of Fields in N\,159
 \label{hstobs}} 
\tablewidth{0pc}
\tablehead{
\colhead{R.A.} & \colhead{Decl.}
& \colhead{} & \colhead{} & 
\multicolumn{1}{c}{Exp. Time}\\
\colhead{(J2000)} & \colhead{(J2000)} & \multicolumn{1}{c}{Program ID/PI} & \multicolumn{1}{c}{Detector/Filter\tablenotemark{a}} & 
\colhead{(s)}
}

\startdata

05 39 10.5 & $-69$ 41 52.1 & 9480/Rhodes & ACS/F775W & 1200.0 \\
05 40 12.8 & $-69$ 44 29.3 & 06535/Heydari-Malayeri & WFPC2/F300W & 51.2 \\
& & & WFPC2/F410M & 172.0 \\
& & & WFPC2/F467M & 139.2 \\
& & & WFPC2/F469N & 1292.0 \\
& & & WFPC2/F487N & 1272.0 \\
& & & WFPC2/F502N & 984.0 \\
& & & WFPC2/F547M & 34.4 \\
& & & WFPC2/F656N & 1280.0 \\
& & & WFPC2/F814W & 21.6 \\

05 40 14.3 & $-69$ 50 06.8 & 09827/Bianchi & WFPC2/F170W & 240.0 \\
& & & WFPC2/F225W & 200.0 \\
& & & WFPC2/F336W & 40.0 \\
& & & WFPC2/F439W & 40.0 \\
& & & WFPC2/F555W & 20.0 \\

05 40 50.3 & $-69$ 55 10.2 & 9827/Bianchi & ACS/F435W & 130.0 \\
& & & ACS/F555W & 100.0 \\
& & & ACS/F658N & 600.0 \\
& & & ACS/F814W & 80.0 \\

\enddata
\tablenotetext{a}{F330W: Wide $U$; F336W: WFPC2 $U$; F410M: Str\"omgren $v$; F435W: Johnson $B$; F439W: WFPC2 $B$; F467M: Str\"omgren $b$; F469N: \heii; F487N: \hb; F502N: \oiii; F547M: Str\"omgren $y$; F555W: Johnson $V$; F656N: \ha; F658N: \ha; F673N: \sii; F675W: WFPC2 $R$; F775W: SDSS $i$; F814W: Broad $I$.}
\end{deluxetable}

\begin{deluxetable}{rcccc}
\tablecolumns{5}
\tablecaption{Parameters for IRAC and MIPS Photometric 
Measurements \label{photpar}}
\tablewidth{0pc}
\tablehead{
\colhead{} & \colhead{Aperture} & \colhead{Background} & 
 \multicolumn{1}{c}{Aperture} & \multicolumn{1}{c}{Zero-Mag.} \\
\colhead{} & \colhead{Radius} & \colhead{Annulus} & 
 \multicolumn{1}{c}{Correction} & \multicolumn{1}{c}{Flux} \\
\colhead{Band} & \colhead{($''$)} & \colhead{($''$)} & 
 \multicolumn{1}{c}{Factor} & \multicolumn{1}{c}{(Jy)} 
}
\startdata

IRAC 3.6 \um\ & 3.6 &  3.6-8.4  & 1.124 & 277.5 \\
 4.5 \um\ & 3.6 &  3.6-8.4      & 1.127 & 179.5 \\
 5.8 \um\ & 3.6 &  3.6-8.4      & 1.143 & 116.6 \\
 8.0 \um\ &  3.6 &  3.6-8.4     & 1.234 & ~\,63.1 \\
MIPS~ 24 \um\  & 6~  &  20-32  & 1.699 & ~\,7.14 \\
 70 \um       & 16~~ &  39-65  & 2.087 & 0.775 \\

\enddata
\end{deluxetable}

\begin{deluxetable}{rrrrrrrrll}
\rotate
\tabletypesize{\scriptsize}
\tablecolumns{10}
\tablecaption{Multi-wavelength Photometry of YSO Candidates Selected from CMD
 Criteria \label{ysoclass}}
\tablewidth{0pc}
\tablehead{
 \colhead{Name} &  \colhead{No}  &
 \colhead{[3.6]} & 
 \colhead{[4.5]} & 
 \colhead{[5.8]} & 
 \colhead{[8.0]} & 
 \colhead{[24]}  & 
 \colhead{[70]}  &
 \colhead{Class.}  &
 \colhead{Remarks} \\
 \colhead{(1)} &  \colhead{(2)}  &
 \colhead{(3)} & 
 \colhead{(4)} & 
 \colhead{(5)} & 
 \colhead{(6)} & 
 \colhead{(7)} & 
 \colhead{(8)} & 
 \colhead{(9)} & \colhead{(10)} 
}

\startdata

053921.21-694409.4 &  51 & 13.76  0.01 & 13.54  0.01 & 12.54  0.04 & 11.35  0.14 & \nodata \nodata & \nodata \nodata  & 	S in DR 	 &  	\\
053929.21-694719.0 &   9 & 11.38  0.01 & 10.69  0.01 &  9.05  0.02 &  7.20  0.02 &  1.12  0.01 & \nodata \nodata  & 	I 	 & ext 	\\
053930.01-694334.9 &  24 & 13.48  0.02 & 13.45  0.02 & 10.85  0.03 &  9.08  0.03 &  5.64  0.05 & \nodata \nodata  & 	II 	 & mul 	\\
053932.52-694357.4 &  17 & 12.82  0.03 & 12.88  0.03 &  9.84  0.03 &  8.19  0.03 & \nodata \nodata & \nodata \nodata  & 	D 	 &  	\\
053932.73-694344.3 &  29 & 13.82  0.05 & 13.53  0.07 & 11.14  0.06 &  9.42  0.08 & \nodata \nodata & \nodata \nodata  & 	D peak 	 &  	\\
053933.86-694701.6 &  27 & 13.42  0.01 & 13.33  0.01 & 10.97  0.01 &  9.29  0.02 & \nodata \nodata & \nodata \nodata  & 	II 	 &  	\\
053935.99-694604.1 &   7 & 10.99  0.01 & 10.53  0.01 &  8.59  0.01 &  6.84  0.01 &  1.60  0.10 & \nodata \nodata  & 	I 	 & mul 	\\
053937.04-694536.7 &   6 & 10.87  0.01 &  9.68  0.01 &  8.01  0.01 &  6.60  0.01 &  0.51  0.01 & \nodata \nodata  & 	I 	 & mul 	\\
053937.37-695120.8 &  49 & 13.95  0.02 & 13.72  0.02 & 13.14  0.07 & 11.33  0.04 &  7.94  0.07 & \nodata \nodata  & 	II 	 & ext 	\\
053937.53-694609.8 &   2 & 10.41  0.01 &  9.65  0.01 &  7.77  0.01 &  5.82  0.01 & -0.28  0.01 & \nodata \nodata  & 	I/II 	 & mul 	\\
053937.56-694525.4 &   4 & 10.93  0.02 &  9.70  0.01 &  8.09  0.03 &  6.13  0.03 & \nodata \nodata & -4.70  0.20  & 	I 	 & ext 	\\
053938.09-694654.2 &  18 & 12.95  0.04 & 13.06  0.06 & 10.08  0.03 &  8.34  0.05 & \nodata \nodata & \nodata \nodata  & 	S in DR 	 &  	\\
053938.80-694436.0 &  38 & 12.92  0.02 & 12.71  0.04 & 12.13  0.06 &  9.88  0.05 & \nodata \nodata & \nodata \nodata  & 	S 	 & LMC X-1 	\\
053939.10-694443.7 &  34 & 13.36  0.05 & 12.92  0.05 & 11.20  0.04 &  9.53  0.04 & \nodata \nodata & \nodata \nodata  & 	D 	 &  	\\
053939.54-694400.2 &  19 & \nodata \nodata & 13.18  0.05 & 10.34  0.03 &  8.58  0.04 & \nodata \nodata & \nodata \nodata  & 	D 	 &  	\\
053940.78-694632.0 &  14 & 12.21  0.02 & 11.49  0.02 &  9.66  0.02 &  8.02  0.02 &  2.81  99.9 & \nodata \nodata  & 	II 	 &  	\\
053941.89-694612.0 &   3 &  9.32  0.01 &  8.24  0.01 &  7.00  0.01 &  5.93  0.01 &  0.04  0.01 & \nodata \nodata  & 	I/II 	 & mul 	\\
053943.74-694540.3 &  39 & 13.43  0.01 & 12.55  0.01 & 11.48  0.04 &  9.91  0.05 &  4.06  99.9 & \nodata \nodata  & 	III 	 & mul 	\\
053944.55-694436.2 &  21 & 13.31  0.04 & 13.16  0.03 & 10.57  0.04 &  8.70  0.04 &  3.55  0.05 & \nodata \nodata  & 	I/II 	 &  	\\
053945.18-694450.4 &  11 & 11.88  0.01 & 11.16  0.01 &  9.36  0.01 &  7.44  0.01 &  1.65  0.01 & \nodata \nodata  & 	III 	 & ext 	\\
053945.20-694508.1 &  22 & 13.05  0.05 & 12.78  0.03 & 10.72  0.06 &  9.00  0.06 & \nodata \nodata & \nodata \nodata  & 	D 	 &  	\\
053945.21-694533.1 &  41 & 14.38  0.04 & 13.58  0.04 & \nodata \nodata & 10.15  0.05 & \nodata \nodata & \nodata \nodata  & 	S in DR 	 &  	\\
053946.39-694435.5 &  20 & 13.26  0.04 & 13.77  0.07 & 10.35  0.03 &  8.62  0.03 & \nodata \nodata & \nodata \nodata  & 	D 	 &  	\\
053947.68-694526.1 &  16 & 12.97  0.02 & 13.12  0.05 & 10.00  0.02 &  8.14  0.03 &  3.51  99.9 & \nodata \nodata  & 	II 	 &  	\\
053948.25-694534.3 &  28 & 14.06  0.04 & 13.66  0.06 & 11.31  0.04 &  9.37  0.04 & \nodata \nodata & \nodata \nodata  & 	D 	 &  	\\
053948.83-694416.7 &  23 & 13.33  0.03 & 13.54  0.03 & 10.85  0.03 &  9.03  0.03 & \nodata \nodata & \nodata \nodata  & 	D peak 	 &  	\\
053949.26-694407.5 &  31 & 14.00  0.07 & 13.80  0.09 & 11.24  0.04 &  9.47  0.05 & \nodata \nodata & \nodata \nodata  & 	D 	 &  	\\
053950.29-694545.4 &  40 & 14.22  0.04 & 13.69  0.10 & 11.74  0.07 &  9.99  0.08 & \nodata \nodata & \nodata \nodata  & 	D 	 &  	\\
053952.39-694518.3 &  12 & \nodata \nodata & 12.10  0.03 & 10.74  0.04 &  7.94  0.03 & \nodata \nodata & \nodata \nodata  & 	D 	 &  	\\
053952.60-694517.0 &  15 & 12.73  0.02 & 12.13  0.04 & 10.85  0.04 &  8.14  0.03 & \nodata \nodata & \nodata \nodata  & 	III 	 & mul 	\\
053956.62-694439.1 &  30 & 13.55  0.03 & 13.10  0.03 & 11.28  0.03 &  9.43  0.03 & \nodata \nodata & \nodata \nodata  & 	D 	 &  	\\
053956.92-695024.5 &  52 & 14.62  0.01 & 13.83  0.01 & 13.49  0.04 & 11.57  0.03 &  8.08  0.11 & \nodata \nodata  & 	II 	 & mul 	\\
053959.34-694526.3 &   5 &  9.29  0.01 &  8.57  0.01 &  7.67  0.01 &  6.54  0.01 &  2.00  0.01 & \nodata \nodata  & 	II 	 & ext 	\\
054000.69-694713.4 &   8 & 10.29  0.01 &  9.71  0.01 &  8.48  0.01 &  6.89  0.01 &  1.72  0.01 & \nodata \nodata  & 	III 	 & ext 	\\
054001.44-694801.6 &  44 & 13.61  0.02 & 13.55  0.02 & 12.01  0.05 & 10.83  0.08 &  6.43  0.07 & \nodata \nodata  & 	I/II 	 &  	\\
054003.49-694355.0 &  32 & \nodata \nodata & 13.15  0.02 & 11.58  0.08 &  9.48  0.05 & \nodata \nodata & \nodata \nodata  & 	YSC 	 &  	\\
054003.55-694710.0 &  42 & 13.89  0.03 & 13.62  0.05 & 12.05  0.05 & 10.34  0.06 &  5.48  0.15 & \nodata \nodata  & 	III 	 &  	\\
054004.13-694532.8 &  43 & 14.85  0.09 & 13.61  0.05 & 12.38  0.06 & 10.55  0.06 & \nodata \nodata & \nodata \nodata  & 	S in DR 	 &  	\\
054004.40-694437.6 &   1 &  9.85  0.01 &  8.96  0.01 &  7.35  0.01 &  5.40  0.01 & \nodata \nodata & -4.79  0.20  & 	III 	 & ext 	\\
054005.91-694451.6 &  26 & 14.30  0.10 & 13.34  0.07 & \nodata \nodata &  9.28  0.05 & \nodata \nodata & \nodata \nodata  & 	D 	 &  	\\
054006.85-694400.6 &  25 & 13.63  0.04 & 13.60  0.05 & 10.90  0.03 &  9.19  0.03 & \nodata \nodata & \nodata \nodata  & 	D 	 &  	\\
054008.53-694530.6 &  35 & 13.58  0.06 & 12.76  0.07 & 11.93  0.11 &  9.55  0.06 & \nodata \nodata & \nodata \nodata  & 	D 	 &  	\\
054009.49-694453.5 &  10 & 12.12  0.01 & 12.06  0.02 &  9.09  0.02 &  7.31  0.02 &  1.46  0.03 & \nodata \nodata  & 	I/II 	 & mul 	\\
054011.47-694504.5 &  13 & 13.17  0.09 & 13.21  0.07 &  9.89  0.03 &  7.96  0.03 & \nodata \nodata & \nodata \nodata  & 	D 	 &  	\\
054013.17-694948.8 &  45 & 13.91  0.01 & 13.07  0.01 & 12.10  0.02 & 10.96  0.04 &  7.77  0.12 & \nodata \nodata  & 	II 	 &  	\\
054014.04-694454.8 &  33 & 13.81  0.03 & 13.43  0.04 & 11.38  0.03 &  9.48  0.05 & \nodata \nodata & \nodata \nodata  & 	S in DR 	 &  	\\
054019.00-694445.6 &  37 & 14.05  0.04 & 12.77  0.02 & 11.13  0.03 &  9.66  0.04 &  4.22  99.9 & \nodata \nodata  & 	II 	 & ext 	\\
054021.33-694944.0 &  36 & 13.87  0.02 & 13.63  0.03 & 11.40  0.02 &  9.65  0.03 &  6.80  0.06 & \nodata \nodata  & 	II 	 & mul 	\\
054037.09-694521.5 &  48 & 14.05  0.01 & 13.83  0.01 & 13.75  0.05 & 11.22  0.01 & \nodata \nodata & \nodata \nodata  & 	G 	 &  	\\
054044.66-694550.9 &  47 & 14.05  0.01 & 13.83  0.01 & 13.76  0.05 & 11.01  0.01 &  8.34  0.12 & \nodata \nodata  & 	G 	 &  	\\
054046.29-694441.1 &  50 & 13.50  0.01 & 13.62  0.01 & 12.58  0.05 & 11.35  0.06 & \nodata \nodata & \nodata \nodata  & 	S in DR 	 &  	\\
054050.85-695001.9 &  46 & 13.72  0.01 & 13.15  0.01 & 12.21  0.01 & 10.98  0.02 &  7.40  0.06 & \nodata \nodata  & 	II 	 &  	\\

\enddata
\tablecomments{Column 1: source name. Column 2: Ranking of the brightness at
 8 \um. Columns 3-8: photometric measurements in 3.6, 4.5, 5.8, 8.0, 24, and
 70 \um\ bands in magnitudes. 
 Measurements with uncertainties of $99.9$ are the upper brightness limits
 as they include fluxes from neighbors or backgrounds.
 The uncertainties listed here are only errors in measurements and do not
 include errors in in flux calibration, i.e., 5\% in 3.6, 4.5, 5.8, and
 8.0 \um, 10\% in 24 \um, and 20\% in 70 \um.  Thus, the total uncertainty
 of a flux is the quadratic sum of the measurement error and the calibration
 error.
 Columns 9 and 10: classification and remarks: D -- diffuse emission,
 DR -- dusty region, ext -- extended source, G -- background galaxy,
 I/II/III -- Type I/II/III YSO, mul -- multiple, S -- star, 
 YSC -- young star cluster.}
\end{deluxetable}

\begin{deluxetable}{rrrrrrrrrrll}
\rotate
\tabletypesize{\scriptsize}
\tablecolumns{12}
\tablecaption{Multi-wavelength Photometry of YSO Candidates Selected from CMD
 Criteria \label{ysoclassb}}
\tablewidth{0pc}
\tablehead{
 \colhead{Name} &  \colhead{No}  &
 \colhead{$U$} & 
 \colhead{$B$} & 
 \colhead{$V$} & 
 \colhead{$I$} & 
 \colhead{$J$} & 
 \colhead{$H$} & 
 \colhead{$K_s$} & 
 \colhead{Flag}  &
 \colhead{Class.}  &
 \colhead{Remarks} \\
 \colhead{(1)} &  \colhead{(2)}  &
 \colhead{(3)} & 
 \colhead{(4)} & 
 \colhead{(5)} & 
 \colhead{(6)} & 
 \colhead{(7)} & 
 \colhead{(8)} & 
 \colhead{(9)} & \colhead{(10)} & \colhead{(11)} & \colhead{(12)}
}

\startdata
053921.21-694409.4 &  51 & \nodata \nodata & 21.88  0.22 & 21.17  0.20 & \nodata \nodata & 16.04  0.11 &  15.30  0.12 &  14.95  0.15 &   0 & 	S in DR 	 &  	\\
053929.21-694719.0 &   9 & \nodata \nodata & \nodata \nodata & \nodata \nodata & \nodata \nodata & 16.99  0.05 &  16.45  0.15 &  14.84  0.08 &  10 & 	I 	 & ext 	\\
053930.01-694334.9 &  24 & 19.45  0.10 & 19.77  0.06 & 19.41  0.06 & 18.67  0.08 & 18.29  0.06 &  17.23  0.07 &  16.25  0.05 &  10 & 	II 	 & mul 	\\
053932.52-694357.4 &  17 & \nodata \nodata & \nodata \nodata & \nodata \nodata & \nodata \nodata & \nodata \nodata &  \nodata \nodata &  \nodata \nodata &   0 & 	D 	 &  	\\
053932.73-694344.3 &  29 & \nodata \nodata & \nodata \nodata & \nodata \nodata & \nodata \nodata & \nodata \nodata &  \nodata \nodata &  \nodata \nodata &   0 & 	D peak 	 &  	\\
053933.86-694701.6 &  27 & 18.16  0.08 & 18.89  0.05 & 18.37  0.06 & 17.61  0.13 & 16.95  0.03 &  16.30  0.04 &  15.51  0.04 &  10 & 	II 	 &  	\\
053935.99-694604.1 &   7 & \nodata \nodata & \nodata \nodata & \nodata \nodata & \nodata \nodata & 17.91  0.10 &  \nodata \nodata &  14.47  0.10 &  20 & 	I 	 & mul 	\\
053937.04-694536.7 &   6 & \nodata \nodata & \nodata \nodata & \nodata \nodata & \nodata \nodata & 18.19  0.09 &  17.04  0.15 &  14.65  0.05 &  10 & 	I 	 & mul 	\\
053937.37-695120.8 &  49 & \nodata \nodata & 19.77  0.05 & 18.71  0.05 & 17.64  0.05 & 16.53  0.05 &  15.85  0.09 &  15.93  0.11 &  10 & 	II 	 & ext 	\\
053937.53-694609.8 &   2 & \nodata \nodata & \nodata \nodata & \nodata \nodata & \nodata \nodata & 17.04  0.10 &  \nodata \nodata &  15.31  0.10 &  20 & 	I/II 	 & mul 	\\
053937.56-694525.4 &   4 & \nodata \nodata & \nodata \nodata & \nodata \nodata & \nodata \nodata & 18.11  0.12 &  17.38  0.16 &  15.68  0.11 &  10 & 	I 	 & ext 	\\
053938.09-694654.2 &  18 & 19.14  0.21 & 19.41  0.12 & 18.63  0.09 & 17.50  0.08 & \nodata \nodata &  \nodata \nodata &  \nodata \nodata &   0 & 	S in DR 	 &  	\\
053938.80-694436.0 &  38 & 13.82  0.09 & 14.54  0.05 & 14.61  0.17 & 13.89  0.09 & 13.70  0.06 &  13.54  0.09 &  13.29  0.06 &   0 & 	S 	 & LMC X-1 	\\
053939.10-694443.7 &  34 & \nodata \nodata & \nodata \nodata & \nodata \nodata & \nodata \nodata & \nodata \nodata &  \nodata \nodata &  \nodata \nodata &   0 & 	D 	 &  	\\
053939.54-694400.2 &  19 & \nodata \nodata & \nodata \nodata & \nodata \nodata & \nodata \nodata & \nodata \nodata &  \nodata \nodata &  \nodata \nodata &   0 & 	D 	 &  	\\
053940.78-694632.0 &  14 & \nodata \nodata & \nodata \nodata & \nodata \nodata & \nodata \nodata & 16.80  0.07 &  16.66  0.06 &  15.55  0.10 &  10 & 	II 	 &  	\\
053941.89-694612.0 &   3 & \nodata \nodata & \nodata \nodata & \nodata \nodata & \nodata \nodata & 18.11  0.10 &  \nodata \nodata &  13.31  0.10 &  20 & 	I/II 	 & mul 	\\
053943.74-694540.3 &  39 & \nodata \nodata & \nodata \nodata & \nodata \nodata & \nodata \nodata & 18.61  0.07 &  17.34  0.08 &  16.08  0.05 &  10 & 	III 	 & mul 	\\
053944.55-694436.2 &  21 & \nodata \nodata & \nodata \nodata & \nodata \nodata & \nodata \nodata & 16.61  0.03 &  15.86  0.03 &  15.33  0.03 &  10 & 	I/II 	 &  	\\
053945.18-694450.4 &  11 & 15.79  0.18 & 16.71  0.11 & 16.13  0.07 & 14.70  0.12 & 15.11  0.03 &  14.83  0.03 &  14.54  0.03 &  10 & 	III 	 & ext 	\\
053945.20-694508.1 &  22 & \nodata \nodata & \nodata \nodata & \nodata \nodata & \nodata \nodata & \nodata \nodata &  \nodata \nodata &  \nodata \nodata &   0 & 	D 	 &  	\\
053945.21-694533.1 &  41 & 15.44  0.05 & 16.45  0.04 & 16.02  0.04 & 15.66  0.06 & 15.53  0.07 &  15.25  0.10 &  15.45  0.23 &   0 & 	S in DR 	 &  	\\
053946.39-694435.5 &  20 & \nodata \nodata & \nodata \nodata & \nodata \nodata & \nodata \nodata & \nodata \nodata &  \nodata \nodata &  \nodata \nodata &   0 & 	D 	 &  	\\
053947.68-694526.1 &  16 & \nodata \nodata & \nodata \nodata & \nodata \nodata & \nodata \nodata & 17.90  0.06 &  17.71  0.09 &  17.28  0.13 &  10 & 	II 	 &  	\\
053948.25-694534.3 &  28 & \nodata \nodata & \nodata \nodata & \nodata \nodata & \nodata \nodata & \nodata \nodata &  \nodata \nodata &  \nodata \nodata &   0 & 	D 	 &  	\\
053948.83-694416.7 &  23 & \nodata \nodata & \nodata \nodata & \nodata \nodata & \nodata \nodata & \nodata \nodata &  \nodata \nodata &  \nodata \nodata &   0 & 	D peak 	 &  	\\
053949.26-694407.5 &  31 & \nodata \nodata & \nodata \nodata & \nodata \nodata & \nodata \nodata & \nodata \nodata &  \nodata \nodata &  \nodata \nodata &   0 & 	D 	 &  	\\
053950.29-694545.4 &  40 & \nodata \nodata & \nodata \nodata & \nodata \nodata & \nodata \nodata & \nodata \nodata &  \nodata \nodata &  \nodata \nodata &   0 & 	D 	 &  	\\
053952.39-694518.3 &  12 & \nodata \nodata & \nodata \nodata & \nodata \nodata & \nodata \nodata & \nodata \nodata &  \nodata \nodata &  \nodata \nodata &   0 & 	D 	 &  	\\
053952.60-694517.0 &  15 & \nodata \nodata & 15.15  0.08 & 14.82  0.09 & 14.07  0.15 & 14.11  0.02 &  13.91  0.02 &  13.78  0.01 &  10 & 	III 	 & mul 	\\
053956.62-694439.1 &  30 & \nodata \nodata & \nodata \nodata & \nodata \nodata & \nodata \nodata & \nodata \nodata &  \nodata \nodata &  \nodata \nodata &   0 & 	D 	 &  	\\
053956.92-695024.5 &  52 & \nodata \nodata & \nodata \nodata & \nodata \nodata & \nodata \nodata & 18.51  0.10 &  17.41  0.13 &  16.36  0.11 &  10 & 	II 	 & mul 	\\
053959.34-694526.3 &   5 & \nodata \nodata & \nodata \nodata & \nodata \nodata & \nodata \nodata & 16.34  0.03 &  14.00  0.02 &  11.96  0.02 &  10 & 	II 	 & ext 	\\
054000.69-694713.4 &   8 & 14.81  0.13 & 15.69  0.18 & 15.07  0.14 & 14.34  0.13 & 14.96  0.03 &  13.78  0.04 &  12.47  0.03 &  10 & 	III 	 & ext 	\\
054001.44-694801.6 &  44 & \nodata \nodata & \nodata \nodata & \nodata \nodata & \nodata \nodata & 18.51  0.08 &  17.43  0.09 &  16.60  0.09 &  10 & 	I/II 	 &  	\\
054003.49-694355.0 &  32 & \nodata \nodata & \nodata \nodata & \nodata \nodata & \nodata \nodata & 18.41  0.14 &  17.32  0.15 &  16.26  0.15 &  10 & 	YSC 	 &  	\\
054003.55-694710.0 &  42 & 18.36  0.24 & 18.86  0.09 & 17.51  0.09 & 16.21  0.07 & 15.49  0.02 &  14.88  0.01 &  14.70  0.02 &  10 & 	III 	 &  	\\
054004.13-694532.8 &  43 & \nodata \nodata & \nodata \nodata & \nodata \nodata & \nodata \nodata & \nodata \nodata &  \nodata \nodata &  \nodata \nodata &   0 & 	S in DR 	 &  	\\
054004.40-694437.6 &   1 & 16.90  0.20 & 14.67  0.12 & 15.55  0.14 & 13.94  0.14 & 14.18  0.04 &  13.99  0.05 &  13.53  0.11 &  10 & 	III 	 & ext 	\\
054005.91-694451.6 &  26 & \nodata \nodata & \nodata \nodata & \nodata \nodata & \nodata \nodata & \nodata \nodata &  \nodata \nodata &  \nodata \nodata &   0 & 	D 	 &  	\\
054006.85-694400.6 &  25 & \nodata \nodata & \nodata \nodata & \nodata \nodata & \nodata \nodata & \nodata \nodata &  \nodata \nodata &  \nodata \nodata &   0 & 	D 	 &  	\\
054008.53-694530.6 &  35 & \nodata \nodata & \nodata \nodata & \nodata \nodata & \nodata \nodata & \nodata \nodata &  \nodata \nodata &  \nodata \nodata &   0 & 	D 	 &  	\\
054009.49-694453.5 &  10 & \nodata \nodata & 21.92  0.31 & 20.76  0.22 & \nodata \nodata & 18.10  0.27 &  17.36  0.09 &  16.57  0.13 &  10 & 	I/II 	 & mul 	\\
054011.47-694504.5 &  13 & \nodata \nodata & \nodata \nodata & \nodata \nodata & \nodata \nodata & \nodata \nodata &  \nodata \nodata &  \nodata \nodata &   0 & 	D 	 &  	\\
054013.17-694948.8 &  45 & \nodata \nodata & \nodata \nodata & \nodata \nodata & \nodata \nodata & 18.72  0.08 &  17.50  0.08 &  16.20  0.04 &  10 & 	II 	 &  	\\
054014.04-694454.8 &  33 & \nodata \nodata & \nodata \nodata & \nodata \nodata & \nodata \nodata & 15.86  0.08 &  15.14  0.10 &  14.75  0.14 &   0 & 	S in DR 	 &  	\\
054019.00-694445.6 &  37 & \nodata \nodata & \nodata \nodata & \nodata \nodata & \nodata \nodata & \nodata \nodata &  \nodata \nodata &  17.90  0.25 &  10 & 	II 	 & ext 	\\
054021.33-694944.0 &  36 & 19.55  0.10 & 19.77  0.05 & 19.27  0.06 & 18.26  0.09 & 17.97  0.09 &  16.69  0.12 &  16.66  0.16 &  10 & 	II 	 & mul 	\\
054037.09-694521.5 &  48 & \nodata \nodata & 21.16  0.12 & 19.66  0.09 & 17.84  0.08 & 16.43  0.13 &  15.55  0.14 &  14.70  0.12 &   0 & 	G 	 &  	\\
054044.66-694550.9 &  47 & \nodata \nodata & 21.02  0.12 & 20.41  0.11 & \nodata \nodata & 16.37  0.14 &  15.34  0.17 &  14.85  0.14 &   0 & 	G 	 &  	\\
054046.29-694441.1 &  50 & 20.64  0.22 & 18.95  0.04 & 17.54  0.05 & 15.93  0.05 & 14.82  0.05 &  13.93  0.05 &  13.79  0.06 &   0 & 	S in DR 	 &  	\\
054050.85-695001.9 &  46 & \nodata \nodata & \nodata \nodata & \nodata \nodata & \nodata \nodata & 18.21  0.08 &  17.05  0.06 &  15.86  0.04 &  10 & 	II 	 &  	\\

\enddata
\tablecomments{Column 1: source name. Column 2: Ranking of the brightness at
 8 \um. Columns 3-9: $UBVIJHK_s$ photometric measurements in magnitudes. 
 Measurements with uncertainties of $99.9$ are the upper brightness limits
 as they include fluxes from neighbors or backgrounds.
 The uncertainties listed here are only errors in measurements and do not
 include errors in in flux calibration, i.e., 10\% in $U$, 5\% in $BV$, and
 10\% in $IJHK_s$. Thus, the total uncertainty of a flux is the quadratic sum
 of the measurement error and the calibration error.
 Column 10: data used for $JHK_s$ photometry: 0 -- $JHK_s$ from 2MASS catalog,
 10 -- $JHK_s$ from IRSF data, 20 -- $JK_s$ from VLT/NACO data.  Column 11
 and 12: classification and remarks: D -- diffuse emission, DR -- dusty region,
 ext -- extended source, G -- background galaxy,
 I/II/III -- Type I/II/III YSO, mul -- multiple, S -- star, 
 YSC -- young star cluster.}
\end{deluxetable}

\begin{deluxetable}{lrrrrrrrrrrrr}
\rotate
\tabletypesize{\scriptsize}
\tablecolumns{13}
\tablecaption{Inferred Physical Parameters from SED Fits to YSOs \label{sedfits}}
\tablewidth{0pc}
\tablehead{
 \colhead{} &
 \colhead{} &
 \colhead{} &
 \colhead{} &
 \colhead{} &
 \colhead{} &
\multicolumn{7}{c}{Physical Parameters of the Best-Fit Model} \\
 \cline{7-13}
 \colhead{} & 
 \colhead{[8.0]} & 
 \colhead{} &
 \colhead{Stage} &
 \colhead{$\bar{M}_{\ast} \pm \Delta {M}_{\ast}$} & 
 \colhead{$\bar{L}_{\rm tot} \pm \Delta {L}_{\rm tot}$} & 
 \colhead{$M_{\ast}$} &
 \colhead{$R_{\ast}$} & 
 \colhead{$T_{\ast}$} &
 \colhead{$\dot{M}_{\rm env}$} & 
 \colhead{$M_{\rm disk}$} & 
 \colhead{$A_V$} &
 \colhead{$L_{\rm tot}$} \\
 \colhead{Source Name} & 
 \colhead{(mag)} &
 \colhead{Type} &
 \colhead{Range} & 
 \colhead{($M_\odot$)} & 
 \colhead{($L_\odot$)} &
 \colhead{($M_\odot$)} & 
 \colhead{($R_\odot$)} & 
 \colhead{(K)} & 
 \colhead{($M_\odot$/yr)} & 
 \colhead{($M_\odot$)} & 
 \colhead{(mag)} &
 \colhead{($L_\odot$)} 
}

\startdata

053937.56-694525.4   &  6.13 & I 	 & 1.0 $\pm$ 0.0 & 34.8 $\pm$ 8.4 & 2.7E+05 $\pm$ 1.5E+05 	 & 29.3 & 270.2 &  7803 & 7.0E-04 & 0.0E+00 &  7.0 & 2.4E+05 \\
053937.04-694536.7   & 6.60 & I & 1.0 $\pm$ 0.0 & 28.6 $\pm$ 5.5 & 1.2E+05 $\pm$ 5.1E+04 &  	25.2 &   6.5 & 38000 & 1.5E-03 & 0.0E+00 &  0.0 & 8.2E+04 	\\
053935.99-694604.1   &  6.84 & I 	 & 1.0 $\pm$ 0.0 & 18.5 $\pm$ 1.4 & 4.5E+04 $\pm$ 3.6E+02 	 & 17.3 &  10.4 & 26100 & 3.8E-03 & 0.0E+00 &  2.5 & 4.5E+04 \\
053929.21-694719.0   &  7.20 & I 	 & 1.0 $\pm$ 0.0 & 26.1 $\pm$ 1.5 & 8.8E+04 $\pm$ 2.5E+04 	 & 28.2 & 129.0 &  9549 & 7.1E-04 & 0.0E+00 &  2.3 & 1.2E+05 \\
\cline{1-13}
053937.53-694609.8   & 5.82 & I/II & 1.0 $\pm$ 0.0 & 31.2 $\pm$ 2.9 & 2.2E+05 $\pm$ 3.8E+04 & 	29.3 & 270.0 &  7800 & 7.0E-04 & 0.0E+00 &  1.3 & 2.4E+05 	\\
053941.89-694612.0   &  5.93 & I/II 	 & 1.0 $\pm$ 0.0 & 33.7 $\pm$ 2.6 & 1.7E+05 $\pm$ 2.9E+04 	 & 34.1 &   7.9 & 41760 & 2.5E-03 & 1.1E-01 & 12.7 & 1.7E+05 \\
053944.55-694436.2   &  8.70 & I/II 	 & 2.0 \nodata & 12.2 \nodata & 1.1E+04 \nodata 	 & 12.2 &   4.4 & 28500 & 1.5E-06 & 7.3E-03 &  0.9 & 1.1E+04 \\
054001.44-694801.6   & 10.83 & I/II 	 & 2.0 $\pm$ 0.0 &  6.4 $\pm$ 0.7 & 1.3E+03 $\pm$ 7.7E+02 	 & 6.3 &   3.1 & 19210 & 1.6E-07 & 1.2E-03 &  1.2 & 1.1E+03 \\
\cline{1-13}
053959.34-694526.3   &  6.54 & II 	 & 1.0 $\pm$ 0.0 & 15.7 $\pm$ 2.3 & 3.0E+04 $\pm$ 9.7E+03 	 & 15.4 &   7.7 & 27940 & 1.1E-04 & 2.7E-02 &  0.6 & 3.2E+04 \\
053940.78-694632.0   &  8.02 & II 	 & 1.0 $\pm$ 0.0 &  9.4 $\pm$ 2.7 & 5.7E+03 $\pm$ 2.8E+03 	 & 7.9 &   8.8 & 15820 & 1.0E-05 & 6.0E-02 &  1.9 & 4.3E+03 \\
053947.68-694526.1   &  8.14 & II 	 & 1.7 $\pm$ 0.5 & 13.0 $\pm$ 2.4 & 1.4E+04 $\pm$ 8.4E+03 	 & 14.8 &   4.8 & 31100 & 0.0E+00 & 1.9E-01 &  0.0 & 2.0E+04 \\
053933.86-694701.6  & 9.29 & II & 2.9 $\pm$ 0.5 & 12.3 $\pm$ 1.1 & 1.2E+04 $\pm$ 2.5E+03 & 	12.9 &   4.5 & 29000 & 0.0E+00 & 7.1E-07 &  2.2 & 1.4E+04 	\\
054021.33-694944.0   &  9.65 & II 	 & 2.0 $\pm$ 0.0 &  7.2 $\pm$ 0.1 & 1.9E+03 $\pm$ 6.5E+01 	 & 7.3 &   3.3 & 21120 & 0.0E+00 & 6.4E-04 &  2.2 & 2.0E+03 \\
054019.00-694445.6   &  9.66 & II 	 & 1.1 $\pm$ 0.2 & 10.8 $\pm$ 1.8 & 5.6E+03 $\pm$ 2.7E+03 	 & 11.2 &  32.1 &  8432 & 2.6E-04 & 6.4E-02 & 15.9 & 4.7E+03 \\
054013.17-694948.8   & 10.96 & II 	 & 1.9 $\pm$ 0.3 &  5.5 $\pm$ 0.6 & 7.1E+02 $\pm$ 1.6E+02 	 & 5.7 &   2.9 & 17860 & 0.0E+00 & 2.3E-02 &  2.8 & 7.6E+02 \\
054050.85-695001.9   & 10.98 & II 	 & 2.0 $\pm$ 0.0 &  5.6 $\pm$ 0.1 & 7.0E+02 $\pm$ 5.7E+01 	 & 5.7 &   2.9 & 17860 & 0.0E+00 & 2.3E-02 &  0.7 & 7.6E+02 \\
053937.37-695120.8   & 11.33 & II 	 & 1.3 $\pm$ 0.5 &  6.0 $\pm$ 0.2 & 5.1E+02 $\pm$ 5.9E+01 	 & 5.8 &  12.5 &  7098 & 3.1E-05 & 1.7E-01 &  0.8 & 4.7E+02 \\
053956.92-695024.5   & 11.57 & II 	 & 2.8 $\pm$ 0.4 &  7.7 $\pm$ 1.0 & 2.6E+03 $\pm$ 9.2E+02 	 & 8.2 &   3.5 & 22860 & 0.0E+00 & 4.9E-06 &  3.8 & 3.1E+03 \\
\cline{1-13}
054004.40-694437.6   &  5.40 & III 	 & 1.0 \nodata & 19.7 \nodata & 5.3E+04 \nodata 	 & 19.7 &  23.3 & 18190 & 5.9E-03 & 0.0E+00 &  0.0 & 5.3E+04 \\
054000.69-694713.4  & 6.89 & III & 2.5 $\pm$ 0.5 & 19.9 $\pm$ 2.6 & 4.7E+04 $\pm$ 1.5E+04 & 	17.2 &   5.3 & 33000 & 1.3E-06 & 7.7E-05 &  0.0 & 3.1E+04 	\\
053945.18-694450.4   &  7.44 & III 	 & 1.9 $\pm$ 0.3 & 16.6 $\pm$ 1.6 & 2.8E+04 $\pm$ 7.3E+03      & 16.6 &   5.2 & 32760 & 4.3E-07 & 2.3E-05 &  0.1 & 2.8E+04 \\
054003.55-694710.0  & 10.34 & III & 1.0 $\pm$ 0.0 &  8.0 $\pm$ 0.5 & 1.2E+03 $\pm$ 3.1E+02 & 	7.8 &  27.0 &  6700 & 3.8E-04 & 4.5E-03 &  1.8 & 1.3E+03 	\\
\cline{1-13}
054009.49-694453.5  & 7.31 & mul(I/II) & 1.0 $\pm$ 0.0 & 20.8 $\pm$ 2.6 & 6.3E+04 $\pm$ 1.6E+04 & 22.7 &  18.0 & 23000 & 4.6E-03 & 0.0E+00 &  2.3 & 8.5E+04 	\\
053952.60-694517.0  & 8.14 & mul(III) & 1.0 $\pm$ 0.0 &  8.6 $\pm$ 0.7 & 2.4E+03 $\pm$ 3.8E+02 & 9.2 &  20.0 &  9200 & 9.6E-05 & 5.5E-01 &  0.1 & 2.8E+03 	\\
053930.01-694334.9  & 9.08 & mul(II) & 1.7 $\pm$ 0.5 &  7.7 $\pm$ 0.6 & 2.6E+03 $\pm$ 7.2E+02 & 7.5 &   3.4 & 22000 & 6.3E-05 & 1.6E-02 &  1.1 & 2.3E+03 	\\
054003.49-694355.0  & 9.48 & YSC & 1.8 $\pm$ 0.9 & 11.2 $\pm$ 1.8 & 8.4E+03 $\pm$ 3.9E+03 & 	9.1 &   4.4 & 23000 & 2.1E-05 & 6.0E-04 &  3.4 & 5.0E+03 	\\
053943.74-694540.3  & 9.91 & mul(III) & 1.3 $\pm$ 0.6 &  9.4 $\pm$ 3.1 & 5.1E+03 $\pm$ 1.3E+04 & 9.0 &  16.0 & 11000 & 1.0E-03 & 1.7E-02 &  8.4 & 3.3E+03 	\\

\enddata
\end{deluxetable}

\begin{deluxetable}{lccccccccc}
\rotate
\tabletypesize{\scriptsize}
\tablecolumns{10}
\tablecaption{Inferred Physical Properties of YSOs in Tests of SED Fits \label{fitcom}}
\tablewidth{0pc}
\tablehead{
 \colhead{Source} & 
 \colhead{$\bar{M}_{\ast} \pm \Delta {M}_{\ast}$} & 
 \colhead{$\bar{L}_{\rm tot} \pm \Delta {L}_{\rm tot}$} & 
 \colhead{$\bar{\dot{M}}_{\rm env} \pm \Delta \dot{M}_{\rm env}$} &
 \colhead{Stage} &  \colhead{} &
 \colhead{$\bar{M}_{\ast} \pm \Delta {M}_{\ast}$} & 
 \colhead{$\bar{L}_{\rm tot} \pm \Delta {L}_{\rm tot}$} & 
 \colhead{$\bar{\dot{M}}_{\rm env} \pm \Delta \bar{\dot{M}}_{\rm env}$} &
 \colhead{Stage} \\
 \colhead{Name} & 
 \colhead{($M_\odot$)} & 
 \colhead{($L_\odot$)} & 
 \colhead{($M_\odot$/yr)} &
 \colhead{Range} & \colhead{} &
 \colhead{($M_\odot$)} & 
 \colhead{($L_\odot$)} & 
 \colhead{($M_\odot$/yr)} &
 \colhead{Range}
 }

\startdata
\multicolumn{1}{c}{} & \multicolumn{9}{c}{Test on SED Fits to YSO with PAH Emission} \\
\multicolumn{1}{c}{} & \multicolumn{4}{c}{Fits to Original SED} &
 \multicolumn{1}{c}{} & \multicolumn{4}{c}{Fits to PAH-corrected SED} \\
\cline{2-5} \cline{7-10} \\
053935.99$-$694604.1 & $18.5\pm1.4$ & 4.5E+04$\pm$3.6E+02 & 2.0E-03$\pm$2.2E-03
 & 1.0$\pm0.0$ & & $18.5\pm3.6$ & 4.1E+04$\pm$2.0E+04 & 1.6E-04$\pm$1.7E-04 & 
   1.0$\pm0.0$ \\
\cline{1-10} \\
\multicolumn{1}{c}{} & \multicolumn{9}{c}{Test on SED Fits to Multiple System with Sources at Similar Evolutionary Stages} \\
\multicolumn{1}{c}{} & \multicolumn{4}{c}{Fits to SED w/ Integrated Fluxes} &
 \multicolumn{1}{c}{} & \multicolumn{4}{c}{Fits to SED of Component a} \\
\cline{2-5} \cline{7-10} \\
053941.89$-$694612.0 & $32.3\pm2.9$ & 1.6E+05$\pm$3.1E+04 & 1.3E-03$\pm$1.0E-03 
 & 1.0$\pm$0.0 & & 26.9$\pm$2.5 & 1.0E+05$\pm$2.0E+04 & 1.1E-03$\pm$9.4E-04 &
   1.0$\pm$0.0 \\
\multicolumn{1}{c}{} & \multicolumn{4}{c}{} &
 \multicolumn{1}{c}{} & \multicolumn{4}{c}{Fits to SED of Component b} \\
\cline{7-10} \\
 & & & & & & 24.8$\pm$2.5 & 8.2E+04$\pm$2.0E+04 & 8.5E-04$\pm$4.9E-04 &
 1.0$\pm$  0.0\\
\cline{1-10} \\
\multicolumn{1}{c}{} & \multicolumn{9}{c}{Test on SED Fits to Multiple System with Sources at Different Evolutionary Stages} \\
\multicolumn{1}{c}{} & \multicolumn{4}{c}{Fits to SED w/ Integrated Fluxes} &
 \multicolumn{1}{c}{} & \multicolumn{4}{c}{Fits to SED w/ {\it HST} $UBVI$ } \\
\cline{2-5} \cline{7-10} \\
054004.40$-$694437.6 & 19.7 \nodata & 5.3E+04 \nodata &  5.9E-03 \nodata &
 1.0 \nodata & & 20.5 \nodata & 5.9E+04 \nodata & 1.4E-04 \nodata &
 1.0 \nodata \\
\multicolumn{1}{c}{} & \multicolumn{4}{c}{} &
 \multicolumn{1}{c}{} & \multicolumn{4}{c}{Fits to mid-IR SED} \\
 & & & & & &  41.2$\pm$9.8 & 3.5E+05$\pm$1.8E+05 &  2.8E-03$\pm$2.0E-03 &
 1.0$\pm$0.0 \\

\enddata
\end{deluxetable}

\begin{deluxetable}{lrlrlcl}
\tabletypesize{\scriptsize}
\tablecolumns{7}
\tablecaption{Comparisons of Classifications of YSOs \label{evo}}
\tablewidth{0pc}
\tablehead{
 \colhead{} & 
 \colhead{[8.0]} & 
 \colhead{} & 
 \colhead{Stage} &

 \colhead{IRS} & 
 \colhead{Silicate} &
 \colhead{Other} \\ 
 \colhead{Name} & 
 \colhead{(mag)} & 
 \colhead{Type} & 
 \colhead{Range} & 
 \colhead{Spec} & 
 \colhead{Absorption} & 
 \colhead{Classification\tablenotemark{a}} 
 }

\startdata

053937.56-694525.4  &  6.13 & I 	 & 1.0 $\pm$ 0.0	& PE & $\surd$  & UCHII \\
053937.04-694536.7  &  6.60 & I 	 & 1.0 $\pm$ 0.0 	& PE & $\surd$  &       \\	
053935.99-694604.1  &  6.84 & I 	 & 1.0 $\pm$ 0.0	& PE & $\times$ & HAeBe \\
053929.21-694719.0  &  7.20 & I 	 & 1.0 $\pm$ 0.0 	& P  & $\surd$  & maser \\
\cline{1-7}
053937.53-694609.8  &  5.82 & I/II 	 & 1.0 $\pm$ 0.0 	& PE & $\times$ & UCHII \\
053941.89-694612.0  &  5.93 & I/II 	 & 1.0 $\pm$ 0.0 	& PE & $\surd$  & HAeBe \\
054009.49-694453.5  &  7.31 & I/II(mul)  & 1.0 $\pm$ 0.0	& PE & $\surd$  &       \\	
053944.55-694436.2  &  8.70 & I/II 	 & 2.0 \nodata          &    &  & \\
054001.44-694801.6  & 10.83 & I/II 	 & 2.0 $\pm$ 0.0        &    &  & \\
\cline{1-7}
053959.34-694526.3  &  6.54 & II 	 & 1.0 $\pm$ 0.0 	& PE & $\times$ &      \\
053940.78-694632.0  &  8.02 & II 	 & 1.0 $\pm$ 0.0 	&    &          & HAeBe \\ 
053947.68-694526.1  &  8.14 & II 	 & 1.7 $\pm$ 0.5        &    &  & \\
053933.86-694701.6  &  9.29 & II 	 & 2.9 $\pm$ 0.5        &    &          & HAeBe \\ 
054021.33-694944.0  &  9.65 & II 	 & 2.0 $\pm$ 0.0        &    &  & \\
054019.00-694445.6  &  9.66 & II 	 & 1.1 $\pm$ 0.2        &    &  & \\
054013.17-694948.8  & 10.96 & II 	 & 1.9 $\pm$ 0.3 	&    &          & HAeBe \\ 
054050.85-695001.9  & 10.98 & II 	 & 2.0 $\pm$ 0.0 	&    &          & HAeBe \\ 
053937.37-695120.8  & 11.33 & II 	 & 1.3 $\pm$ 0.5 	&    &          & HAeBe \\ 
053956.92-695024.5  & 11.57 & II 	 & 2.8 $\pm$ 0.4        &    &  & \\ 
\cline{1-7}
054004.40-694437.6  &  5.40 & III 	 & 1.0 \nodata 	& PE & $\times$ & UCHII	      \\
054000.69-694713.4  &  6.89 & III 	 & 2.5 $\pm$ 0.5 	& PE & $\times$ & HAeBe \\
053945.18-694450.4  &  7.44 & III 	 & 1.9 $\pm$ 0.3 	& PE & $\times$ &       \\
053943.74-694540.3  &  9.91 & III(mul) 	 & 1.3 $\pm$ 0.6	&    &          & HAeBe \\
054003.55-694710.0  & 10.34 & III 	 & 1.0 $\pm$ 0.0        &    &  & \\

\enddata
\tablenotetext{a}{Classifications of UCHII, HAeBe, and maser are from \citet{IJC04}, \citet{Naetal05}, and \citet{Lazetal02}.}
\end{deluxetable}

\begin{deluxetable}{llcccllcc}
\rotate
\tabletypesize{\scriptsize}
\tablecolumns{9}
\tablecaption{Physical Properties of YSOs with Maser and UCHIIs \label{uch2}}
\tablewidth{0pc}
\tablehead{
 \colhead{} & 
 \colhead{} & 
 \colhead{} & 
 \colhead{Stage} & 
 \colhead{$M_\star$} & 
 \colhead{Spec.} &
 \colhead{Spec.\tablenotemark{a}} & 
 \colhead{$\dot{M}_{\rm env}$} &
 \colhead{$\dot{M}_{\rm crit}$\tablenotemark{b}} \\
 \colhead{Maser/UCHII} & 
 \colhead{YSO ID} & 
 \colhead{Type} & 
 \colhead{Range} & 
 \colhead{($M_\odot$)} & 
 \colhead{Type} & 
 \colhead{Type} &
 \colhead{($M_\odot$/yr)} & 
 \colhead{($M_\odot$/yr)} 
 }

\startdata

maser               &  053929.21$-$694719.0 & I   & I     & 26.1$\pm$1.5 & O7.5V  & \nodata & 5.1E-04$\pm$1.5E-04 & \nodata \\
B0540$-$6946(1)     &  054004.40$-$694437.6 & III  & I     & 19.7\tablenotemark{c} & O9 V 
 & O6 V   & 5.9E-03 & 3.2E-05 \\  
                    &                      &      & I     & 41.2$\pm$9.8\tablenotemark{d} & O6 V
 &        & 2.8E-03$\pm$2.0E-03 & \\ 
B0540$-$6946(4)     &  053937.56$-$694525.4 & I    & I     & 34.8$\pm$8.4 & O6 V
 & O5.5 V   & 2.3E-03$\pm$1.9E-03 & 6.8E-05 \\
B0540$-$6946(5)     &  053937.53$-$694609.8 & I/II & I     &  31.2 $\pm$ 2.9 & O7 V  & O7.5 V   & 2.2E-03$\pm$2.1E-03  & 4.1E-05 \\

\enddata
\tablenotetext{a}{The spectral type is determined from radio observations 
 \citep{IJC04}.}
\tablenotetext{b}{The critical infalling rate is adopted from \citet{CE02}.}
\tablenotetext{c}{Parameters are inferred from model fits to the SED composed 
 of all datapoints from $U$ to 70 \um.}
\tablenotetext{d}{Parameters are inferred from model fits to the SED composed 
 of {\it Spitzer} mid-IR datapoints.}
\end{deluxetable}

\begin{deluxetable}{l|cccccc}
\tabletypesize{\scriptsize}
\tablecolumns{8}
\tablecaption{Physical and Star Formation Properties of GMCs in N\,159 and N\,44\label{gmc}}
\tablewidth{0pc}
\tablehead{
 \multicolumn{1}{c}{} &
 \colhead{N\,159-E} & 
 \colhead{N\,159-W} & 
 \colhead{N\,159-S} &
 \colhead{N\,44-C} &
 \colhead{N\,44-S} &
 \colhead{N\,44-N} \\
 \multicolumn{1}{c}{} &
 \colhead{} & 
 \multicolumn{1}{c}{+N\,159-2} & 
 \multicolumn{1}{c}{+N\,159-3} &
 \colhead{} &
 \colhead{} &
 \colhead{} 
 }

\startdata

$V_{\rm lsr}$  (km~s$^{-1}$) & 234.1 & 236.1$\pm$2.4 & 233.0$\pm$1.6 & 282.5 & 279.4 & 283.8\\
$\Delta V$     (km~s$^{-1}$) &   7.6 &   5.7$\pm$0.3 &   7.7$\pm$0.8 & 7.2   &  15.8 &   3.8\\
Size           (pc)          &  19.2 &  15.8$\pm$ 1.0 & 19.5$\pm$2.4 & 86    &  136  &  \nodata \\
$M_{\rm vir}$  ($10^4 M_\odot$) & 17  &  16   & 36   &  37 & 210 & \nodata \\
$M_{\rm lum}$  ($10^4 M_\odot$) & $21\pm13$ & $26\pm17$ & $39\pm25$ & $46\pm13$ & $145\pm41$ & \nodata \\
$M_{\rm vir} / M_{\rm lum}$     & $0.8^{+1.3}_{-0.3}$ & $0.6^{+1.2}_{-0.2}$ & $0.9^{+1.7}_{-0.3}$ & $0.8^{+0.3}_{-0.2}$ & $1.4^{+0.6}_{-0.3}$ & \nodata \\ 
$t_{\rm ff}$  (Myr)             & 1.0 & 1.1 & 1.1 &  6.6 & 5.5 & \nodata \\
$N_{\rm YSO}$($M_{\rm u1}-M_{\rm u2}$)\tablenotemark{a}  & 3(41-15) & 6(35-18) &  4(8-5)  & 5(45-17) & 4(25-12) & 5(17-8) \\
$M_{\rm YSO}^{\rm total}$($M_{\rm u}-M_{\rm l}$)\tablenotemark{b}   ($M_\odot$) & 440$^{+330}_{-150}$(41-1) & 1380$^{+690}_{-290}$(35-1) & 200$^{+220}_{-70}$(8-1)  & 890$^{+490}_{-30}$(45-1) & 475$^{+70}_{-70}$(25-1) & 310$^{+25}_{-40}$(17-1)\\
SFE$_{\rm YSO}$       &  2.6$^{+2.0}_{-0.9}$E-3 & 8.6$^{+4.3}_{-1.8}$E-3 & 5.6$^{+6.1}_{-1.9}$E-4 & 2.4$^{+1.3}_{-0.1}$E-3 & 2.3$^{+0.3}_{-0.3}$E-4 & \nodata \\
SFR$_{\rm YSO}$ ($M_\odot$~yr$^{-1}$) & 4.4$^{+3.3}_{-1.5}$E-4 & 1.4$^{+0.7}_{-0.3}$E-3 & 2.0$^{+2.2}_{-0.7}$E-4 & 8.9$^{+4.9}_{-0.3}$E-4 & 4.8$^{+0.7}_{-0.7}$E-4 & 3.1$^{+0.3}_{-0.4}$E-4 \\
$\epsilon_{\rm ff}$    &  2.6$^{+2.0}_{-0.9}$E-3 & 8.6$^{+4.3}_{-1.8}$E-3 & 5.6$^{+6.1}_{-1.9}$E-4 & 1.6$^{+0.5}_{-0.2}$E-2 & 1.2$^{+0.2}_{-0.2}$E-4 & \nodata \\
\cline{1-7}  
$L$(\ha )$_{\rm obs}$ ($\times 10^{37}$ ergs~s$^{-1}$)  & 8.9$\pm$0.9 & 4.0$\pm$0.4 & 0.16$\pm$0.02 & 160$\pm$20  & 110$\pm$10 & 0.49$\pm$0.05 \\
$L$(24 \um)    ($\times 10^{38}$ ergs~s$^{-1}$)  & 46$\pm$5 & 28$\pm$3 & 0.55$\pm$0.06 & 23$\pm$2  & 14$\pm$1 & 1.8$\pm$0.2 \\
Aperture Radius ($''$) & 110 & 90 & 120 &  140 & 270 &  180 \\
SFR$_{H\alpha+24}$  ($M_\odot$~yr$^{-1}$) & 1.20$\pm$0.16E-3 & 6.57$\pm$0.96E-4 & 1.72$\pm$0.21E-5 & 1.21$\pm$0.11E-3 & 8.06$\pm$0.75E-4 & 5.46$\pm$0.65E-5 \\
$\Sigma_{{\rm H}_2}$ ($M_\odot$~pc$^{-2}$) &  35 & 62 & 62 & 59 & 42 & 25 \\
$\Sigma_{\rm HI}$ ($M_\odot$~pc$^{-2}$) & 90 & 99 & 117 & 54 & 72 & 45 \\
SFR$_{H\alpha+24}$ ($M_\odot$~yr$^{-1}$~kpc$^{-2}$) & 0.54$\pm$0.07 & 0.44$\pm$0.07 & 6.4$\pm$0.8E-3 & 0.33$\pm$0.04 & 0.060$\pm$0.005 & 9.1$\pm$1.1E-3 \\
SFR$_{\rm YSO}$ ($M_\odot$~yr$^{-1}$~kpc$^{-2}$) & 0.20$^{+0.15}_{-0.07}$ & 0.92$^{+0.41}_{-0.19}$ & 0.075$^{+0.083}_{-0.026}$ & 0.24$^{+0.13}_{-0.01}$ & 0.035$^{+0.005}_{-0.005}$ & 0.051$^{+0.004}_{-0.007}$ \\
SFR$_{\Sigma}$ ($M_\odot$~yr$^{-1}$~kpc$^{-2}$) & 0.22 & 0.32 & 0.36 & 0.19 & 0.19 & 0.10 \\

\enddata
\tablenotetext{a}{Number of YSOs with $\bar{M}_{\ast}$ in the mass range (u1-u2).}
\tablenotetext{b}{Total mass of YSOs extrapolated for the mass range (u-l).}
\end{deluxetable}

\end{document}